\documentclass[twocolumn,showpacs,amsmath,amssymb,superscriptaddress]{revtex4-1}
\usepackage{dcolumn}
\usepackage{color}
\usepackage{graphicx}
\usepackage{amsmath}
\usepackage{bm}

\begin{document}

\title{Microscopic description of octupole shape-phase transitions
in light actinides and rare-earth nuclei}

\author{K.~Nomura}
\affiliation{Grand Acc\'el\'erateur National d'Ions Lourds,
CEA/DSM-CNRS/IN2P3, B.P.~55027, F-14076 Caen Cedex 5, France}

\author{D.~Vretenar}
\author{T.~Nik\v si\'c}
\affiliation{Physics Department, Faculty of Science, University of
Zagreb, 10000 Zagreb, Croatia}

\author{Bing-Nan Lu}
\affiliation{Institut f\"ur Kernphysik, Institute for Advanced Simulation, and J\"ulich
Center for Hadron Physics, Forschungszentrum J\"ulich, D-52425 J\"ulich,
Germany}

\date{\today}

\begin{abstract}
A systematic analysis of low-lying quadrupole and
 octupole collective states is presented, based on the microscopic energy density
 functional framework. By mapping the deformation constrained self-consistent
 axially symmetric mean-field energy surfaces onto the equivalent Hamiltonian of the
 $sdf$ interacting boson model (IBM), that is, onto the energy expectation value in the
 boson condensate state, the Hamiltonian parameters are determined. The
 study is based on the global relativistic energy density functional
 DD-PC1. 
The resulting IBM Hamiltonian is used to calculate excitation spectra
 and transition rates for the positive- and negative-parity collective
 states 
in four isotopic chains characteristic for two regions
of octupole deformation and collectivity: Th, Ra, Sm and Ba. 
Consistent with the empirical trend, the microscopic calculation 
based on the systematics of
 $\beta_{2}$-$\beta_{3}$ energy maps, the resulting low-lying
 negative-parity bands and transition rates show evidence of a shape transition 
 between stable octupole 
 deformation and octupole vibrations characteristic for $\beta_{3}$-soft
 potentials. 
 
\end{abstract}



\maketitle

\section{Introduction}

The study of equilibrium shapes and shape transitions presents a 
recurrent theme in nuclear structure physics \cite{BM,RS,CasBook}. 
Even though most deformed medium-heavy and heavy nuclei exhibit 
quadrupole, reflection-symmetric equilibrium shapes, there are 
regions in the mass table 
where octupole deformations (reflection-asymmetric, pear-like shapes) 
occur (see \cite{butler96} for a review). In particular, nuclei 
with neutron (proton) number $N$ ($Z$) $\approx 34$, 56, 88 and 134
\cite{butler96}. 
Reflection-asymmetric shapes are characterized by the presence 
of negative-parity bands, and by pronounced electric dipole and octupole
transitions. In the case of static octupole deformation, for instance,
the lowest positive-parity even-spin states and the negative-parity
odd-spin states form an alternating-parity band, with states connected
by the enhanced E1 transitions. 

In a simple microscopic picture octupole deformation is expected
to develop through a coupling of orbitals in the
vicinity of the Fermi surface with quantum numbers ($l$,
$j$) and an intruder unique-parity orbital with ($l+3$, $j+3$)
\cite{butler96}. 
For instance, in the case of heavy ($Z\approx 88$ and $N\approx 134$)
nuclei in the region of light actinides, the coupling of the neutron
orbitals $g_{9/2}$ and $j_{15/2}$, and that of the proton single-particle
states $f_{7/2}$ and $i_{13/2}$,  can lead to octupole mean-field
deformations. 
Structure phenomena related to reflection-asymmetric nuclear shapes 
have been extensively investigated in numerous experimental studies 
\cite{butler91,butler96}. In particular, very recently clear evidence
for pronounced octupole deformation in the region $Z\approx 88$ and $N\approx 134$, 
e.g. in $^{220}$Rn and $^{224}$Ra, has been reported in
Coulomb excitation experiments with radioactive ion beams
\cite{gaffney13}. 
Also in the rare-earth region ($Z\approx 56$ and $N\approx 88$), 
a recent experiment \cite{garrett09} has revealed low-energy
negative-parity bands in $^{152}$Sm. 
The renewed interest in studies of reflection asymmetric nuclear shapes  
using accelerated radioactive beams \cite{gaffney13} point to the 
significance of a timely systematic theoretical analysis of 
quadrupole-octupole collective states of nuclei in several mass regions
of the nuclear chart where octupole shapes are expected to occur.  

A variety of theoretical methods have been applied to studies 
of reflection asymmetric shapes and the evolution of the corresponding
negative-parity collective states. 
These include self-consistent mean-field models 
\cite{marcos83,naza84b,naza85,bonche86,bonche88,egido91,robledo10,robledo11,rayner12,robledo13}, 
algebraic (or interacting boson) models 
\cite{scholten78,engel87,taka88,kusnezov88,cottle98}, phenomenological
collective models 
\cite{bizzeti04,bonatsos05,lenis06,bizzeti08,bizzeti10,jolos12,minkov12,bizzeti13},
and cluster models \cite{iachello82,daley86a,shneidman02}. In particular,  
a great number of self-consistent mean-field calculations 
of nuclei with static or dynamic octupole deformations have been reported, e.g., based on the Nilsson-Strutinsky method \cite{naza84b}, 
Skyrme \cite{bonche86,bonche88} and Gogny 
\cite{egido91,robledo10,robledo11,rayner12,robledo13} effective
interactions, and relativistic mean field (RMF) models
\cite{zhang10,lu12,lu14}.

Nuclear energy density functionals (EDFs) enable a complete and accurate
description of ground-state properties and collective excitations over
the whole nuclide chart \cite{ben03rev}. 
Both non-relativistic \cite{Skyrme,VB,Gogny} and relativistic
\cite{Vre05,Nik11rev} EDFs have been successfully applied to the
description of the evolution of single-nucleon shell-structures and the
related shape-transition and shape-coexistence phenomena. 
To compute excitation spectra and transition rates, however, the
EDF framework has to be extended to take into account the restoration of
symmetries broken in the mean-field approximation, and
fluctuations in the collective coordinates. 
In this study we employ a recently developed method \cite{Nom08} for
determining the Hamiltonian of the interacting boson model (IBM) \cite{IBM},
starting with a microscopic, EDF-based self-consistent mean-field calculation of 
deformation energy surfaces. 
By mapping the deformation constrained self-consistent
energy surfaces onto the equivalent Hamiltonian of the
IBM, that is, onto the energy expectation value in the
boson condensate state, the Hamiltonian parameters are determined. The 
resulting IBM Hamiltonian is used to calculate excitation spectra and transition 
rates. This technique has been extended and applied to study moments of
inertia of deformed rotational nuclei \cite{Nom11rot}, to analyze
the $\gamma$-softness in medium-heavy and heavy nuclei \cite{Nom12tri}, and
to describe coexistence and mixing of different intrinsic shapes \cite{Nom12sc}. 
More recently the method of \cite{Nom08} has been applied to a study 
of the octupole shape-phase transition in the Th isotopic chain
\cite{nom13oct}. 

This work extends the study of Ref.~\cite{nom13oct} and presents  
a microscopic analysis of octupole
shape transitions in four isotopic chains characteristic for two regions
of octupole deformation and collectivity: Th, Ra, Sm and Ba. 
The study is based on the microscopic 
framework of nuclear energy density functionals (EDFs), and the
Hamiltonian of the $sdf$ IBM \cite{arima78,engel87} 
is determined from axial quadrupole-octupole 
deformation energy surfaces calculated employing the relativistic 
Hartree-Bogoliubov model based on the universal energy density functional
DD-PC1 \cite{DDPC1}. The mapped $sdf$ IBM Hamiltonian is used to calculate 
low-energy spectra and transition rates for both positive- and
negative-parity states of the four sequences of isotopes. 
The semi-microscopic relativistic functional DD-PC1 was adjusted to 
the experimental masses of a set of 64 deformed nuclei in the mass regions 
$A \approx 150-180$ and $A \approx 230-250$, and 
further tested in a number of mean-field and beyond-mean-field calculations 
in different mass regions.

The article is organized as follows: in Sec.~\ref{sec:model} we describe
the theoretical framework. 
The systematics of self-consistent mean-field results and the quality of
the mapping onto the IBM are described in Sec.~\ref{sec:mapping}. 
Spectroscopic properties calculated with the mapped $sdf$ IBM
Hamiltonian are discussed in Sec.~\ref{sec:results}, including the
systematics of low-lying 
positive- and negative-parity states, E1, E2 and E3 transitions, and 
detailed level schemes of selected nuclei. Section~\ref{sec:summary}
contains the summary and concluding remarks.

\section{Description of the model \label{sec:model}}

The analysis starts by performing
constrained self-consistent relativistic mean-field calculations for 
axially symmetric shapes in the ($\beta_{2}$,$\beta_{3}$) plane, with 
constraints on the mass quadrupole $Q_{20}$ and octupole $Q_{30}$ moments. 
The dimensionless shape variables $\beta_{\lambda}$ ($\lambda=2,3$) are defined in terms of 
the multipole moments $Q_{\lambda 0}$: 
\begin{eqnarray}
\label{eq:beta}
\beta_{\lambda}=\frac{4\pi}{3AR^{\lambda}}
Q_{\lambda 0} 
\end{eqnarray}
with $R=1.2A^{1/3}$ fm. 
The relativistic Hartree-Bogoliubov (RHB) model \cite{Vre05} is used to calculate 
constrained energy surfaces, the 
functional in the particle-hole channel is DD-PC1, and pairing correlations are taken into account 
by employing an interaction that is separable in momentum space, and is completely determined 
by two parameters adjusted to reproduce the empirical bell-shaped pairing gap in symmetric 
nuclear matter \cite{tian09,Nik11rev}. 
For technical details of the calculation of self-consistent RHB energy
surfaces, the reader is referred to Refs.~\cite{lu12,lu14}. 
In this work the energy surface in RHB calculations refers to the
total energy of the nuclear system as a function of deformation
parameters. 

A quantitative study of low-lying quadrupole and octupole collective states must go 
beyond a simple mean-field 
calculation of energy surfaces and take into account collective correlations 
related to restoration of symmetries broken by the mean field and 
quantum fluctuations in deformation variables. 
To this end we employ the interacting boson model (IBM) \cite{IBM} to 
analyze spectroscopic properties associated to both quadrupole and
octupole collective degrees of freedom. 
The building blocks of the most standard version of the IBM that includes
quadrupole degrees of freedom are the monopole $s$ and quadrupole $d$ bosons,  
that correspond to $J^{\pi}=0^{+}$ and $2^{+}$ collective pairs of
valence nucleons, respectively \cite{OAI}. 
To describe reflection-asymmetric deformations and the corresponding
negative-parity states, in addition to these positive-parity bosons the 
model space must include the octupole ($J^{\pi}=3^{-}$)
$f$-boson \cite{arima78,engel87}.  

A general IBM Hamiltonian of the $sdf$ system contains interaction terms
acting in the $sd$ and $f$ boson spaces, and the coupling between the
two spaces \cite{arima78}: 
\begin{eqnarray}
\label{eq:ham}
 \hat H=\hat H_{sd}+\hat H_{f}+\hat H_{sdf}. 
\end{eqnarray}
In the present analysis we employ the
following terms: 
\begin{eqnarray}
\label{eq:sd}
 \hat H_{sd}=\epsilon_{d}\hat n_{d}+\kappa_{2}\hat Q_{sd}\cdot\hat
  Q_{sd}+\alpha\hat L_{d}\cdot\hat L_{d}
\end{eqnarray}
with $\hat n_{d}=d^{\dagger}\cdot\tilde d$, $\hat Q_{sd}=s^{\dagger}\tilde
d+d^{\dagger}s+\chi_{d}[d^{\dagger}\times\tilde
d]^{(2)}$ and $\hat L_{d}=\sqrt{10}[d^{\dagger}\times\tilde d]^{(1)}$
denoting the $d$-boson number operator, the quadrupole operator, and the
angular momentum operator for the $sd$
boson space, respectively. 
\begin{eqnarray}
\label{eq:f}
 \hat H_{f}=\epsilon_{f}\hat n_{f}+\kappa_{2}^{\prime}\hat Q_{f}\cdot\hat Q_{f}
\end{eqnarray}
$\hat n_{f}=f^{\dagger}\cdot\tilde f$ is the $f$-boson
number operator, and $\hat Q_{f}=\chi_{f}[f^{\dagger}\times\tilde f]^{(2)}$ denotes 
the quadrupole $f$-boson interaction. Finally, 
\begin{eqnarray}
\label{eq:sdf}
 \hat H_{sdf}=\kappa_{2}^{\prime\prime}\hat Q_{sd}\cdot\hat Q_{f}+\kappa_{3}:\hat V_{3}^{\dagger}\cdot\hat V_{3}:
\end{eqnarray}
where the last term is the octupole-octupole interaction written in the normal-ordered form with $\hat
V_{3}^{\dagger}=s^{\dagger}\tilde f+\chi_{3}[d^{\dagger}\times\tilde
f]^{(3)}$. 
In the present calculation
$\kappa_{2}^{\prime}=\kappa_{2}^{\prime\prime}/2=\kappa_{2}$. 
The Hamiltonians of Eqs.~(\ref{eq:sd}) and (\ref{eq:f}) have been
used in a number of phenomenological IBM studies of low-energy quadrupole and
octupole collective states. 
The coupling Hamiltonian Eq.~(\ref{eq:sdf}) is similar to
the one used in Ref.~\cite{barfield88}, and can be derived from a microscopic 
octupole-octupole interaction between neutron and the proton bosons by
mapping the totally-symmetric state of the IBM-2 system to the
corresponding state in the IBM-1 \cite{barfield88}. 
In this work, however, the dipole-dipole interaction term $\hat
L_{d}\cdot\hat L_{f}$ (with $\hat
L_{f}=\sqrt{28}[f^{\dagger}\times\tilde f]^{(1)}$) is not included, as
it has been shown \cite{cottle98} to be of little relevance for
low-lying collective states. 
In contrast to previous $sdf$ IBM phenomenological studies, in which the
maximum number of $f$ bosons $N_{f}^{max}$ has been kept constant to reduce
the model space and thus computing time ($N_{f}^{max}=1$ in most cases),
in the present analysis both positive- and negative-parity bosons are
treated in the same way, that is, there is no truncation specific for  
the $f$-boson number. The total number of bosons $s+d+f$ is determined
by the number of valence proton and neutron pairs.

We also note that some empirical IBM studies suggested the importance of
including the $p$ ($J^{\pi}=1^-$) boson to improve the agreement with 
experimental results (for instance, 
\cite{taka86,taka88,zamfir01}), particularly for the E1 properties. 
From the algebraic point of view it was 
shown \cite{engel87} that the $p$-boson is necessary to obtain the SU(3)
dynamical symmetry of the U(16) group. 
On the other hand, the microscopic origin of the $p$-boson has been related either to the 
spurious center-of-mass motion \cite{engel87}, or to the giant dipole
resonance \cite{sugita96}. 
Here we do not include the $p$-boson in the model space, as in fact neither 
of these degrees of freedom are particularly relevant for an analysis
based on microscopic mean-field calculations. 

For each nucleus the parameters of the Hamiltonian: $\epsilon_{d}$,
$\epsilon_{f}$, $\kappa_{2}$, $\kappa_{3}$, $\chi_{d}$, $\chi_{f}$ and
$\chi_{3}$, are determined from the  microscopic RHB energy surfaces
employing the procedure of Ref.~\cite{Nom08,Nom10}: the microscopic
constrained self-consistent mean-field energy surface is mapped onto the
equivalent IBM energy surface, that is, on the expectation value of the
IBM Hamiltonian $\langle\phi|\hat H|\phi\rangle$ in the boson 
condensate state $|\phi\rangle$ \cite{GK}: 
\begin{eqnarray}
\label{eq:coherent}
 |\phi\rangle=\frac{1}{\sqrt{N_{B}!}}(\lambda^{\dagger})^{N_{B}}|-\rangle
\quad{\textnormal{with}}\quad
\lambda^{\dagger}=s^{\dagger}+\bar\beta_{2}d^{\dagger}_{0}+\bar\beta_{3}f^{\dagger}_{0}\;. \nonumber \\
\end{eqnarray}
$N_{B}$ and $|-\rangle$ denote the total number of bosons, that is, half the number of
valence nucleons \cite{OAI}, and the boson vacuum (inert core),
respectively. 
$\bar\beta_{2}$ and $\bar\beta_{3}$ represent the axial quadrupole and octupole 
shape variables in the boson system, respectively, equivalent to the deformation parameters 
of Eq.~(\ref{eq:beta}) that are used to characterize the RHB energy
surfaces. However, since the model spaces of the RHB and the IBM are
different, one finds that generally $\bar\beta_{\lambda}$ is also
different from $\beta_{\lambda}$ \cite{GK}. 
For $\lambda=2$, the proportionality 
$\bar\beta_{2}\equiv C_{2}\beta_{2}$, with $C_2$ being a coefficient,
appears to be a good approximation \cite{GK,Nom08}. 
Here we further assume that, independently of the $\lambda=2$
deformation, a similar relation also holds for $\lambda=3$, that is, 
$\bar\beta_{3}\equiv C_{3}\beta_{3}$. 
The coefficients $C_{\lambda=2,3}$ are determined basically by adjusting
the location of the minimum on the ($\beta_2,\beta_3$) energy surface. 
In the present study, the doubly-magic nuclei $^{208}$Pb and $^{132}$Sn
are taken as inert 
cores for the considered regions of Th-Ra and Sm-Ba nuclei, respectively. 
Therefore, $N_{B}$ takes values from 6 to 12, from 5 to 10, from 7 to
12, and from 4 to 9 for $^{220-232}$Th, $^{218-228}$Ra, $^{146-156}$Sm
and $^{140-150}$Ba, respectively. 
The analytic expression for the expectation value of the IBM Hamiltonian
reads: 
\begin{eqnarray}
\label{eq:oct.pes}
 E(\bar\beta_{2},
  \bar\beta_{3})
&=&
\frac{N_{B}}{1+\bar\beta_{2}^{2}+\bar\beta_{3}^{2}}
\Big(
\epsilon_{s}^{\prime}+
\epsilon_{d}^{\prime}\bar\beta_{2}^{2}+\epsilon_{f}^{\prime}\bar\beta_{3}^{2}
\Big)
\nonumber \\
&&
+\frac{N_{B}(N_{B}-1)}{(1+\bar\beta_{2}^{2}+\bar\beta_{3}^{2})^2}\times
\nonumber \\
&&\Big[
\kappa_{2}\Big(
2\bar\beta_{2}-\sqrt{\frac{2}{7}}\chi_{d}\bar\beta_{2}^{2}-\frac{2}{\sqrt{21}}\chi_{f}\bar\beta_{3}^{2}
\Big)^{2} \nonumber \\
&&+4\kappa_{3}
\Big(\bar\beta_{3}-\frac{2}{\sqrt{15}}\chi_{3}\bar\beta_{2}\bar\beta_{3}
\Big)^{2}
\Big],
\end{eqnarray}
with
\begin{eqnarray}
\label{eq:eps-prime}
&&\epsilon_{s}^{\prime}=5\kappa_{2},\quad
 \epsilon_{d}^{\prime}=\epsilon_{d}+6\alpha+(1+\chi_{d}^{2})\kappa_{2}
\nonumber \\
&&{\textnormal{and}}
\quad
\epsilon_{f}^{\prime}=\epsilon_{f}+\frac{5}{7}\kappa_{2}\chi_{f}^{2}.
\end{eqnarray}
By equating the energy expectation value as a function of the 
quadrupole and octupole deformation parameters 
to the microscopic energy surface in the
neighborhood of the energy minimum (typically up to 2 MeV from the minimum), 
the IBM Hamiltonian 
parameters can be determined without invoking any phenomenological
adjustment to experiment. 

The coupling constant of the $\hat L_{d}\cdot\hat L_{d}$ term, 
$\alpha$, is adjusted separately in such a way
that the cranking moment of inertia calculated in the boson intrinsic
state on the $\beta_{3}=0$ axis becomes identical to the corresponding one obtained from the
mean-field model \cite{Nom11rot}.

The resulting $sdf$ IBM Hamiltonian is diagonalized employing the code
OCTUPOLE \cite{OCTUPOLE}. 
This generates the excitation spectra and, subsequently, electromagnetic
transition rates, that is, the reduced transition  
probabilities $B$(E$\lambda ;J\rightarrow J^{\prime}$): 
\begin{eqnarray}
 B({\textnormal{E}}\lambda ;J\rightarrow J^{\prime})=\frac{1}{2J+1}|\langle
  J^{\prime}||\hat T^{({\textnormal{E}}\lambda)}||J\rangle|^{2}, 
\end{eqnarray}
with $|J\rangle$ ($|J^{\prime}\rangle$) being the wave function for
the initial (final) state with spin $J$ ($J^{\prime}$). 
The E1, E2, and E3 operators, denoted here $\hat
T^{(\textnormal{E1})}$, $\hat T^{(\textnormal{E2})}$ and $\hat
T^{(\textnormal{E3})}$, respectively, are defined as follows:
\begin{eqnarray}
\label{eq:E1}
 \hat T^{(\textnormal{E1})}=e_{1}(d^{\dagger}\times \tilde
  f+f^{\dagger}\times\tilde d)^{(1)}, 
\end{eqnarray}
\begin{eqnarray}
\label{eq:E2}
 \hat T^{(\textnormal{E2})}=e_{2}\hat Q, 
\end{eqnarray}
where $\hat Q=\hat Q_{sd}+\hat Q_{f}$, and the parameters
$\chi_{d}$ and $\chi_{f}$ are consistently the same as the ones used in the
Hamiltonian and, finally,  
\begin{eqnarray}
\label{eq:E3}
 \hat T^{(\textnormal{E3})}=e_{3}(\hat V^{\dagger}_3+\hat V_3). 
\end{eqnarray}
The operator $\hat V^{\dagger}_3$ is defined in Eq.~(\ref{eq:sdf}), and
again the same value for $\chi_{3}$ is consistently used in the
Hamiltonian and E3 operator. 
$e_{1}$, $e_{2}$ and $e_{3}$ denote the corresponding effective charges. 
Their values are $e_{1}=0.01$ $e$b$^{1/2}$
(taken from \cite{babilon05}) for all nuclei considered, $e_{2}=0.19$ $e$b
(from \cite{taka88}) and 0.13 $e$b (from \cite{scholten78}) for Th-Ra
and Sm-Ba isotopes, respectively. The values $e_{3}=0.15$ 
$e$b$^{3/2}$ for Th-Ra and $e_{3}=0.09$ $e$b$^{3/2}$ for Sm-Ba isotopes are 
adjusted to reproduce the overall trend of experimental results. 

From the calculated $B(\textnormal{E}\lambda;J\rightarrow J^{\prime})$ 
values, one obtains the transition intrinsic moments
$Q_{\lambda}(J\rightarrow J^{\prime})$: 
\begin{eqnarray}
\label{eq:qt}
Q_{\lambda}(J\rightarrow
 J^{\prime})=\sqrt{\frac{16\pi}{2\lambda+1}\frac{B(\textnormal{E}\lambda;J\rightarrow
 J^{\prime})}{(J\lambda 00|J^{\prime}0)^{2}}}\; , 
\end{eqnarray}
where $(J\lambda 00|J^{\prime}0)$ denotes the Clebsch-Gordan coefficient.

\section{Mapping the self-consistent mean-field results onto the IBM space\label{sec:mapping}}

\subsection{Systematics of deformation energy surfaces \label{sec:pes}}

\begin{figure*}[ctb!]
\begin{center}
\begin{tabular}{ccc}
\includegraphics[width=5.7cm]{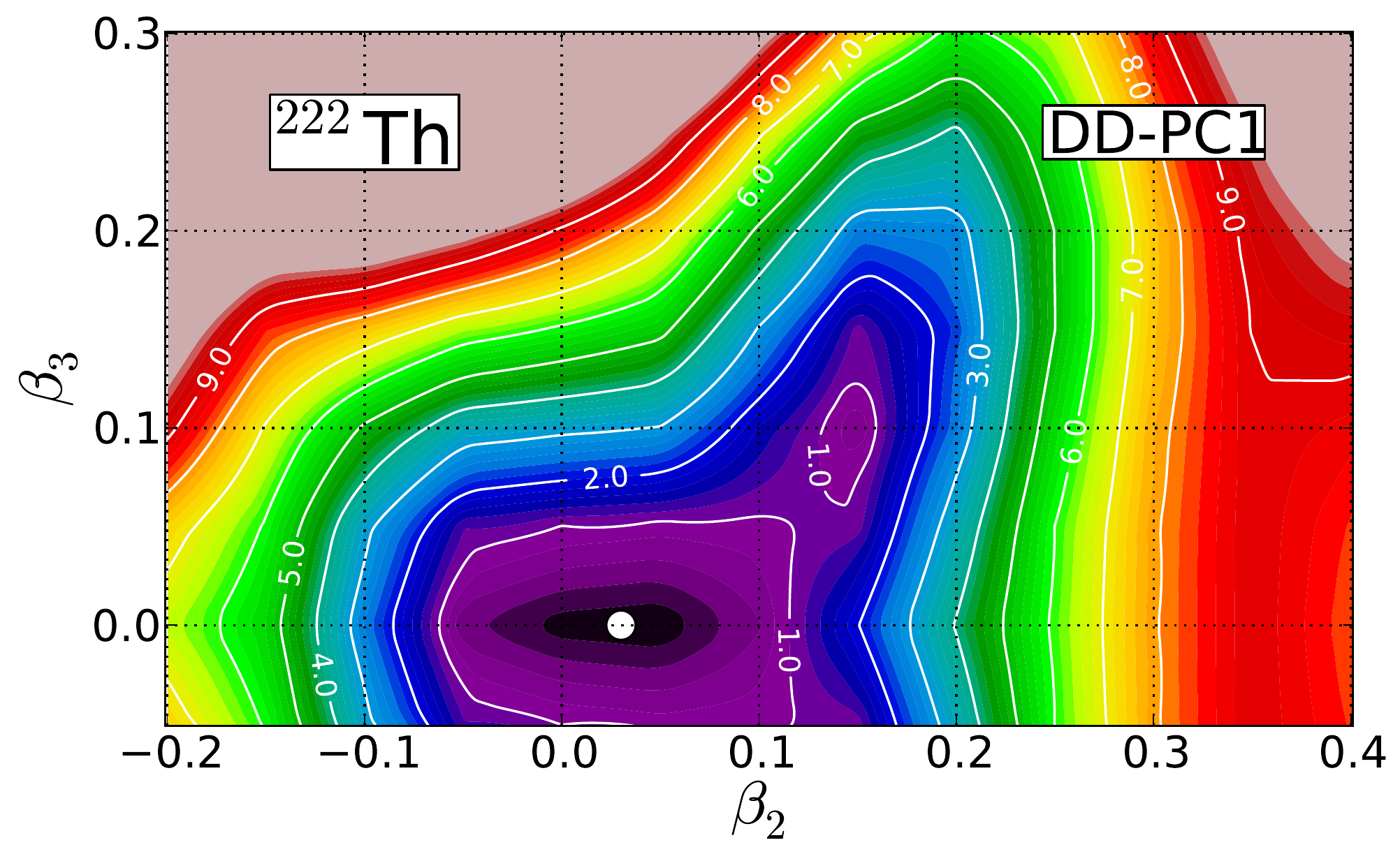} &
\includegraphics[width=5.7cm]{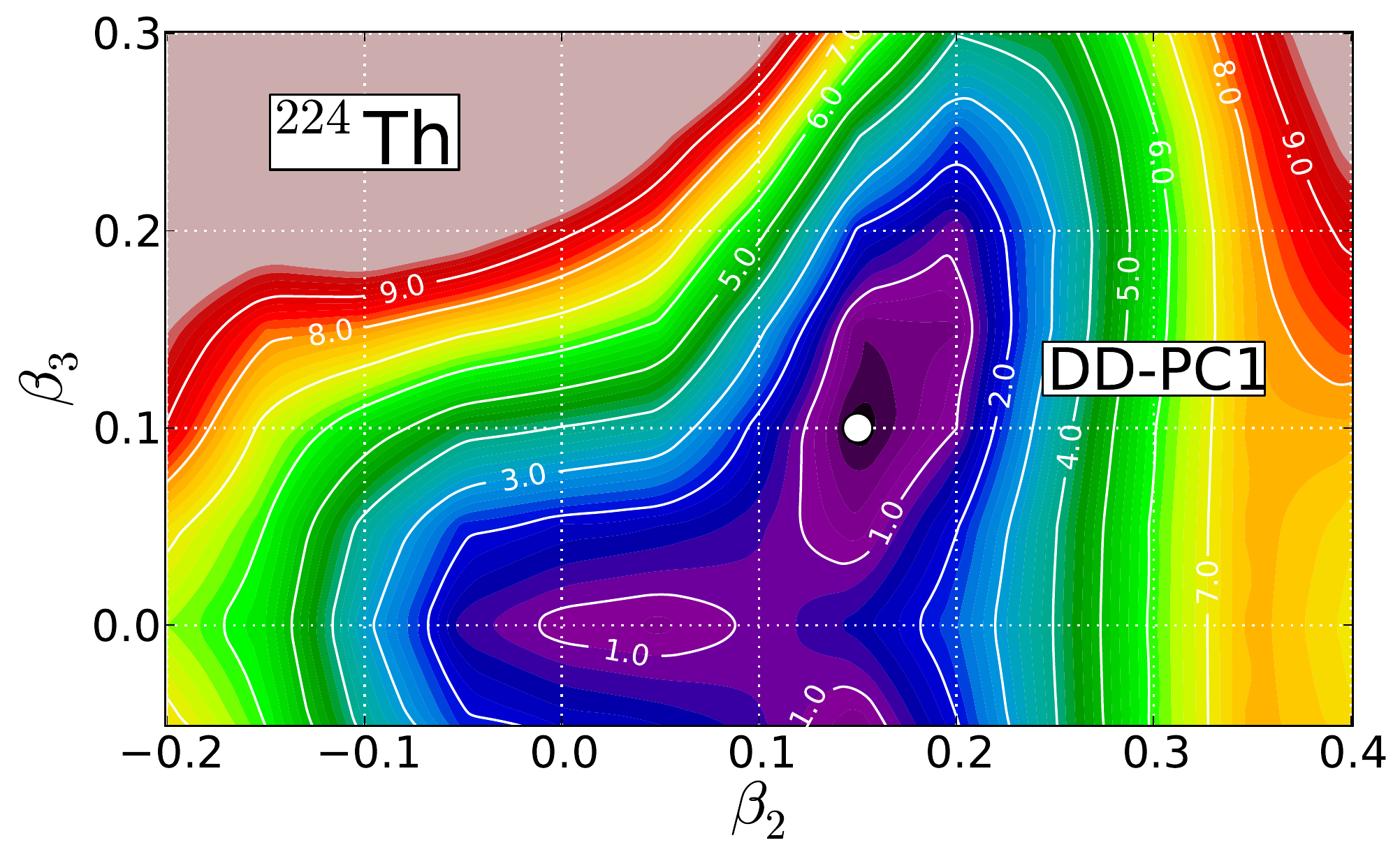} &
\includegraphics[width=5.7cm]{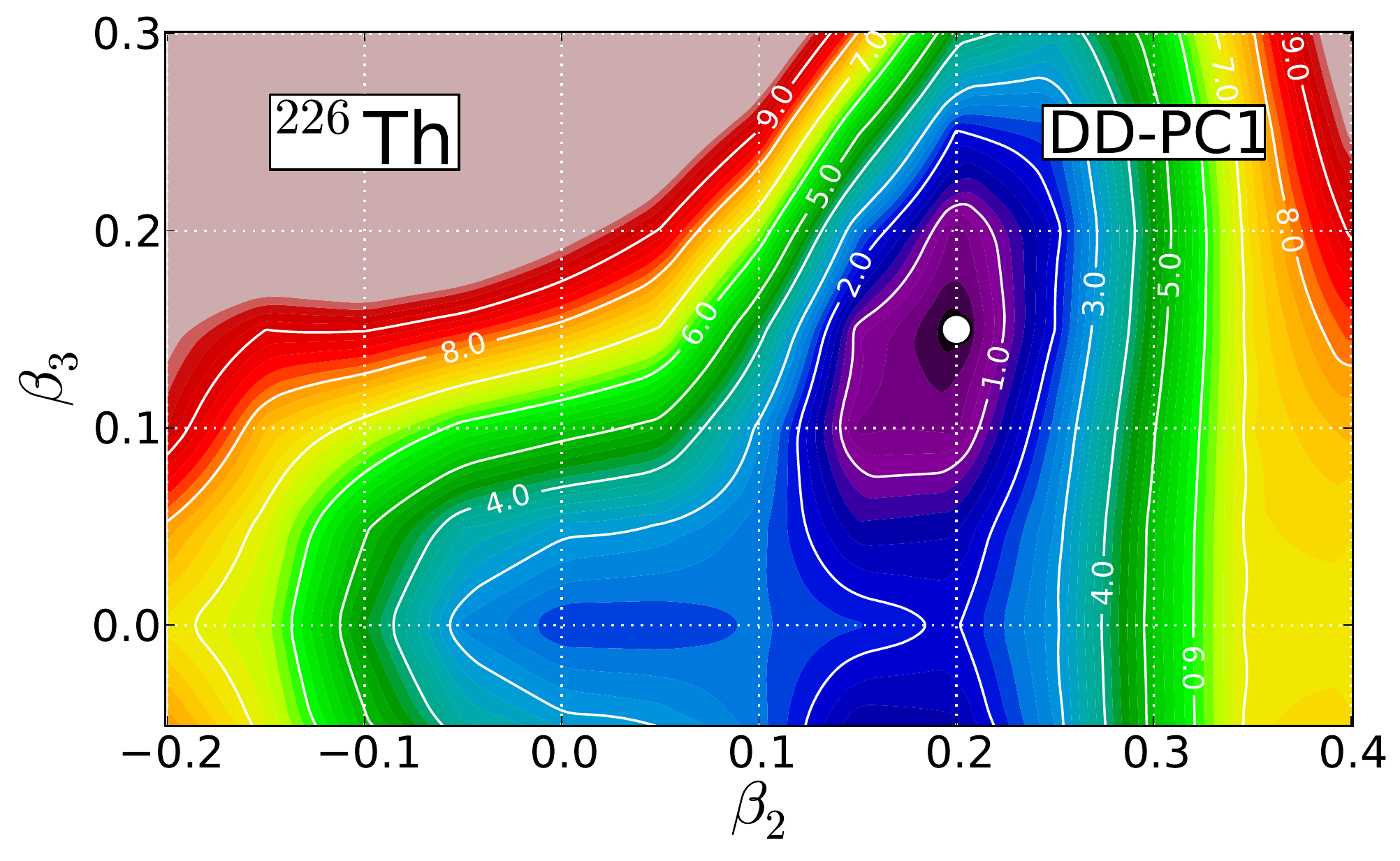} \\
\includegraphics[width=5.7cm]{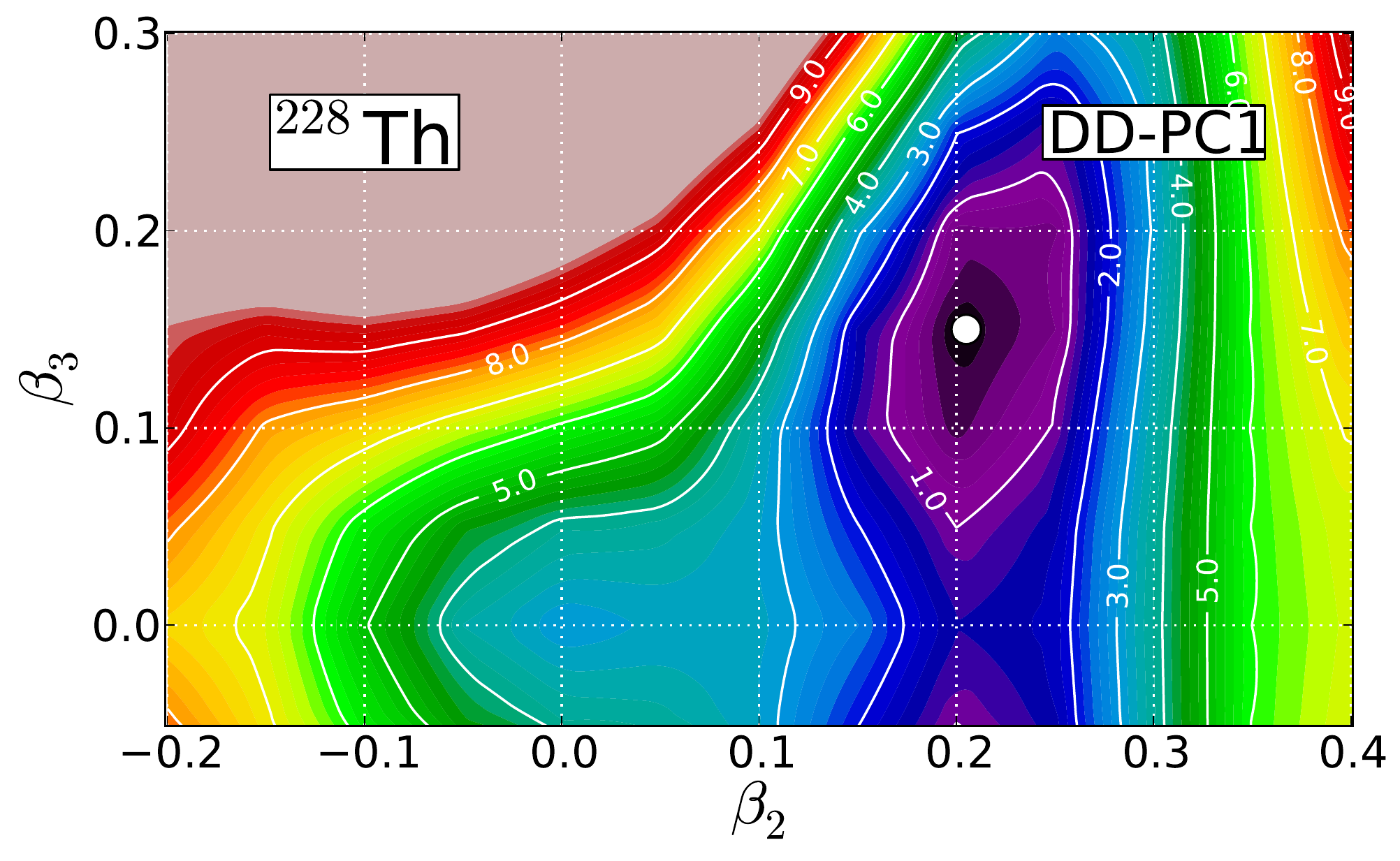} &
\includegraphics[width=5.7cm]{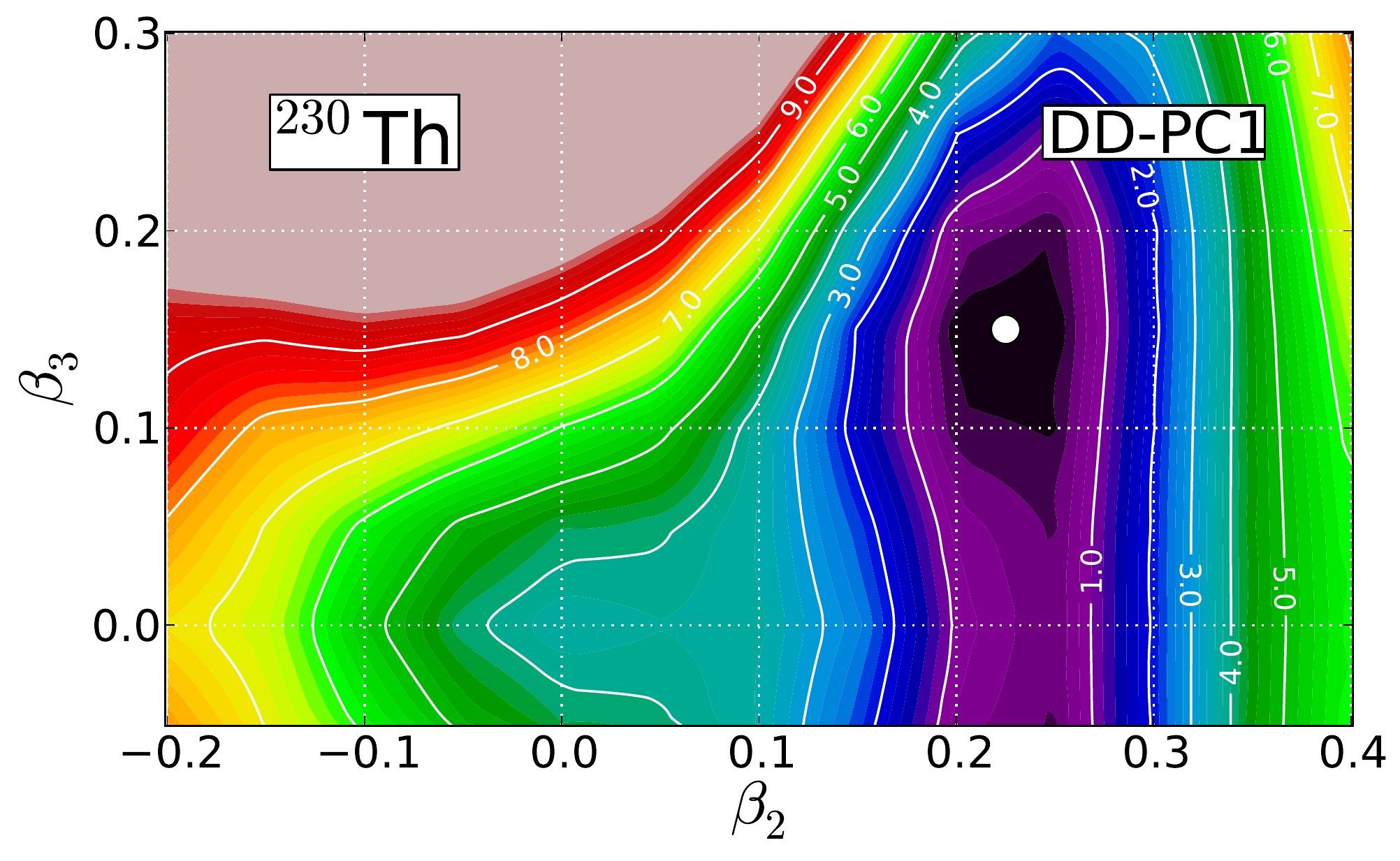} &
\includegraphics[width=5.7cm]{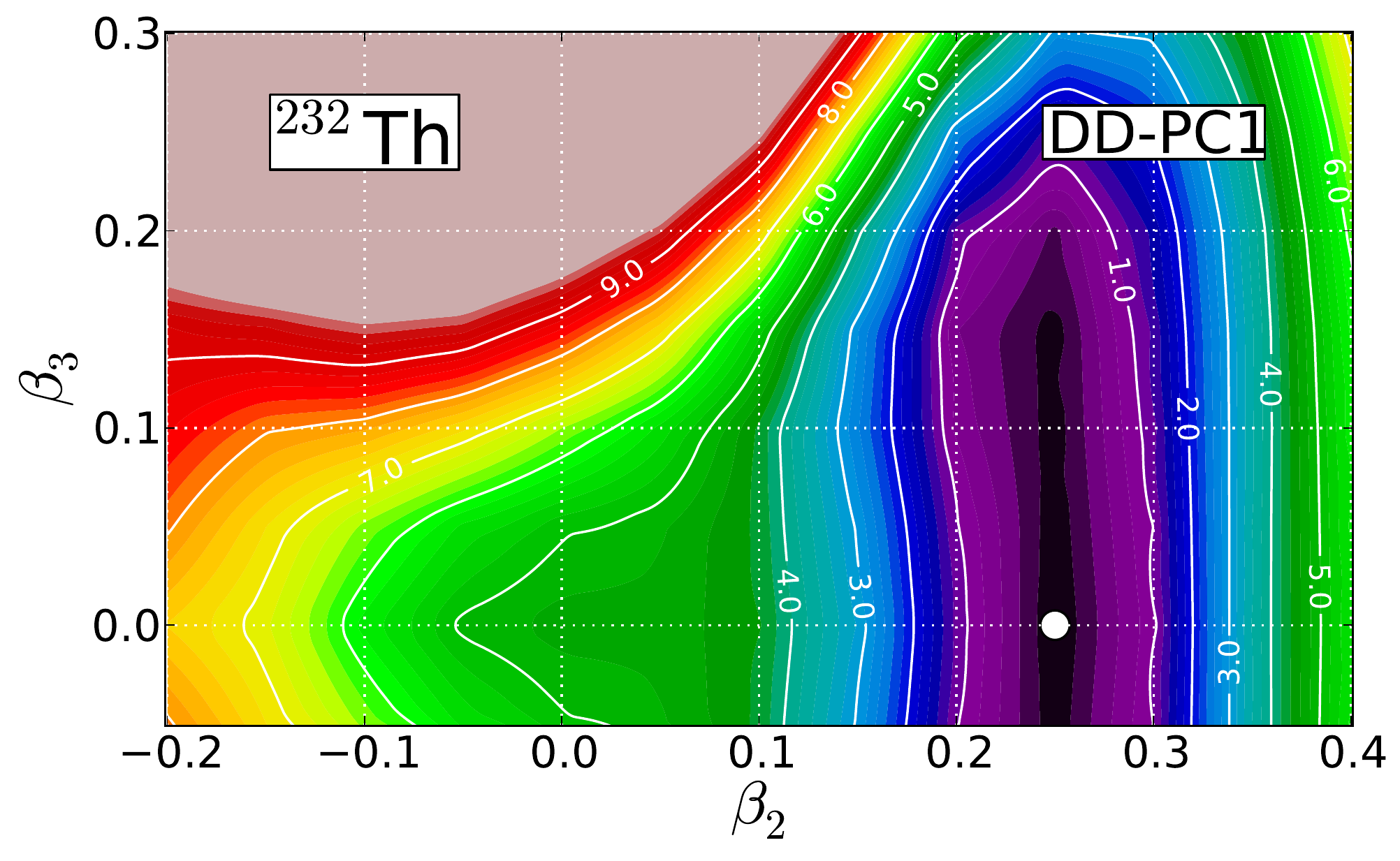} \\
\end{tabular}
\caption{(Color online) Axially symmetric energy surfaces of the isotopes
 $^{222-232}$Th in the $(\beta_{2},\beta_{3})$ 
 plane, calculated using the self-consistent RHB model with the 
 microscopic functional DD-PC1. The contours join points on the surface
 with the same energy (in MeV), and the color scale varies in steps of 0.2 MeV. 
The energy difference between neighboring contours is 1 MeV.
Note that energy surfaces are symmetric with respect to the 
 $\beta_{3}=0$ axis. Open circles denote the absolute energy minima. }
\label{fig:pes_th}
\end{center}
\end{figure*}

\begin{figure*}[ctb!]
\begin{center}
\begin{tabular}{ccc}
\includegraphics[width=5.7cm]{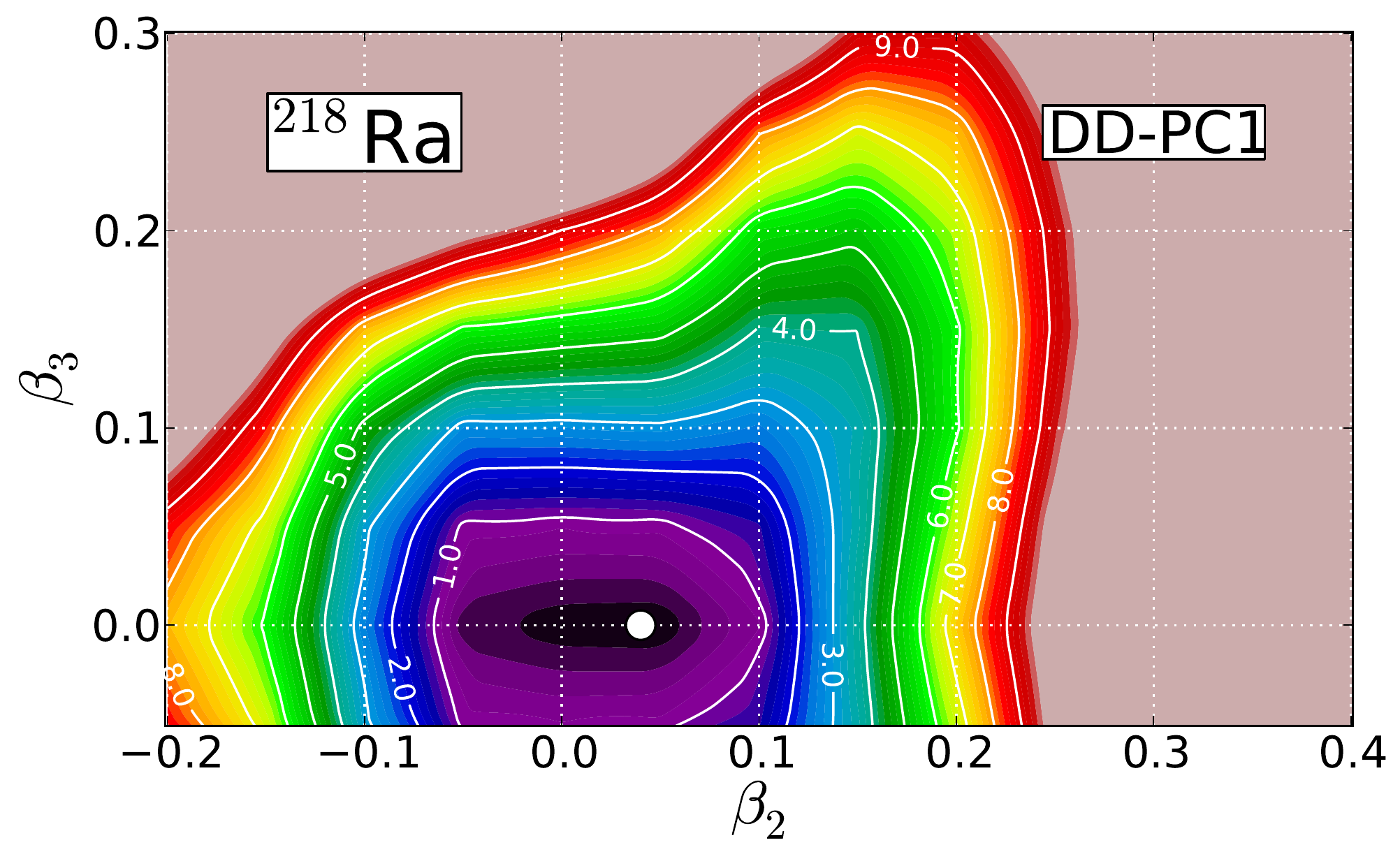} &
\includegraphics[width=5.7cm]{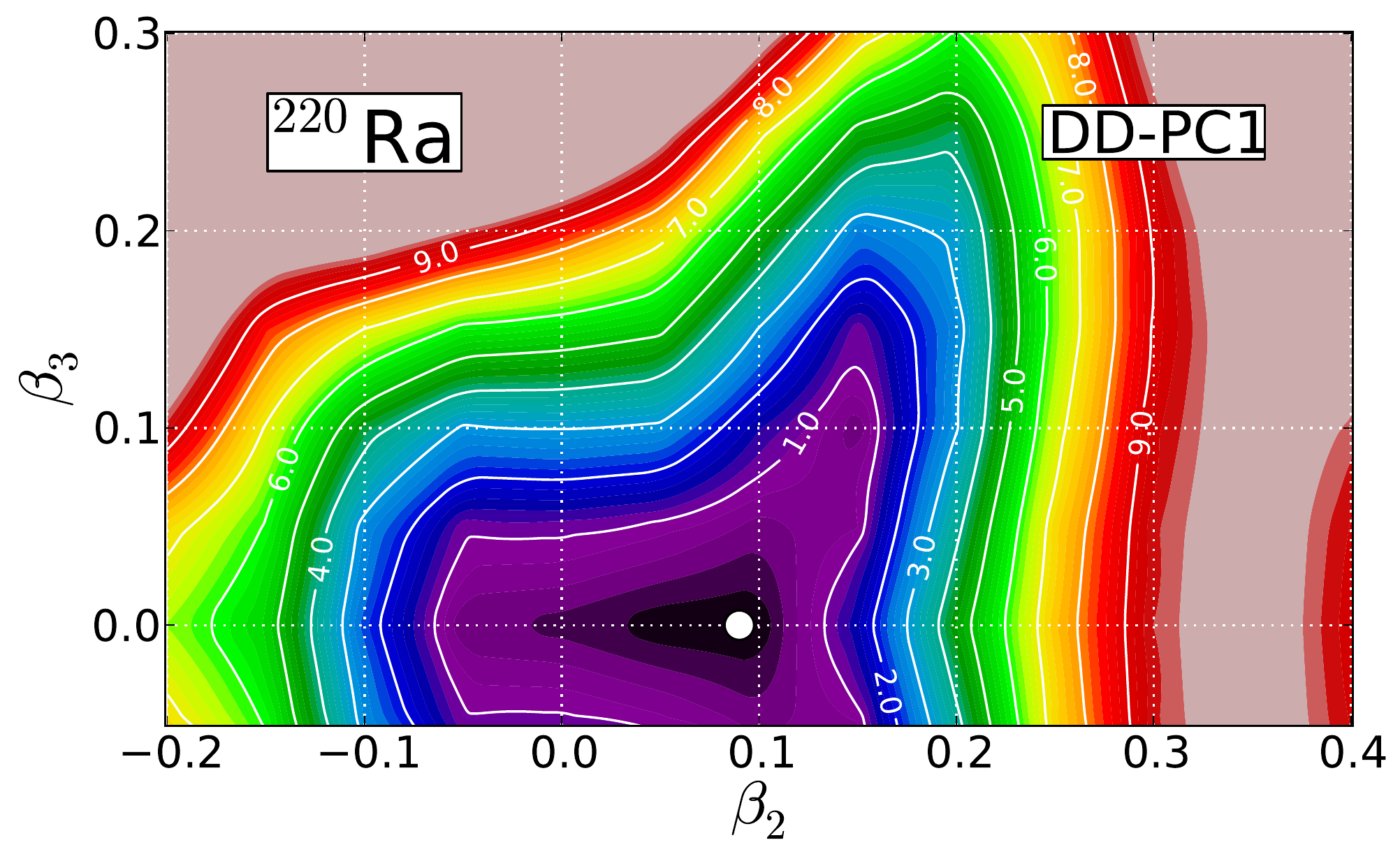} &
\includegraphics[width=5.7cm]{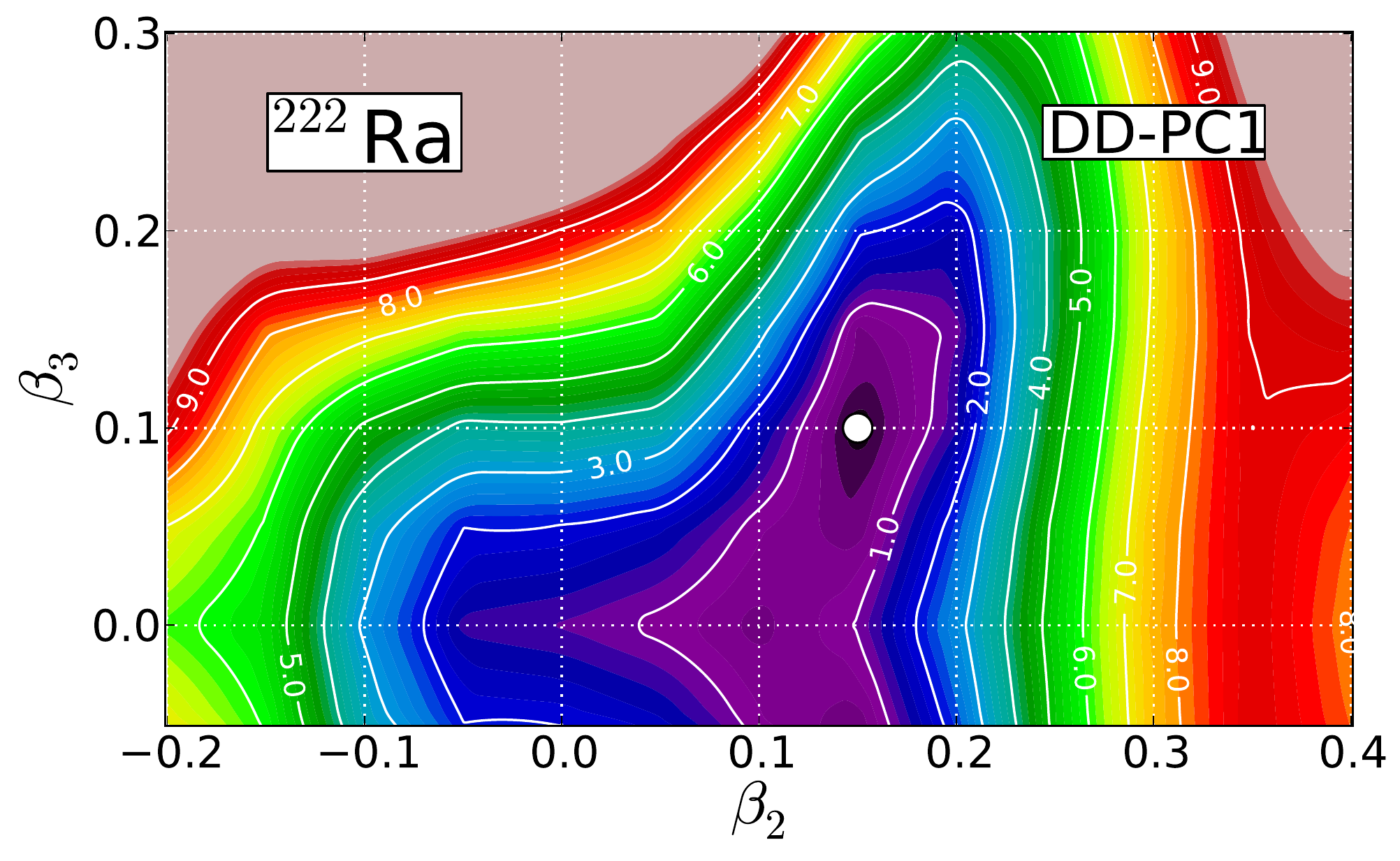} \\
\includegraphics[width=5.7cm]{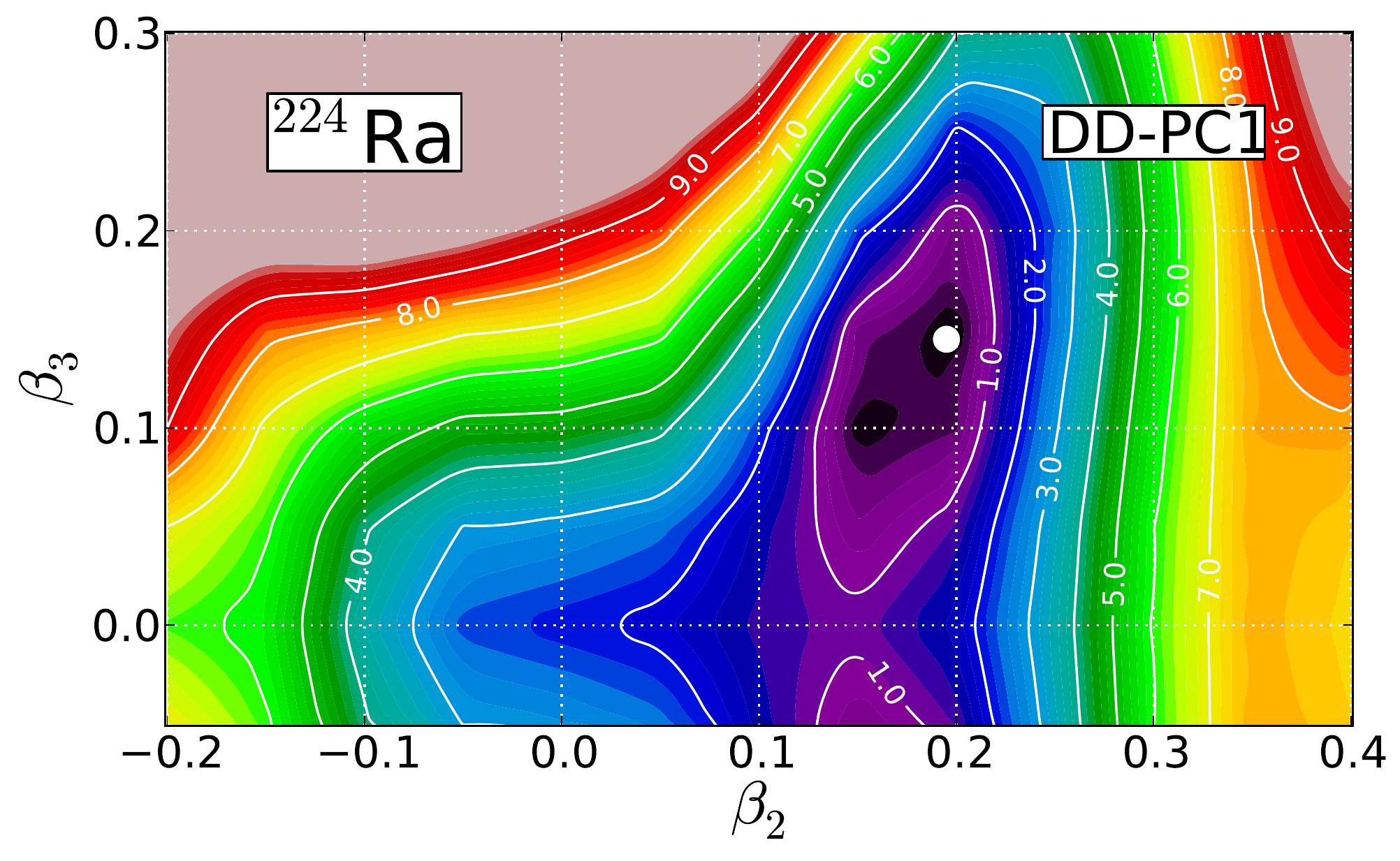} &
\includegraphics[width=5.7cm]{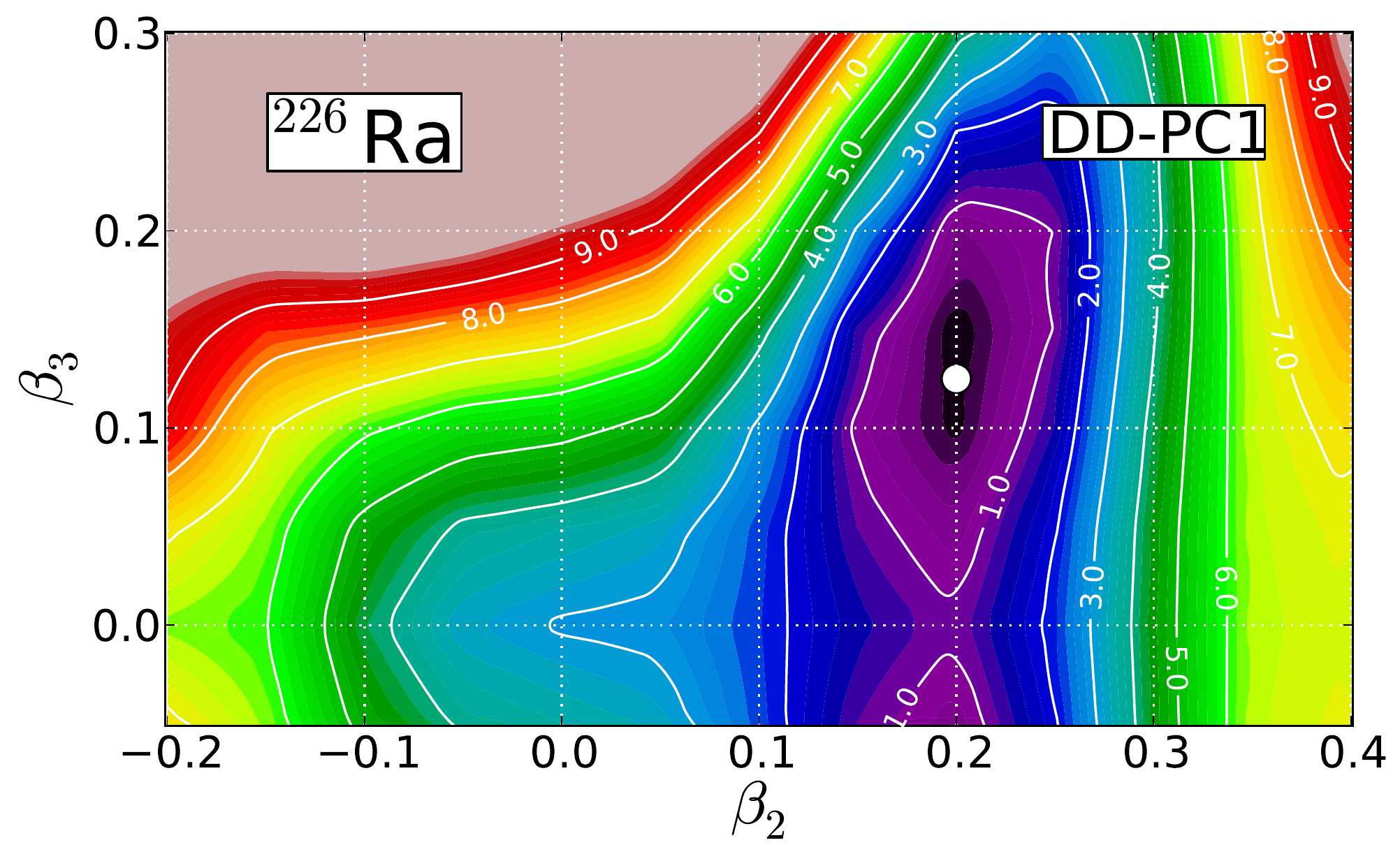} &
\includegraphics[width=5.7cm]{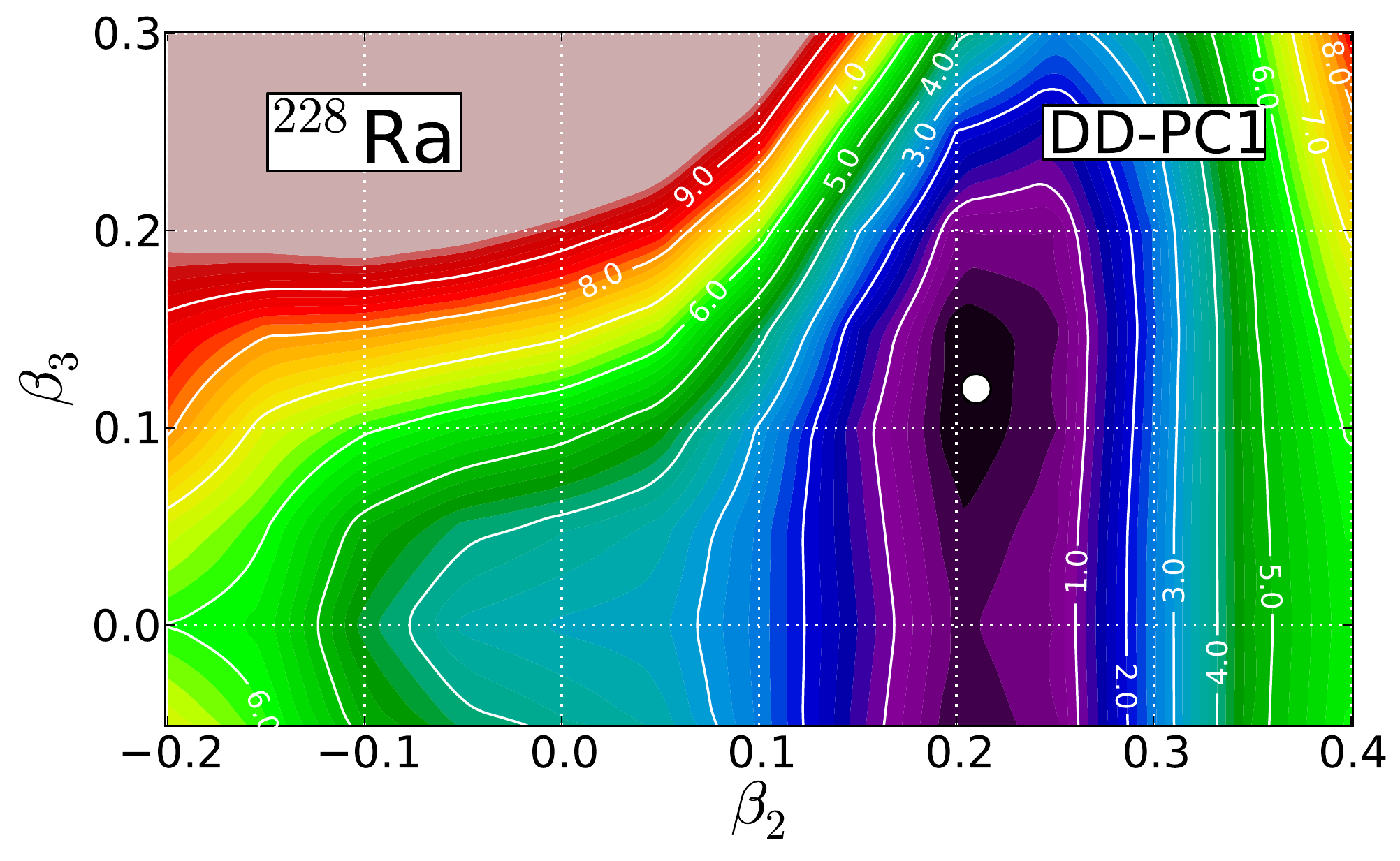}
\end{tabular}
\caption{(Color online) Same as the caption to Fig.~\ref{fig:pes_th},
 but for $^{218-228}$Ra. }
\label{fig:pes_ra}
\end{center}
\end{figure*}

\begin{figure*}[ctb!]
\begin{center}
\begin{tabular}{ccc}
\includegraphics[width=5.7cm]{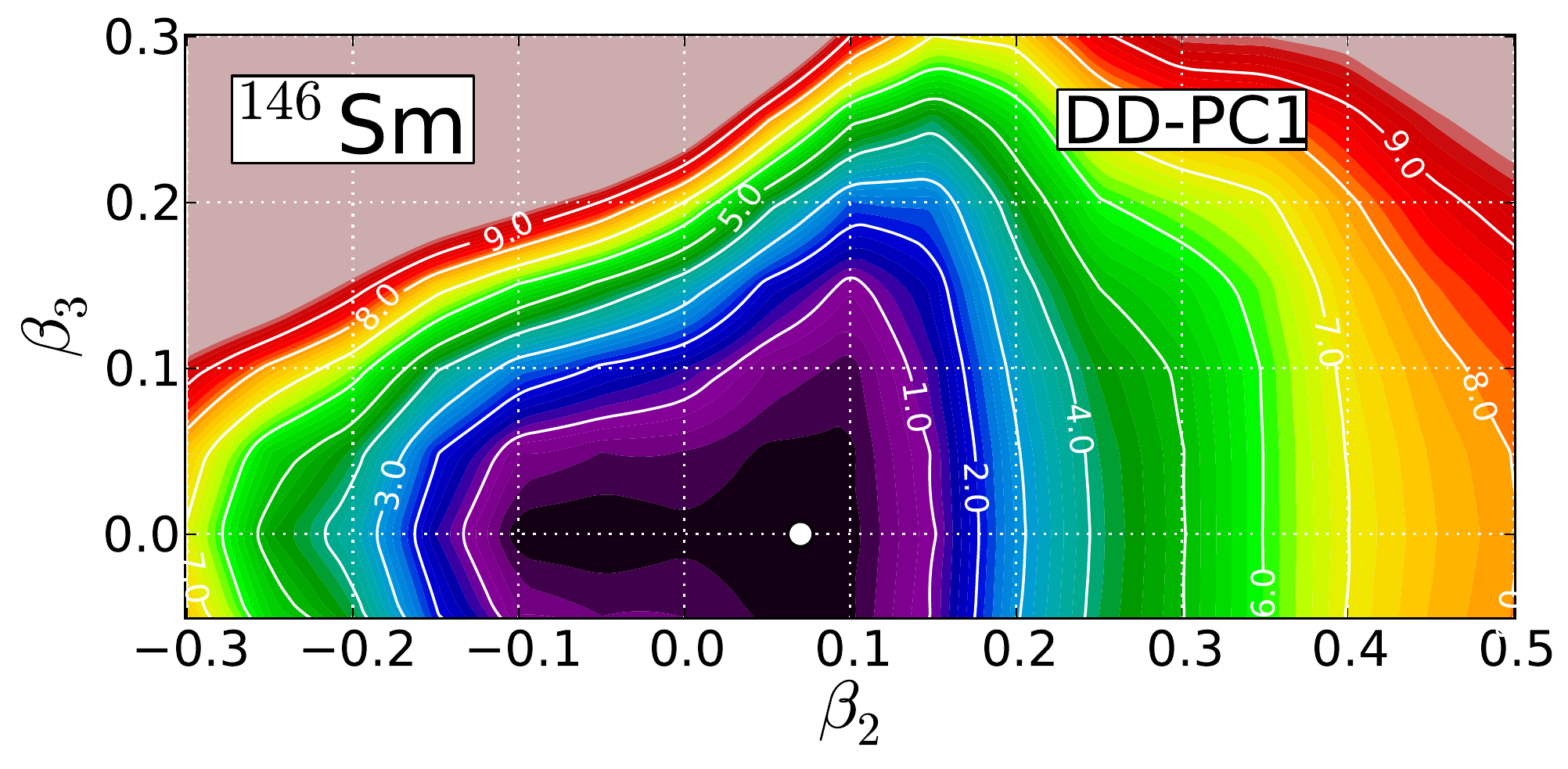} &
\includegraphics[width=5.7cm]{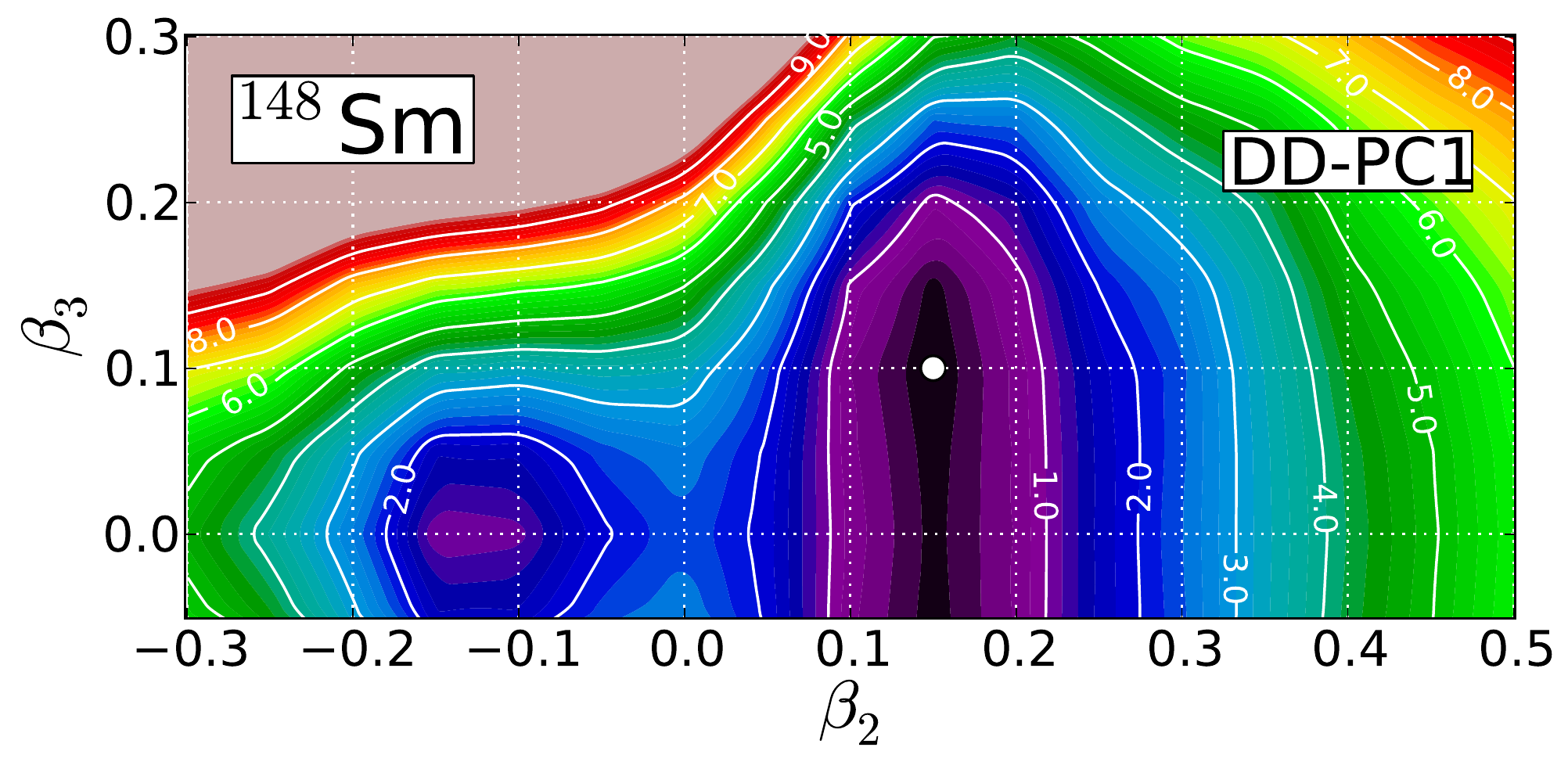} &
\includegraphics[width=5.7cm]{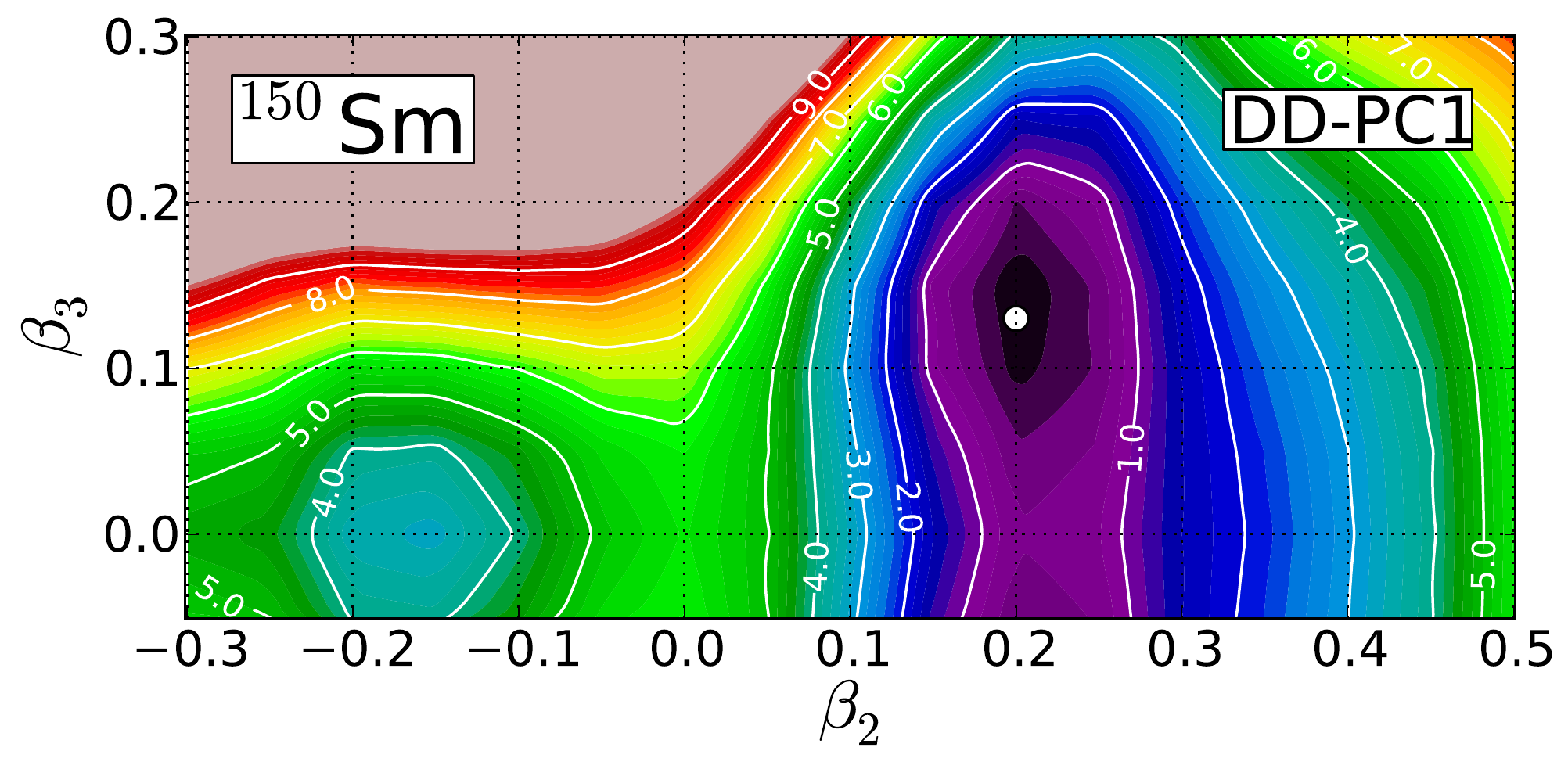} \\
\includegraphics[width=5.7cm]{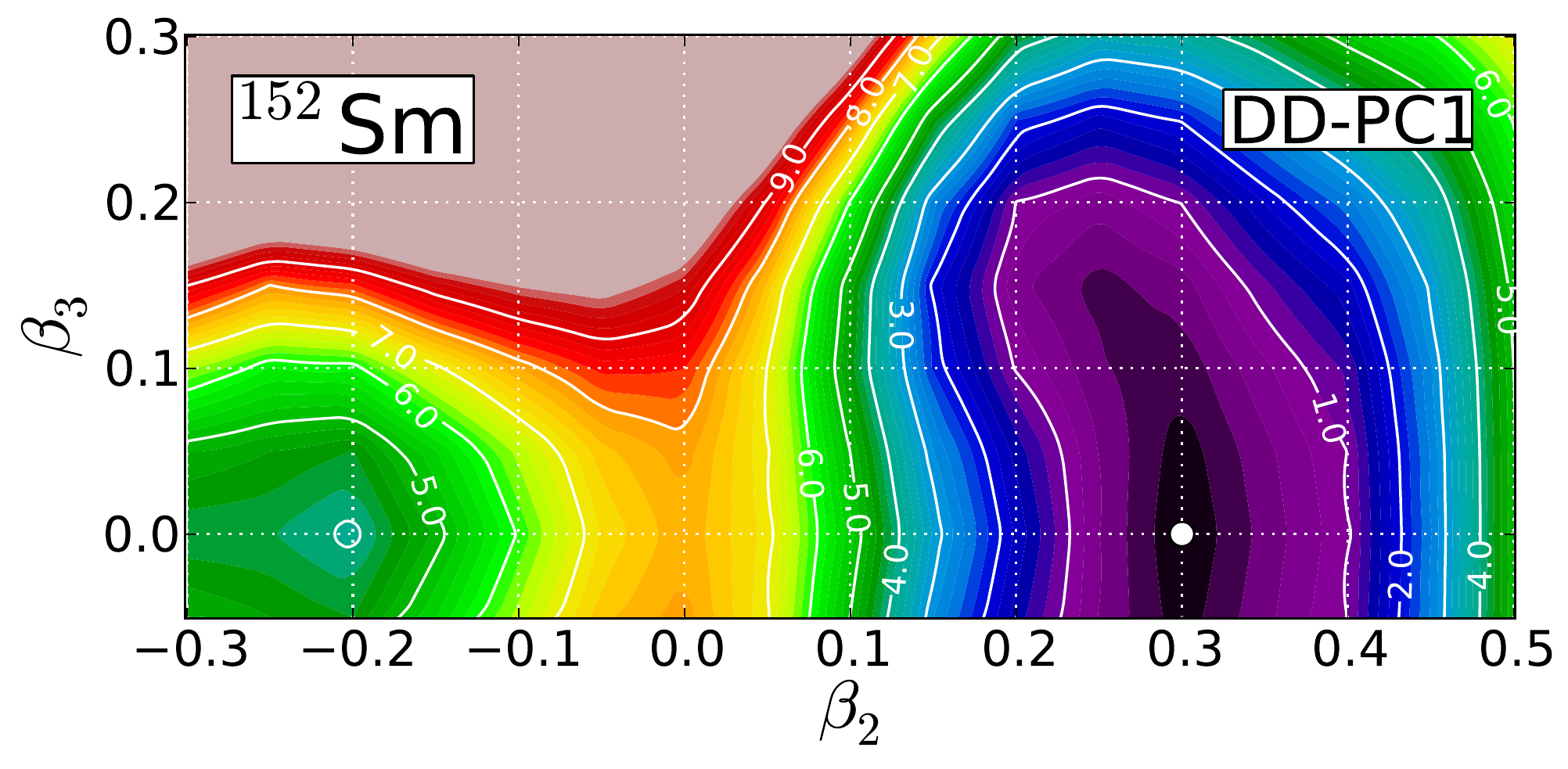} &
\includegraphics[width=5.7cm]{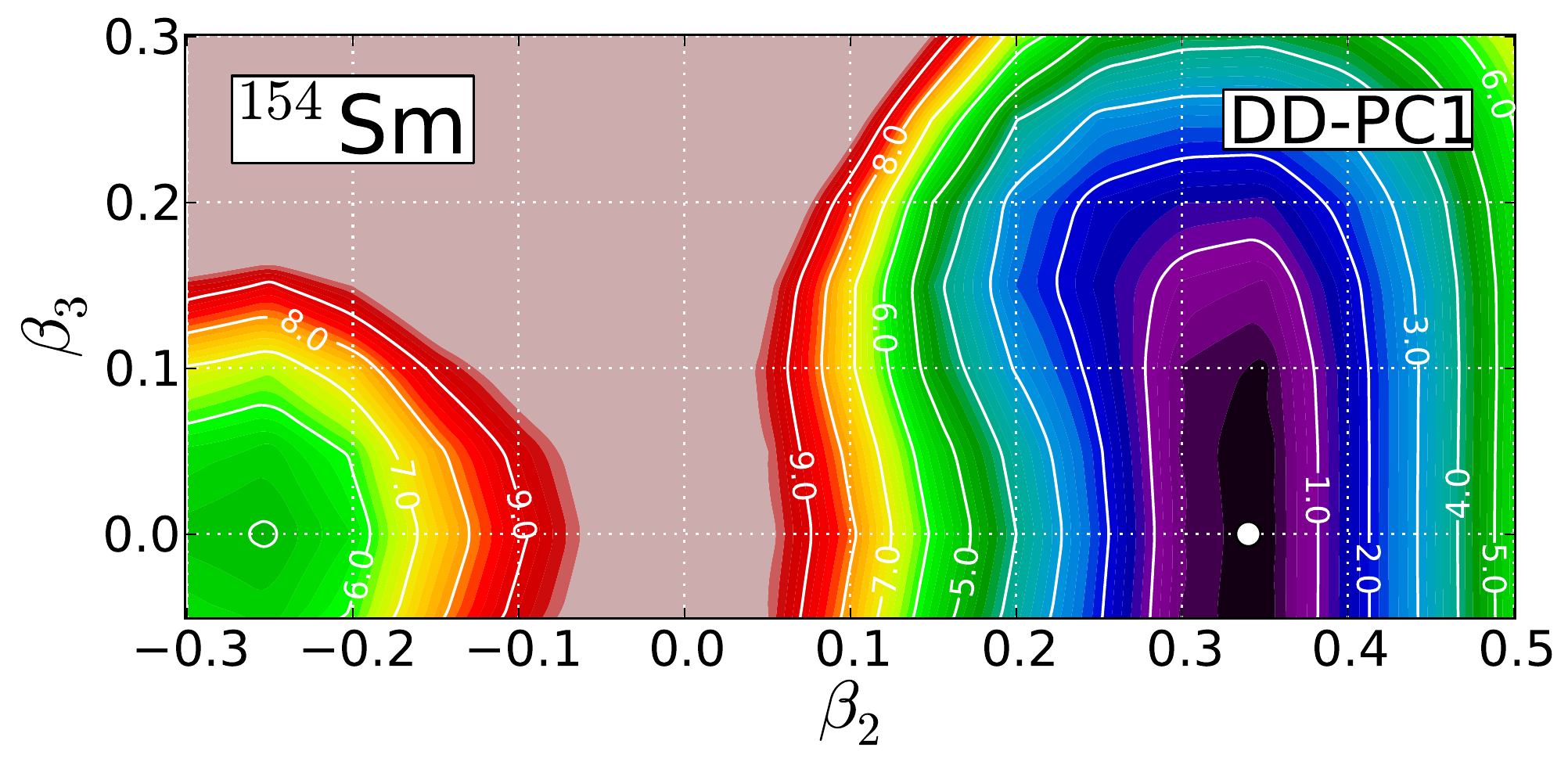} &
\includegraphics[width=5.7cm]{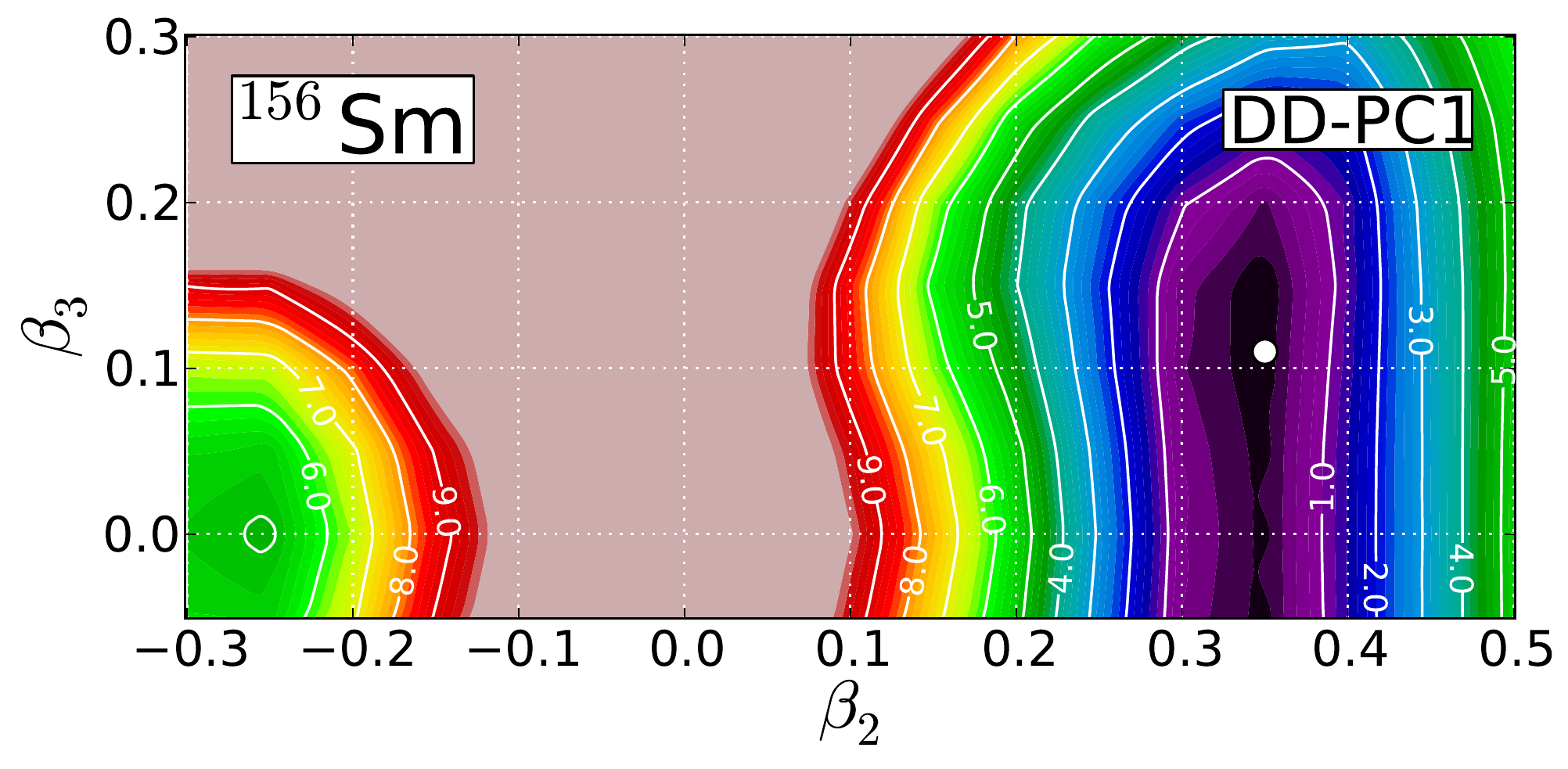} \\
\end{tabular}
\caption{(Color online) Same as the caption to Fig.~\ref{fig:pes_th},
 but for $^{146-156}$Sm. }
\label{fig:pes_sm}
\end{center}
\end{figure*}

\begin{figure*}[ctb!]
\begin{center}
\begin{tabular}{ccc}
\includegraphics[width=5.7cm]{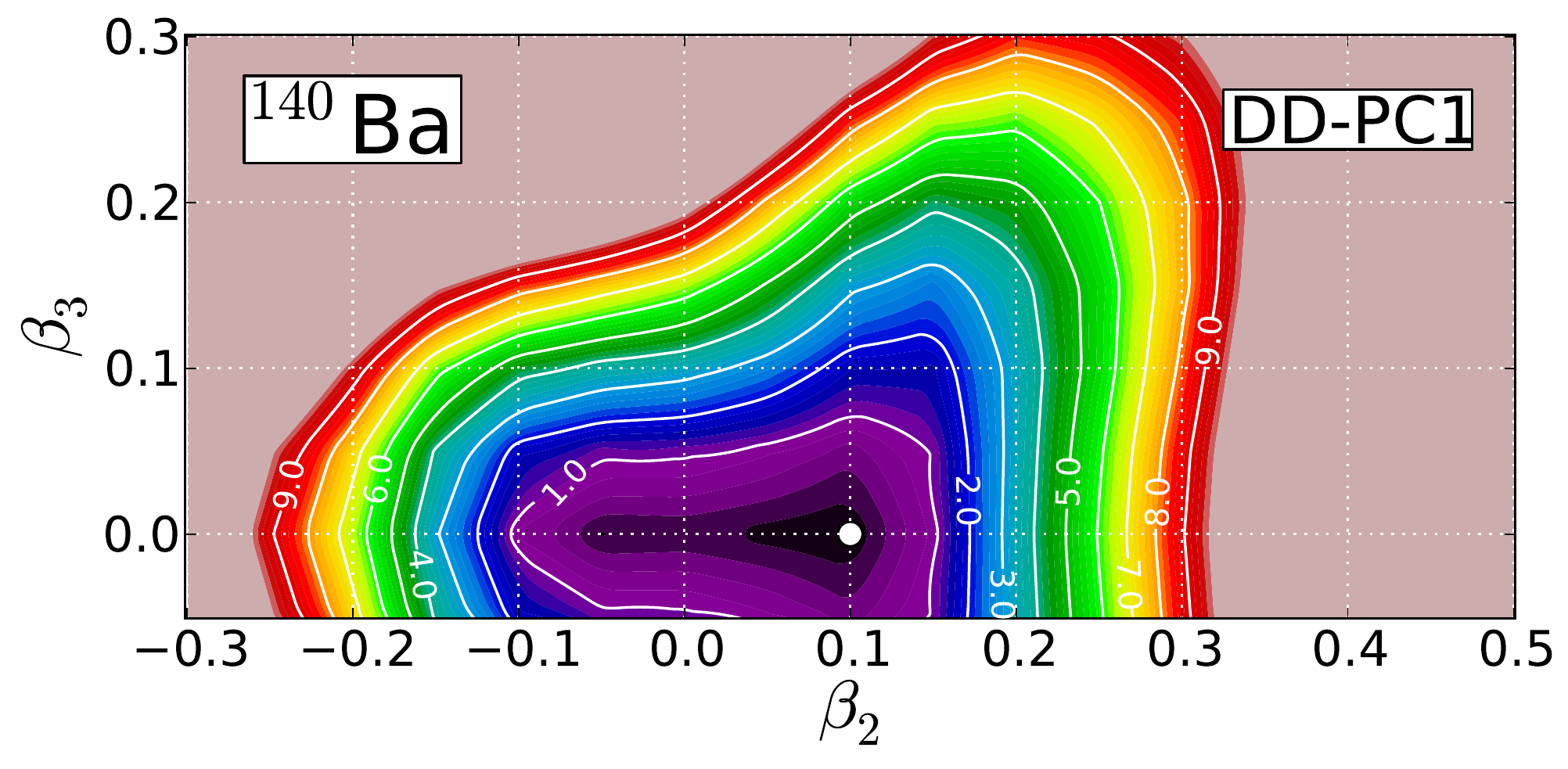} &
\includegraphics[width=5.7cm]{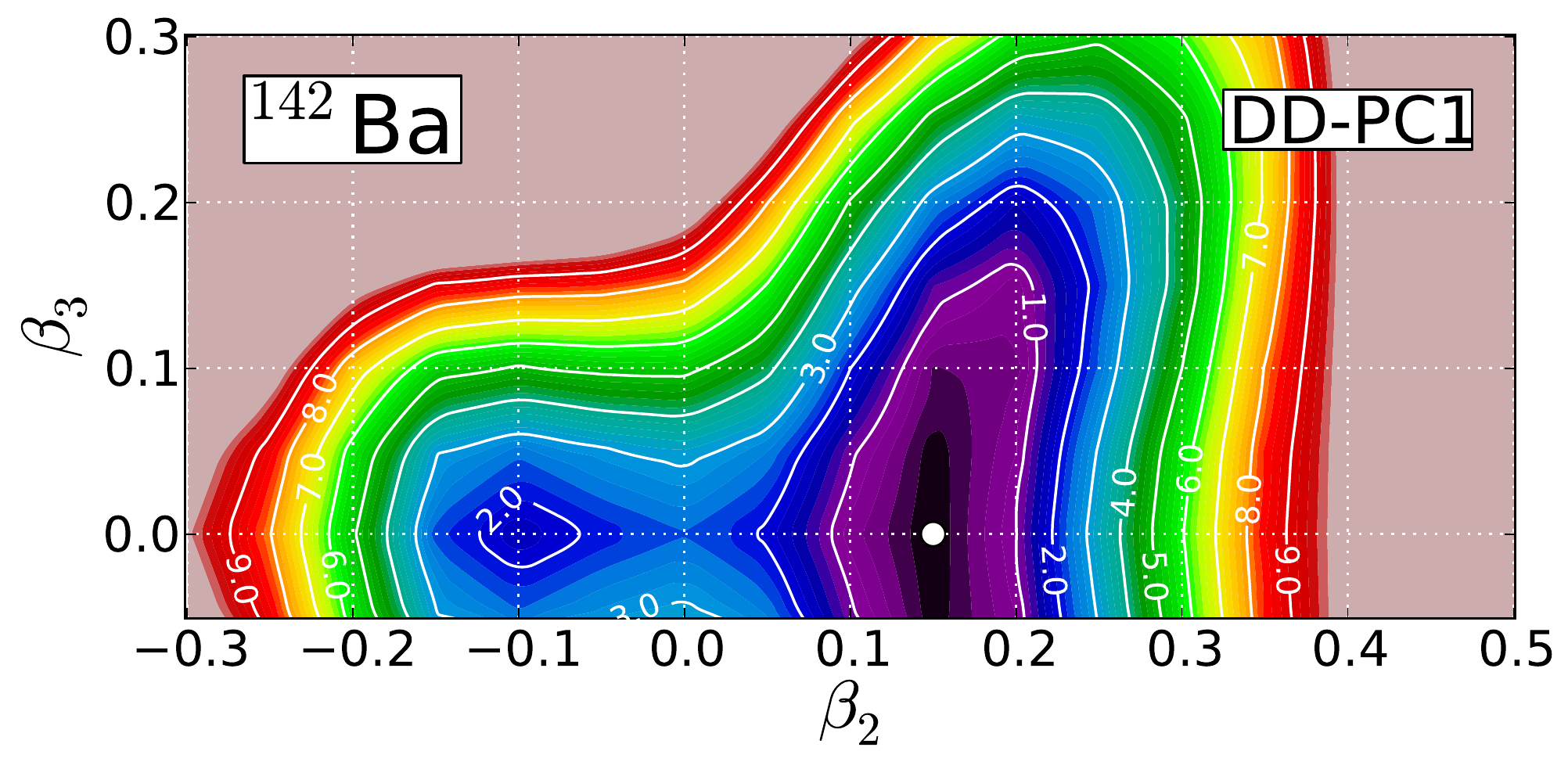} &
\includegraphics[width=5.7cm]{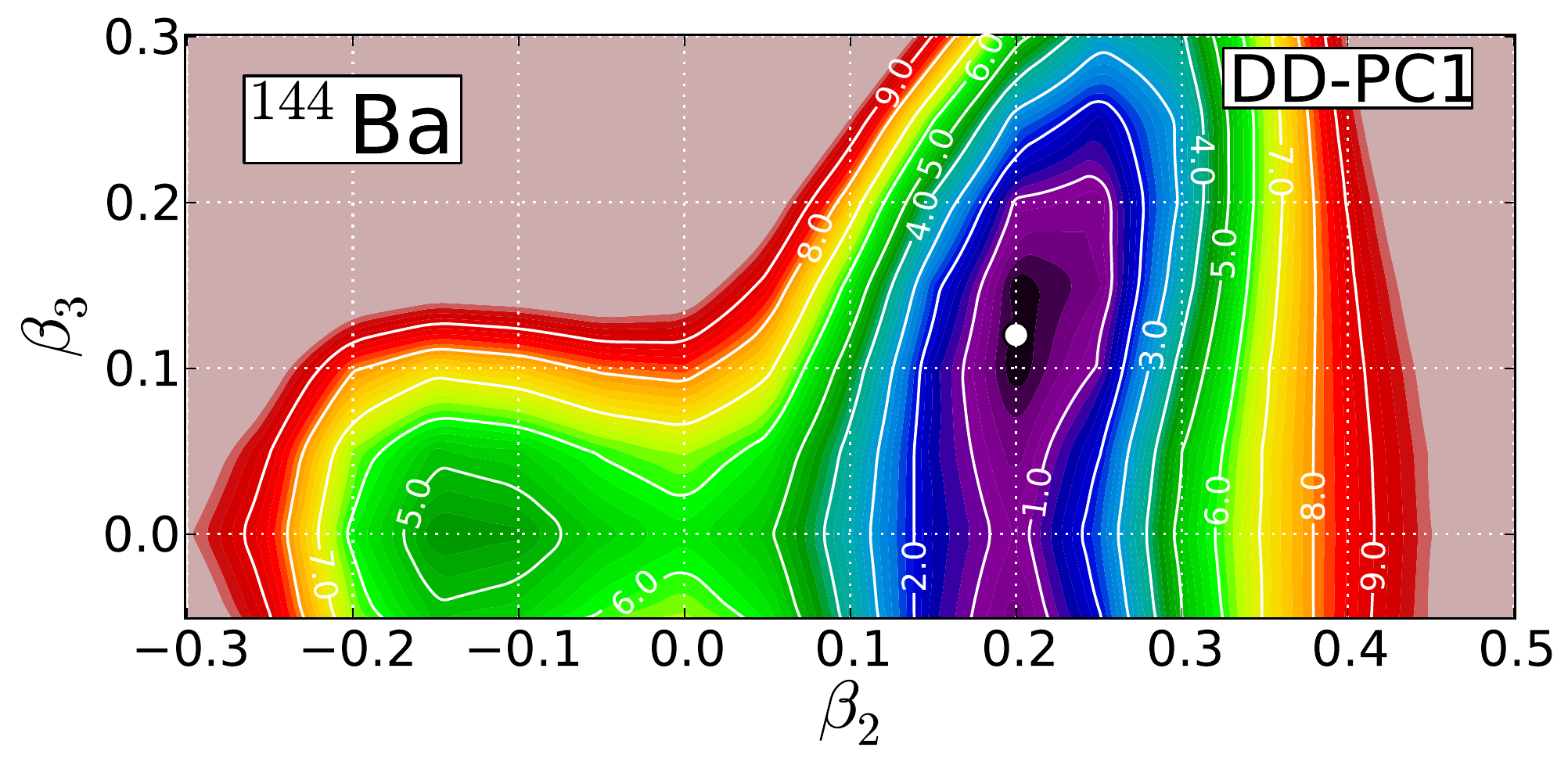} \\
\includegraphics[width=5.7cm]{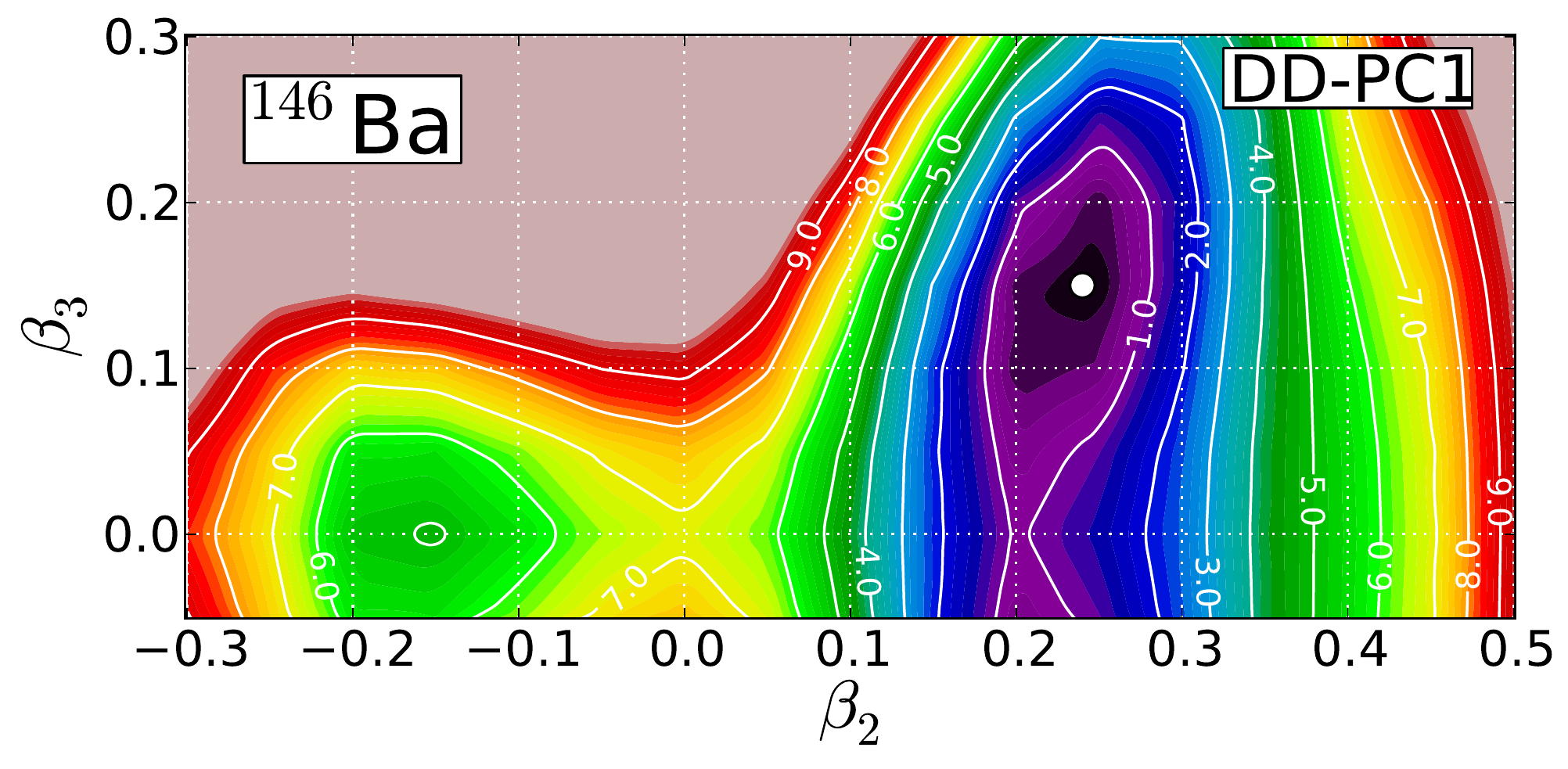} &
\includegraphics[width=5.7cm]{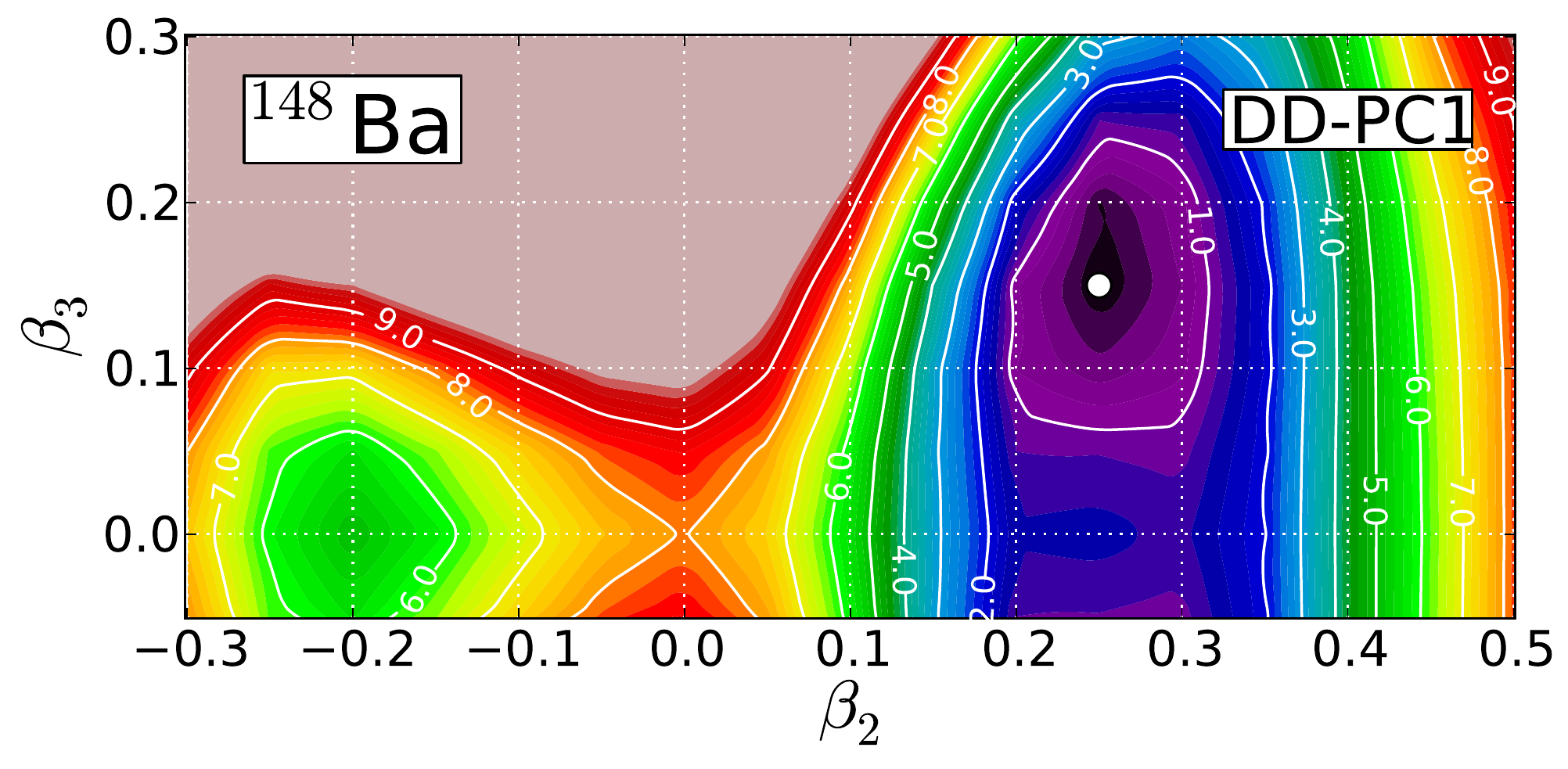} &
\includegraphics[width=5.7cm]{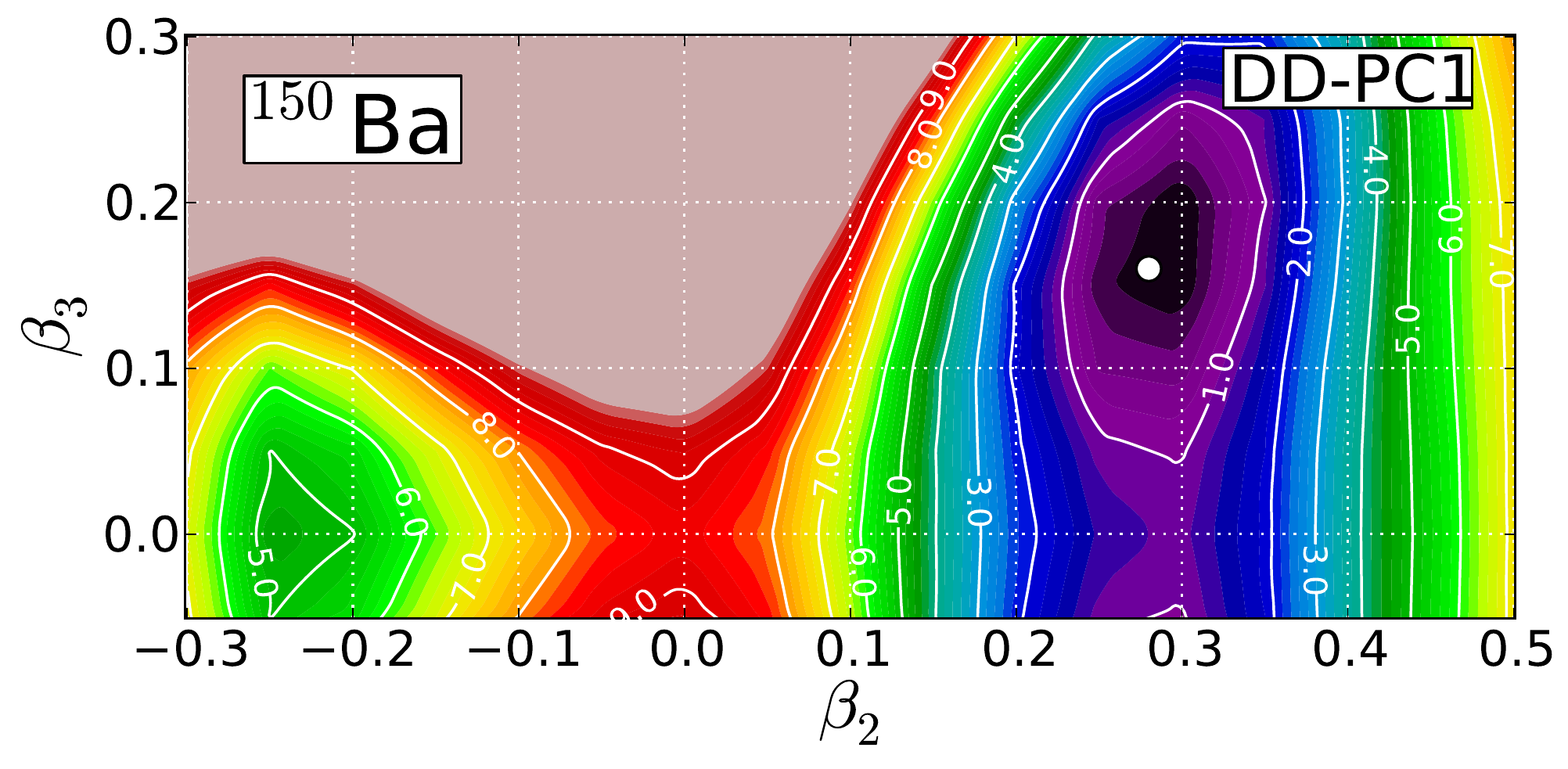} 
\end{tabular}
\caption{(Color online) Same as the caption to Fig.~\ref{fig:pes_th},
 but for $^{140-150}$Ba. }
\label{fig:pes_ba}
\end{center}
\end{figure*}

\begin{figure*}[ctb!]
\begin{center}
\begin{tabular}{ccc}
\includegraphics[width=5.7cm]{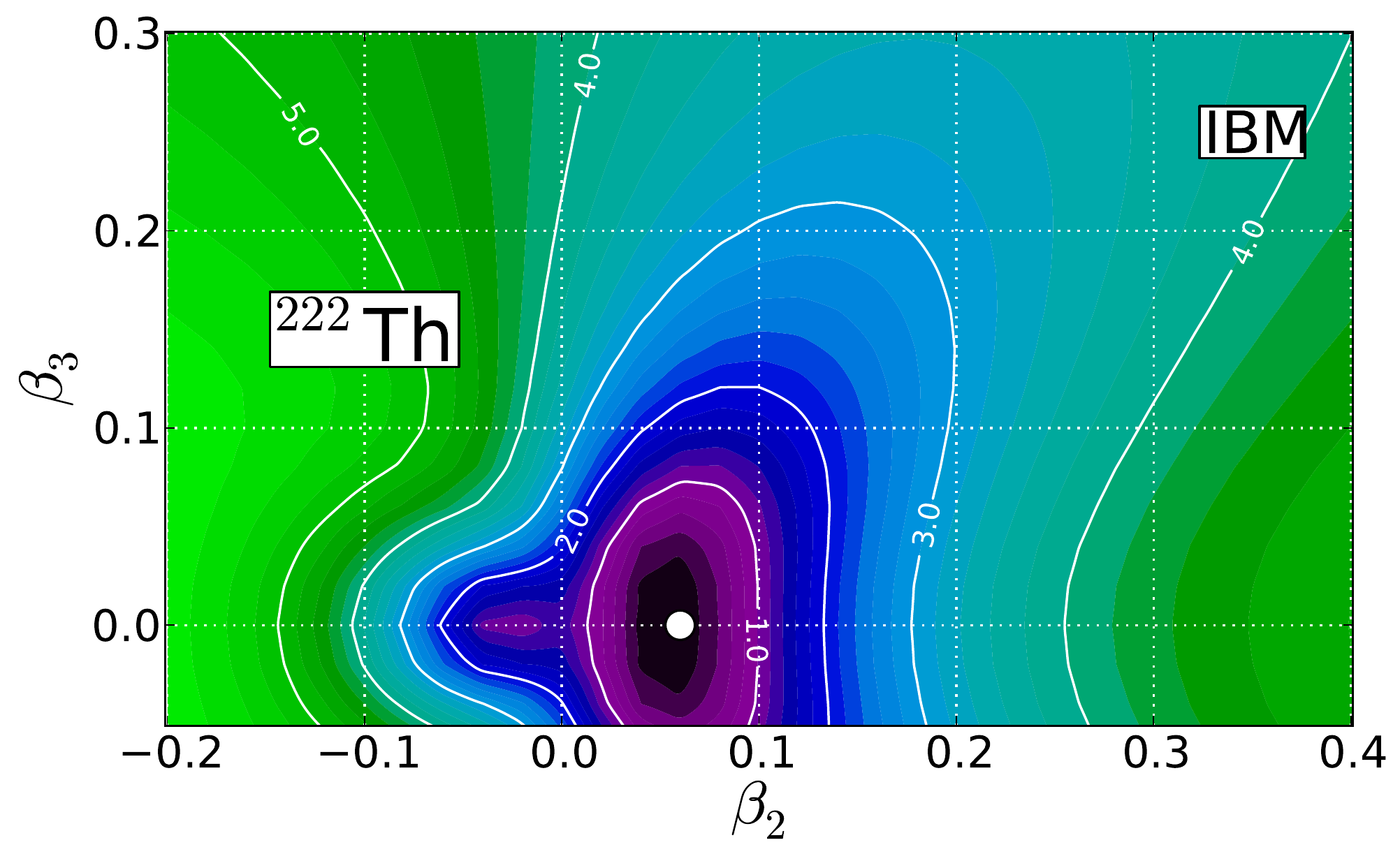} &
 \includegraphics[width=5.7cm]{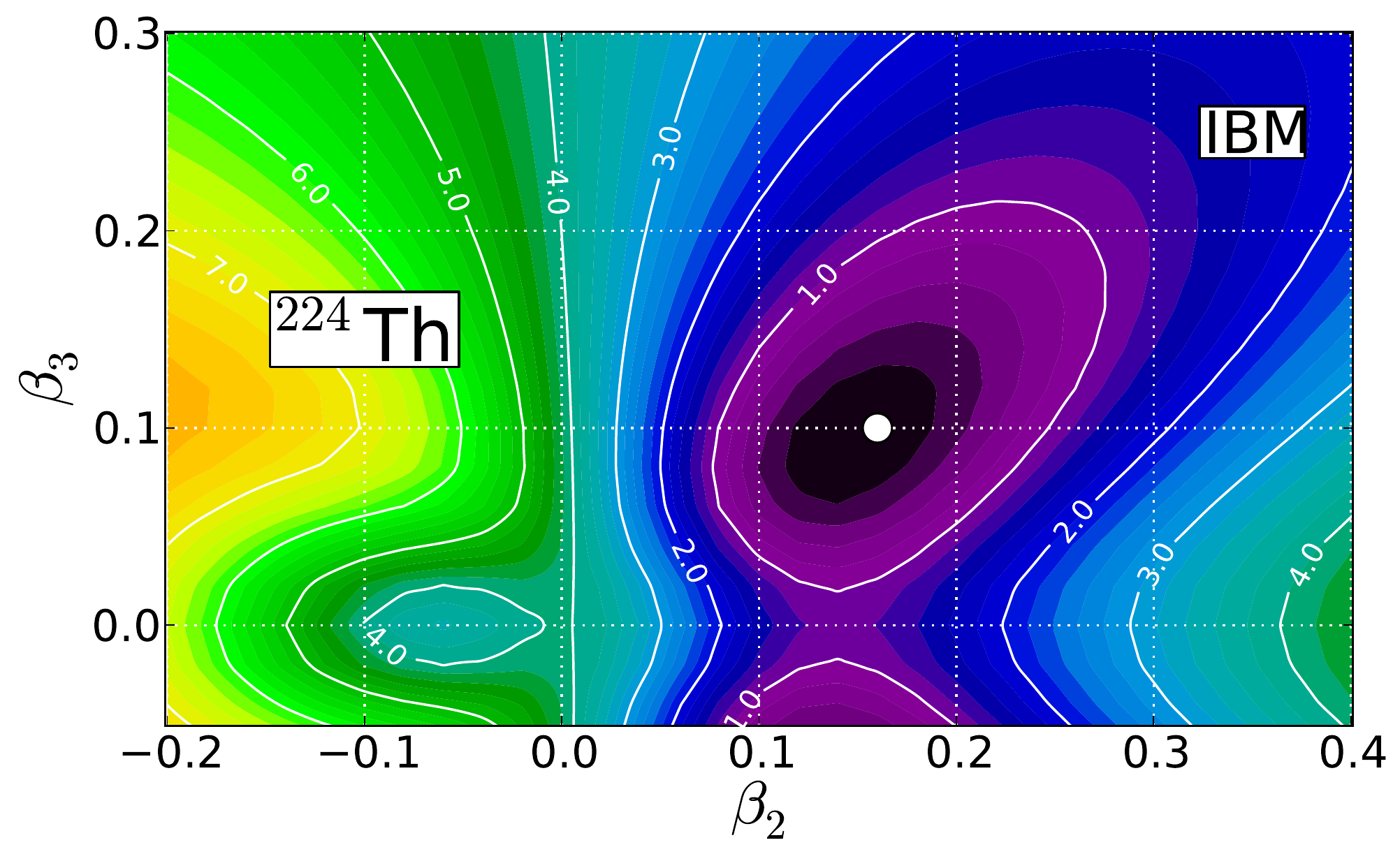} &
\includegraphics[width=5.7cm]{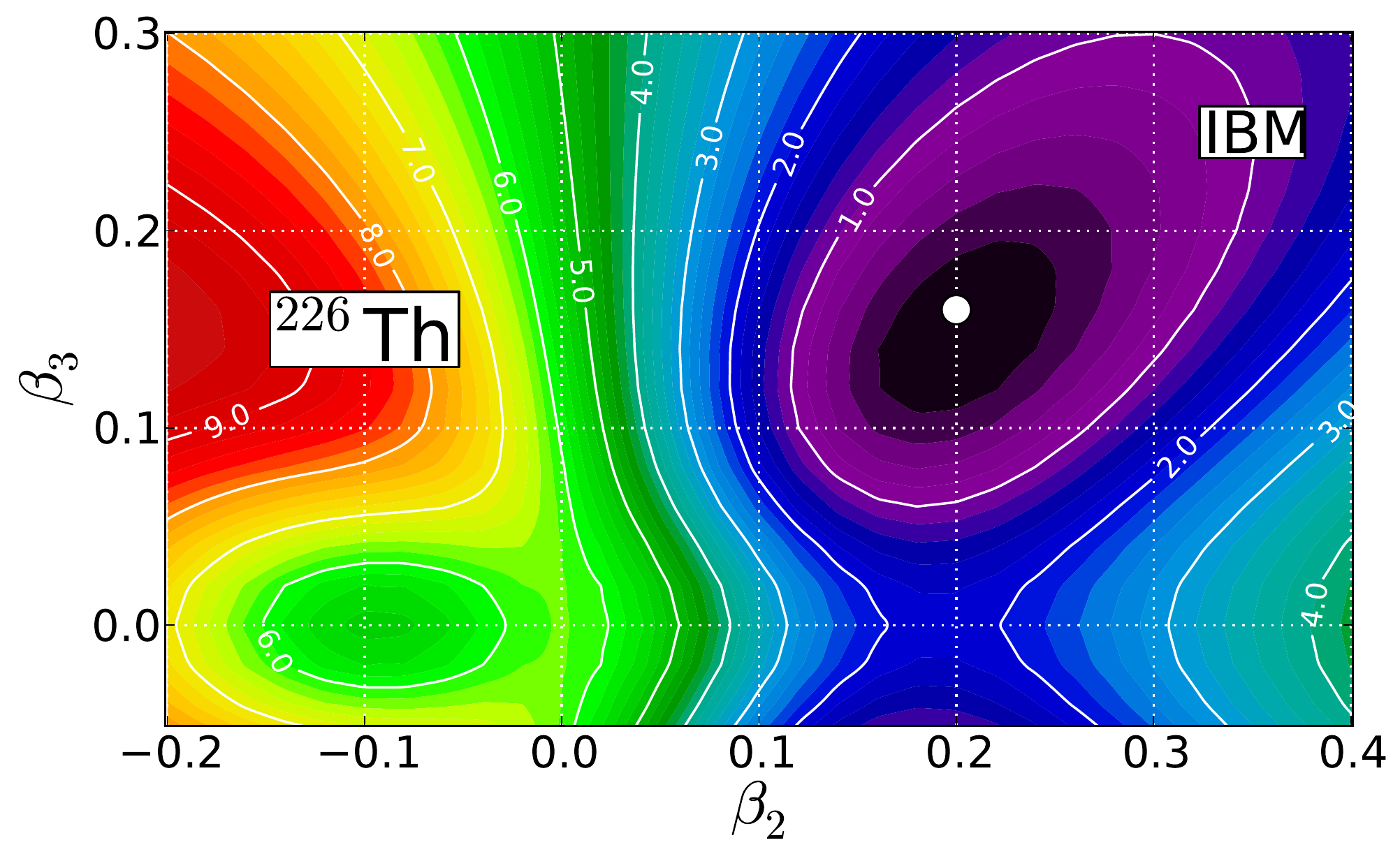} \\
\includegraphics[width=5.7cm]{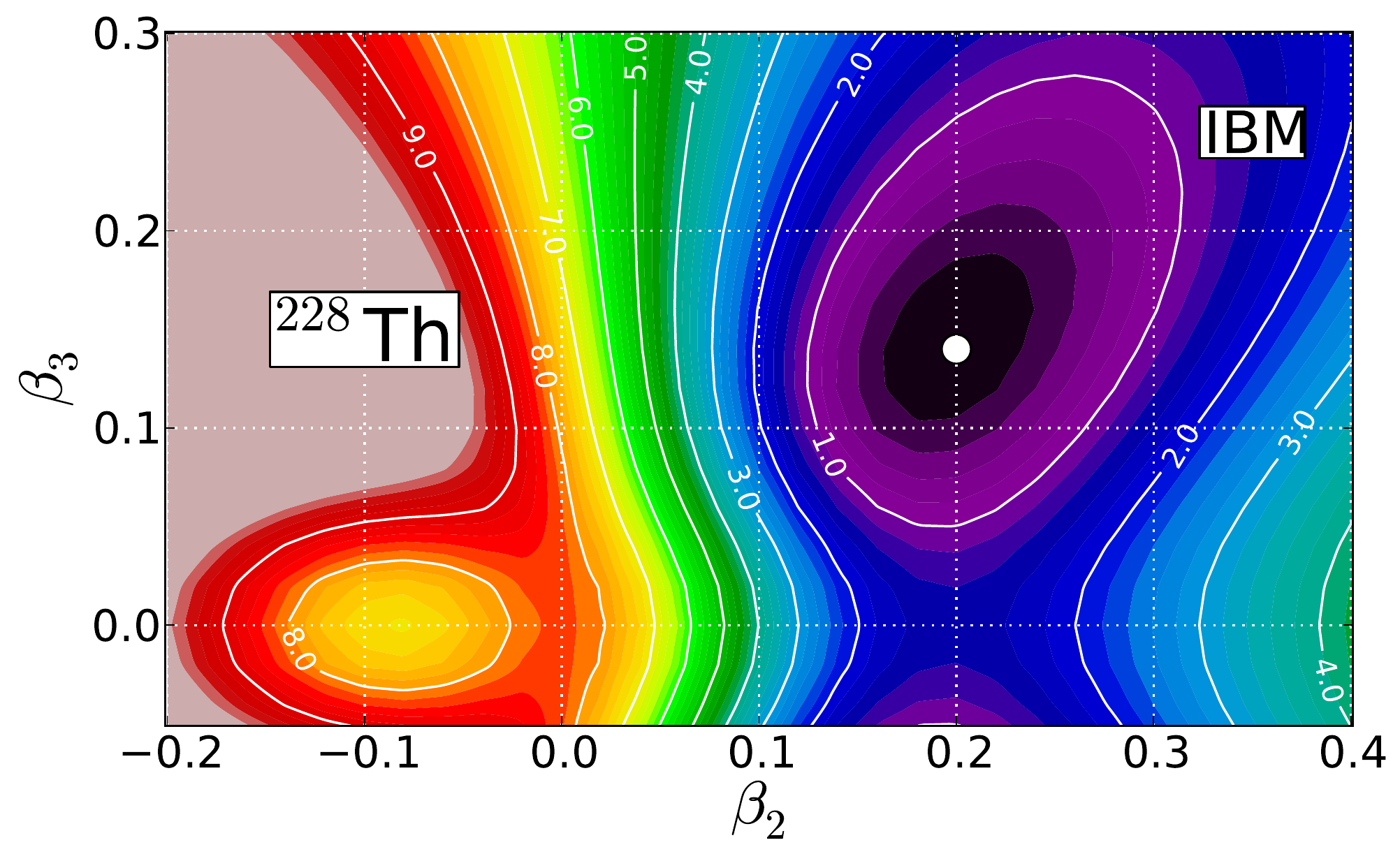} &
\includegraphics[width=5.7cm]{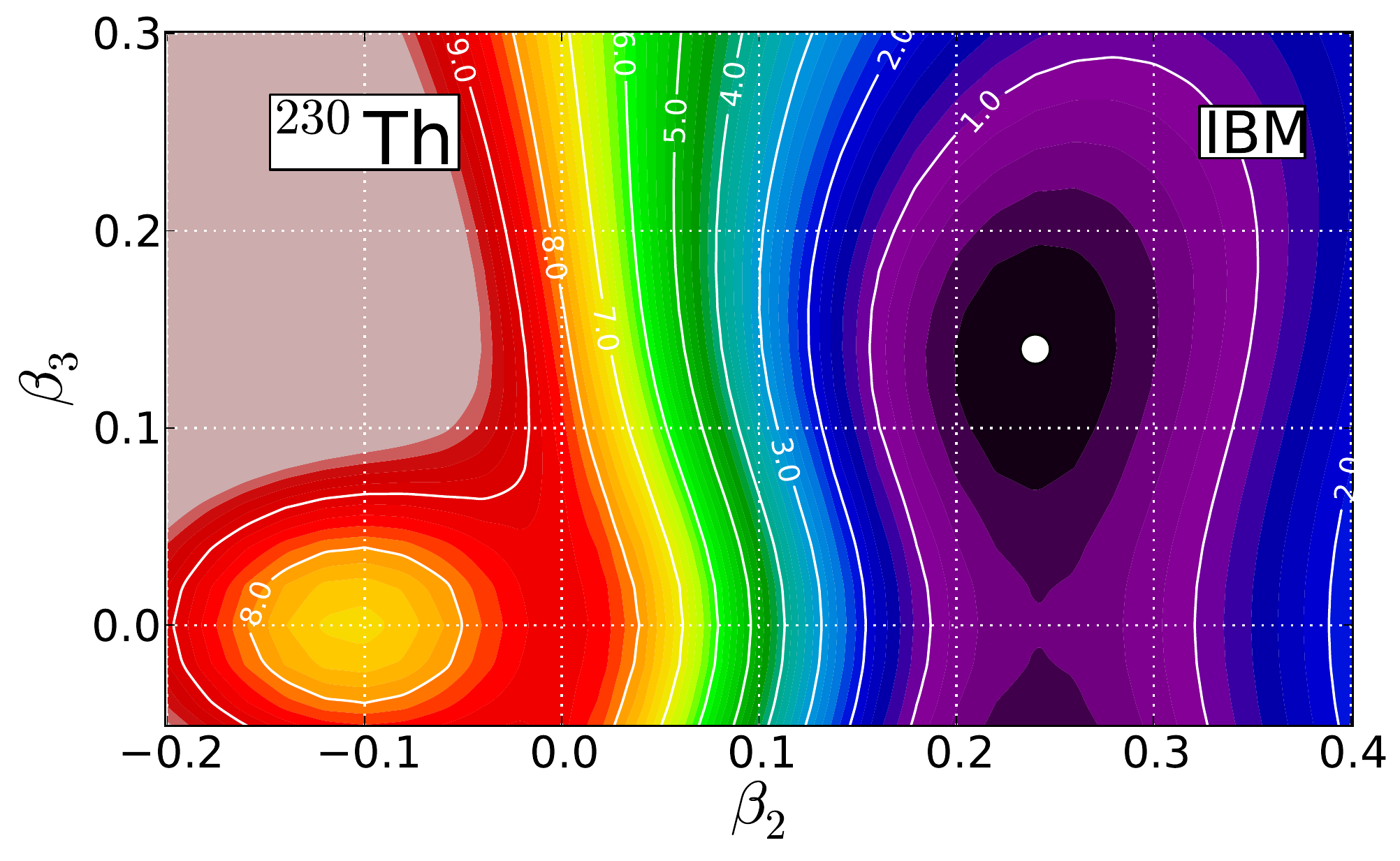} &
\includegraphics[width=5.7cm]{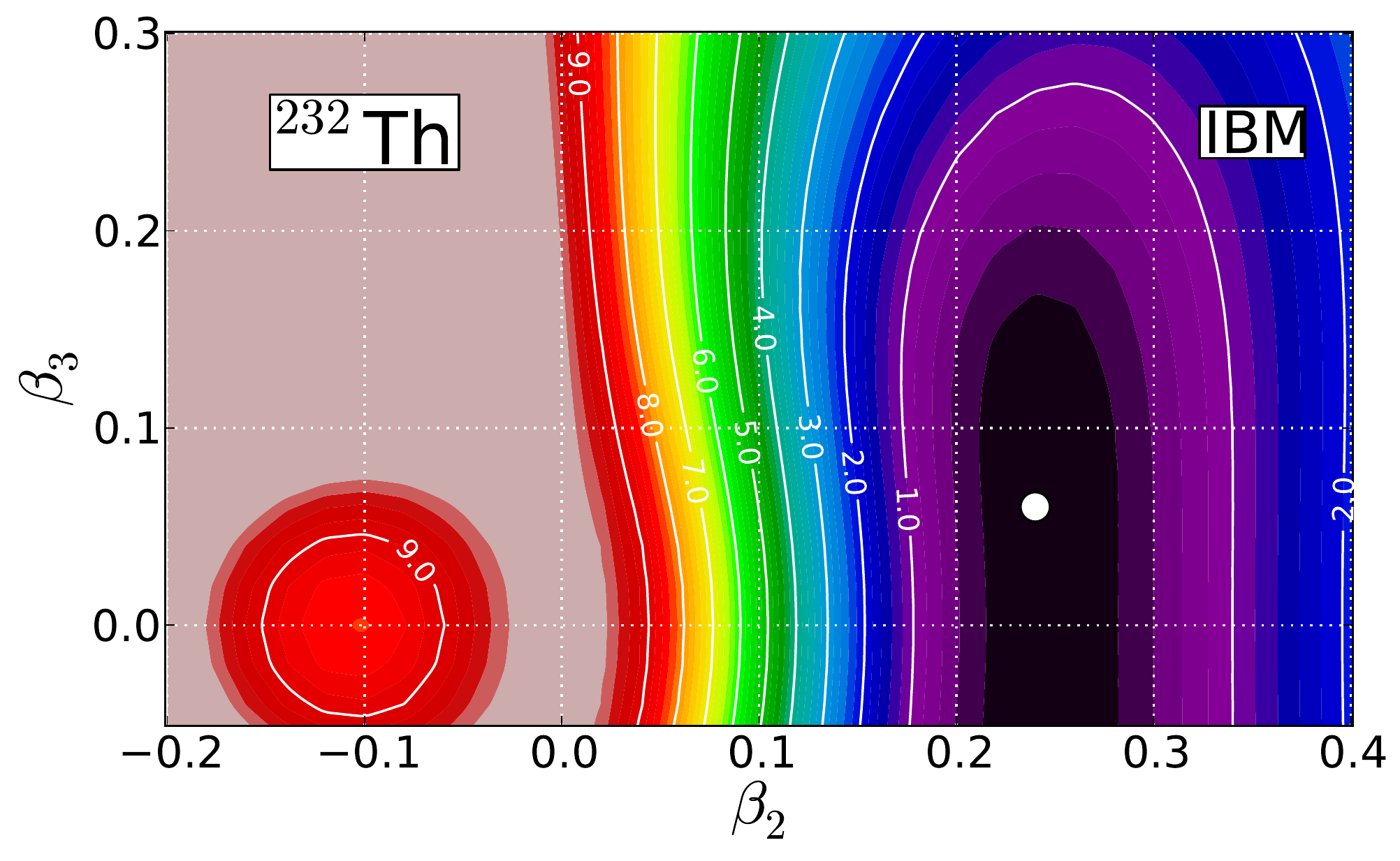}
\end{tabular}
\caption{(Color online) Same as the caption to Fig.~\ref{fig:pes_th},
 but for the mapped IBM energy surfaces of $^{222-232}$Th. }
\label{fig:pes_th_mapped}
\end{center}
\end{figure*}

\begin{figure*}[ctb!]
\begin{center}
\begin{tabular}{ccc}
\includegraphics[width=5.7cm]{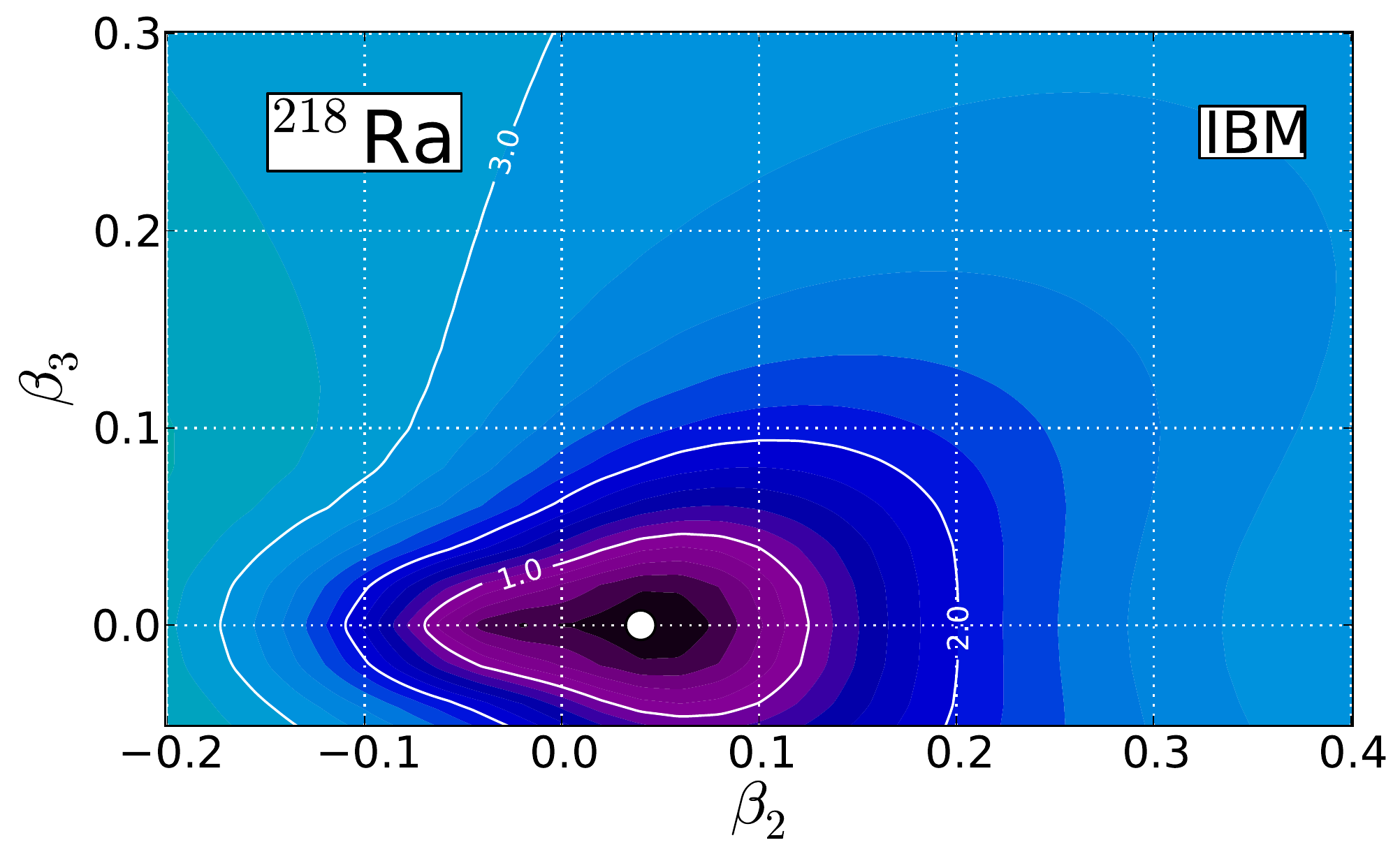} &
\includegraphics[width=5.7cm]{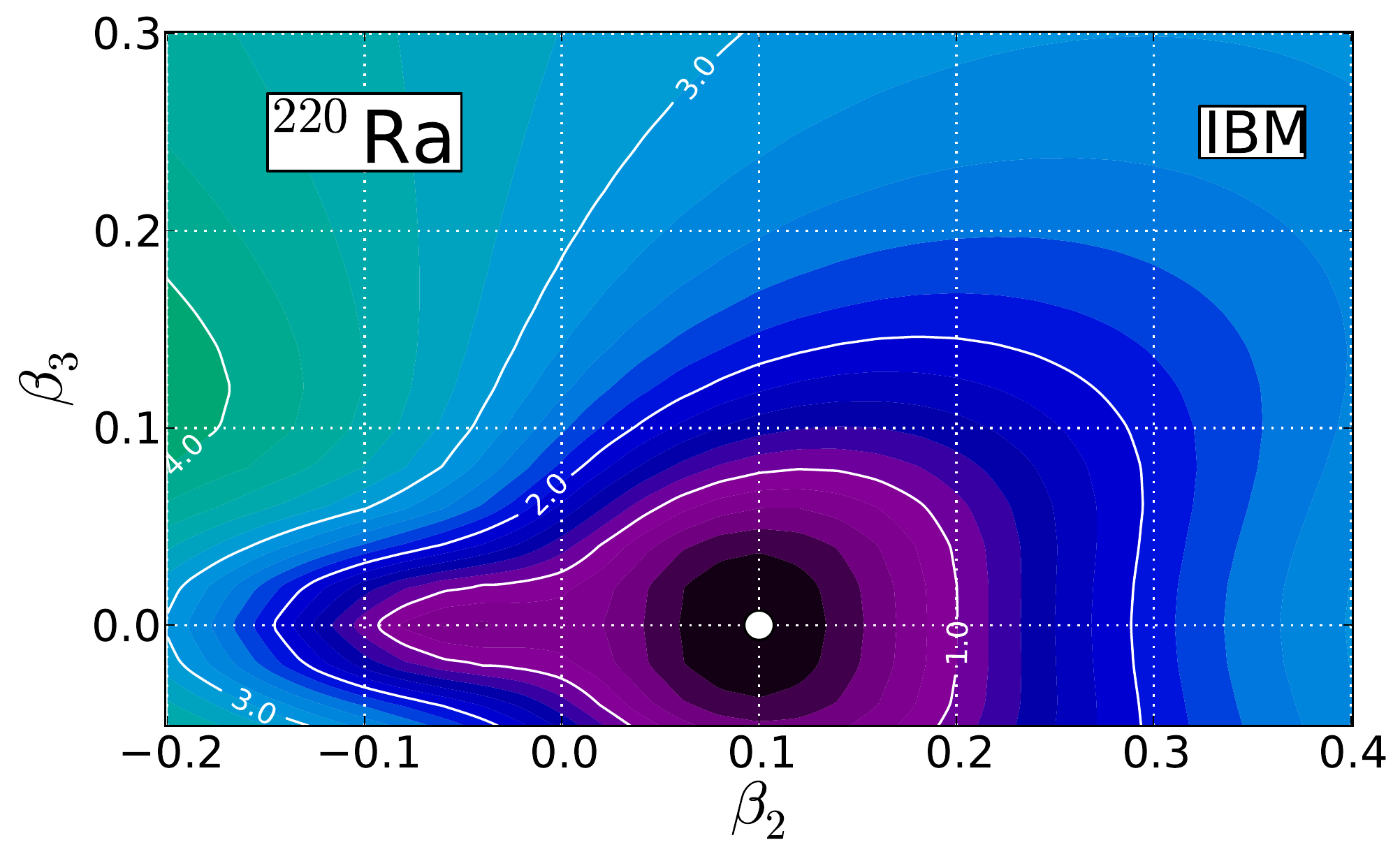} &
\includegraphics[width=5.7cm]{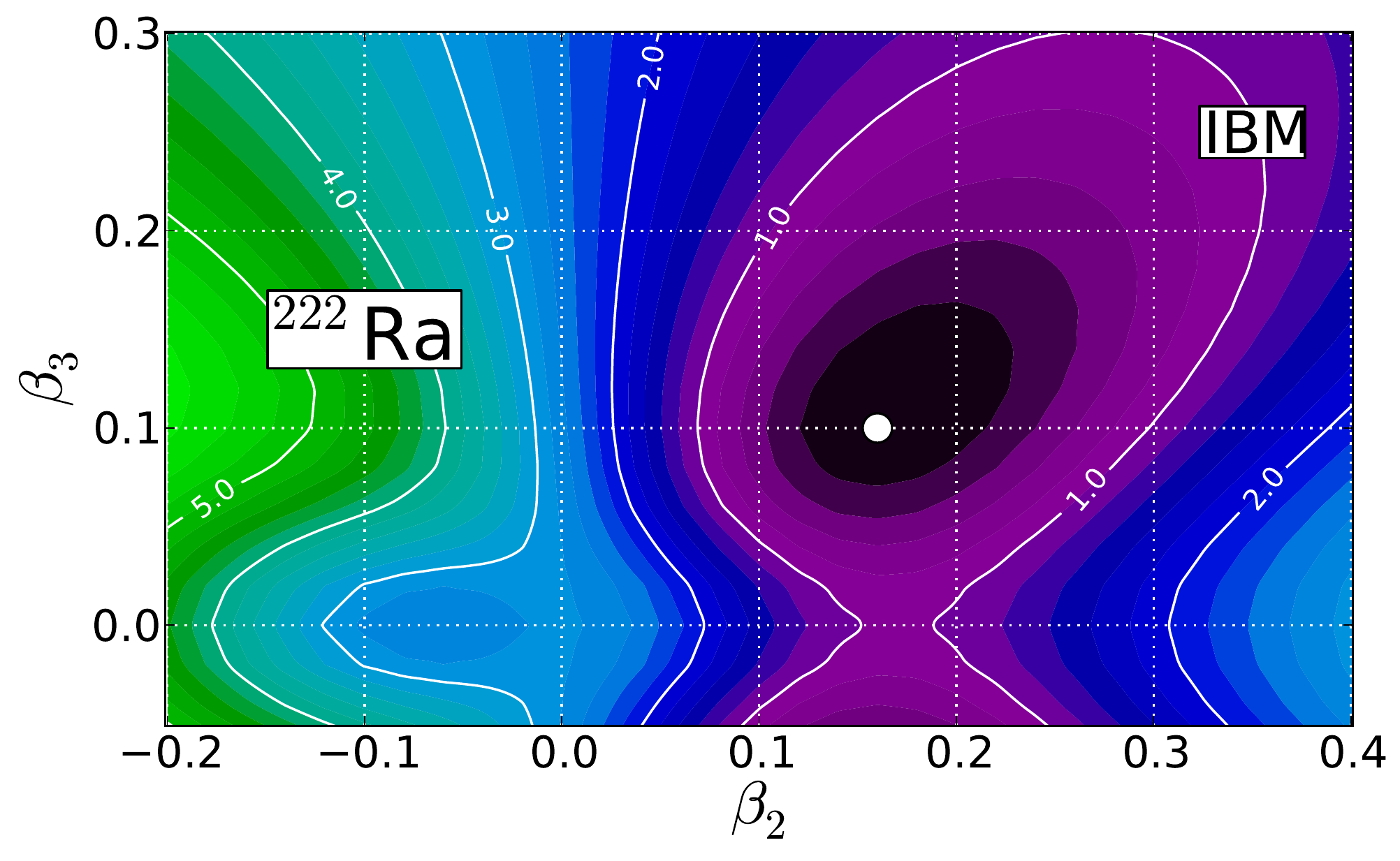} \\
\includegraphics[width=5.7cm]{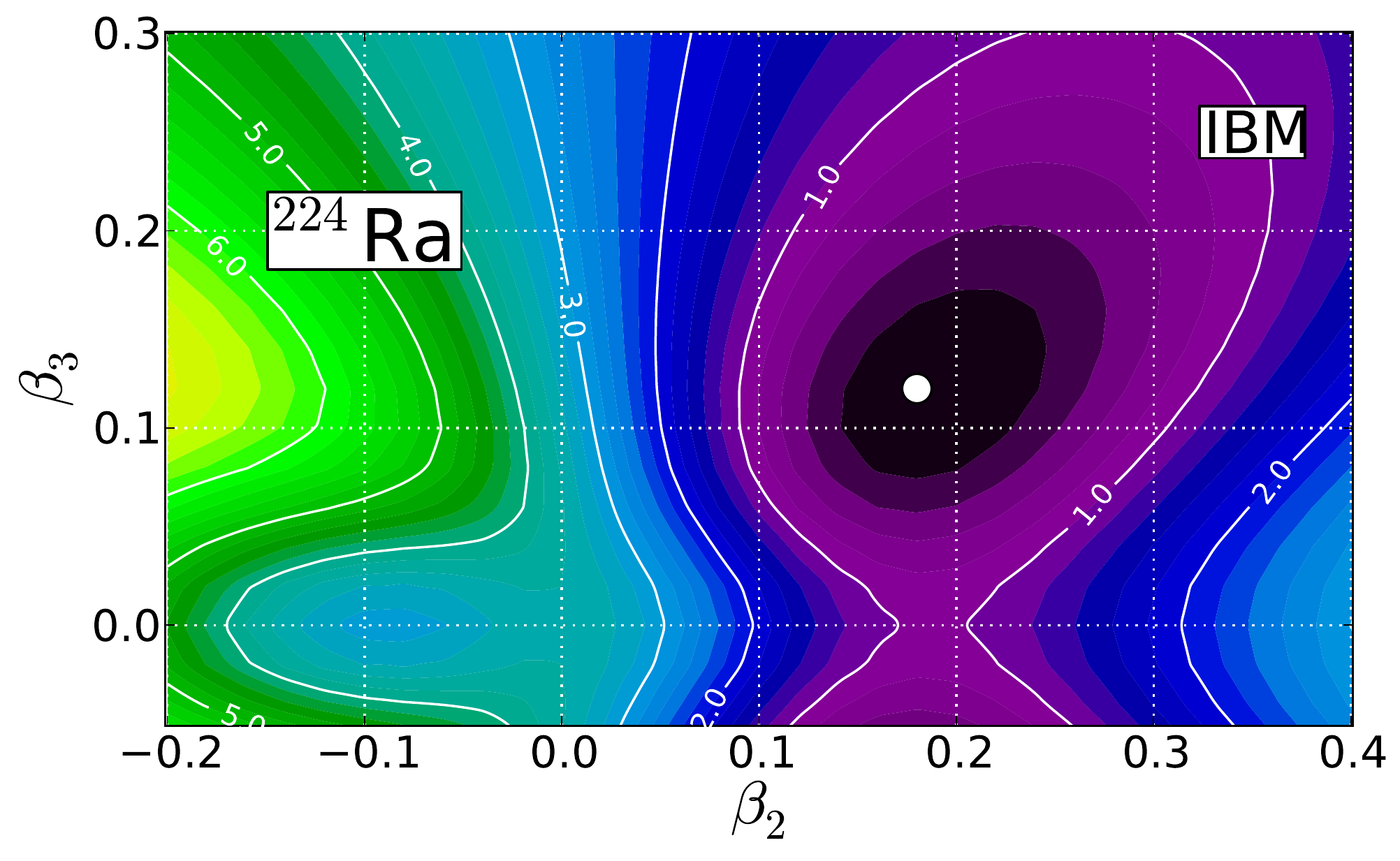} &
\includegraphics[width=5.7cm]{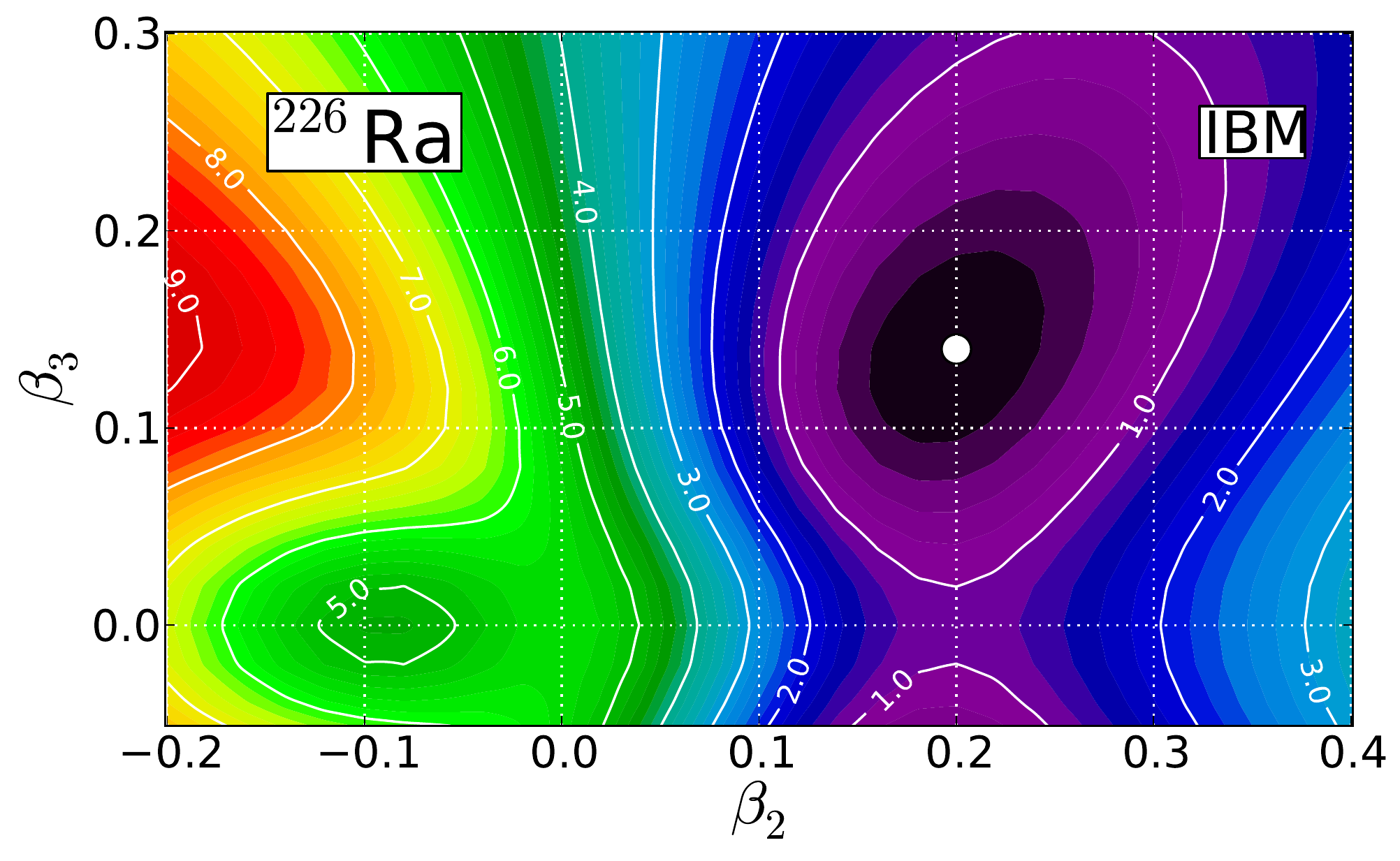} &
\includegraphics[width=5.7cm]{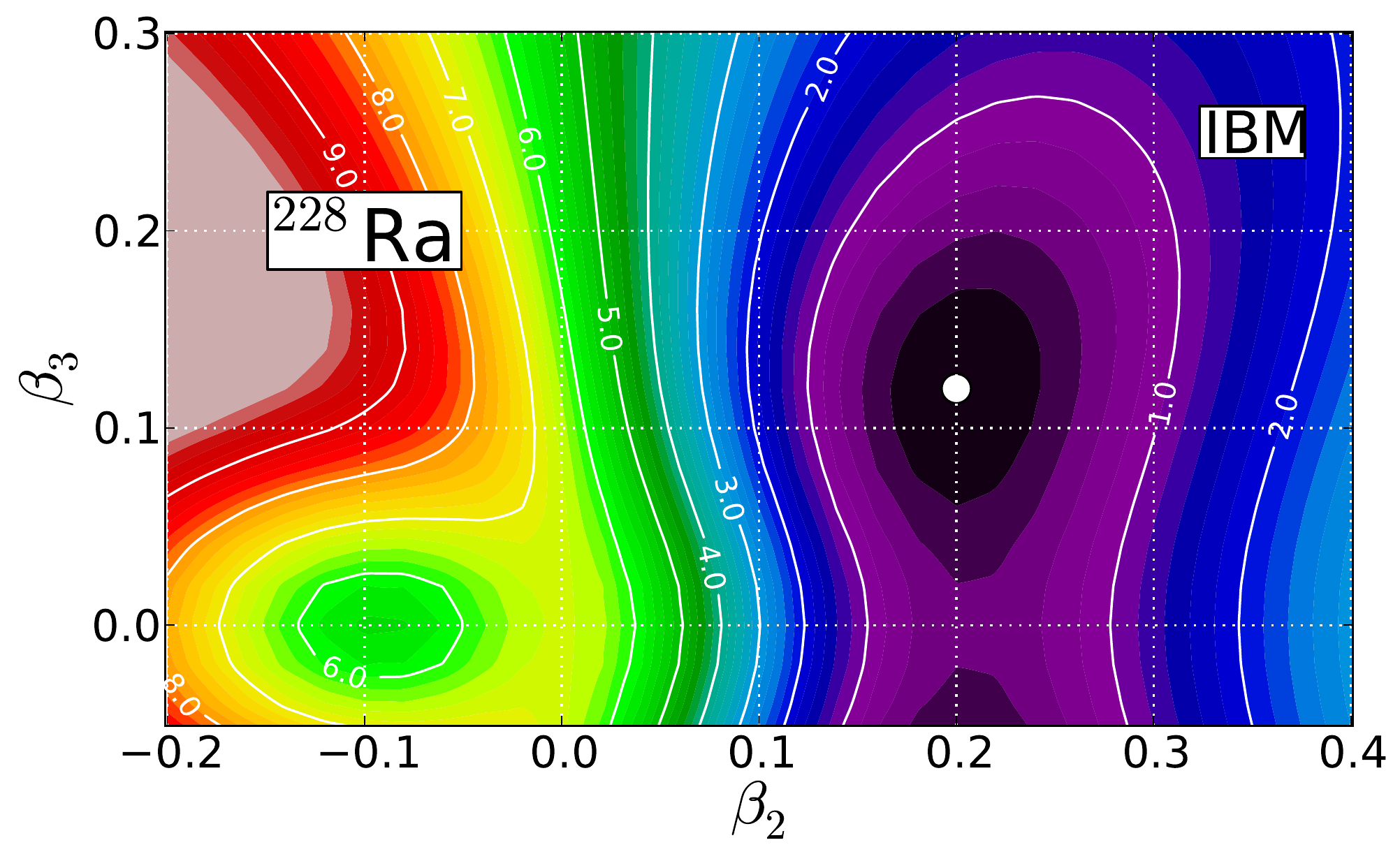}
\end{tabular}
\caption{(Color online) Same as the caption to Fig.~\ref{fig:pes_th},
 but for the mapped IBM energy surfaces of $^{218-228}$Ra. }
\label{fig:pes_ra_mapped}
\end{center}
\end{figure*}

\begin{figure*}[ctb!]
\begin{center}
\begin{tabular}{ccc}
\includegraphics[width=5.7cm]{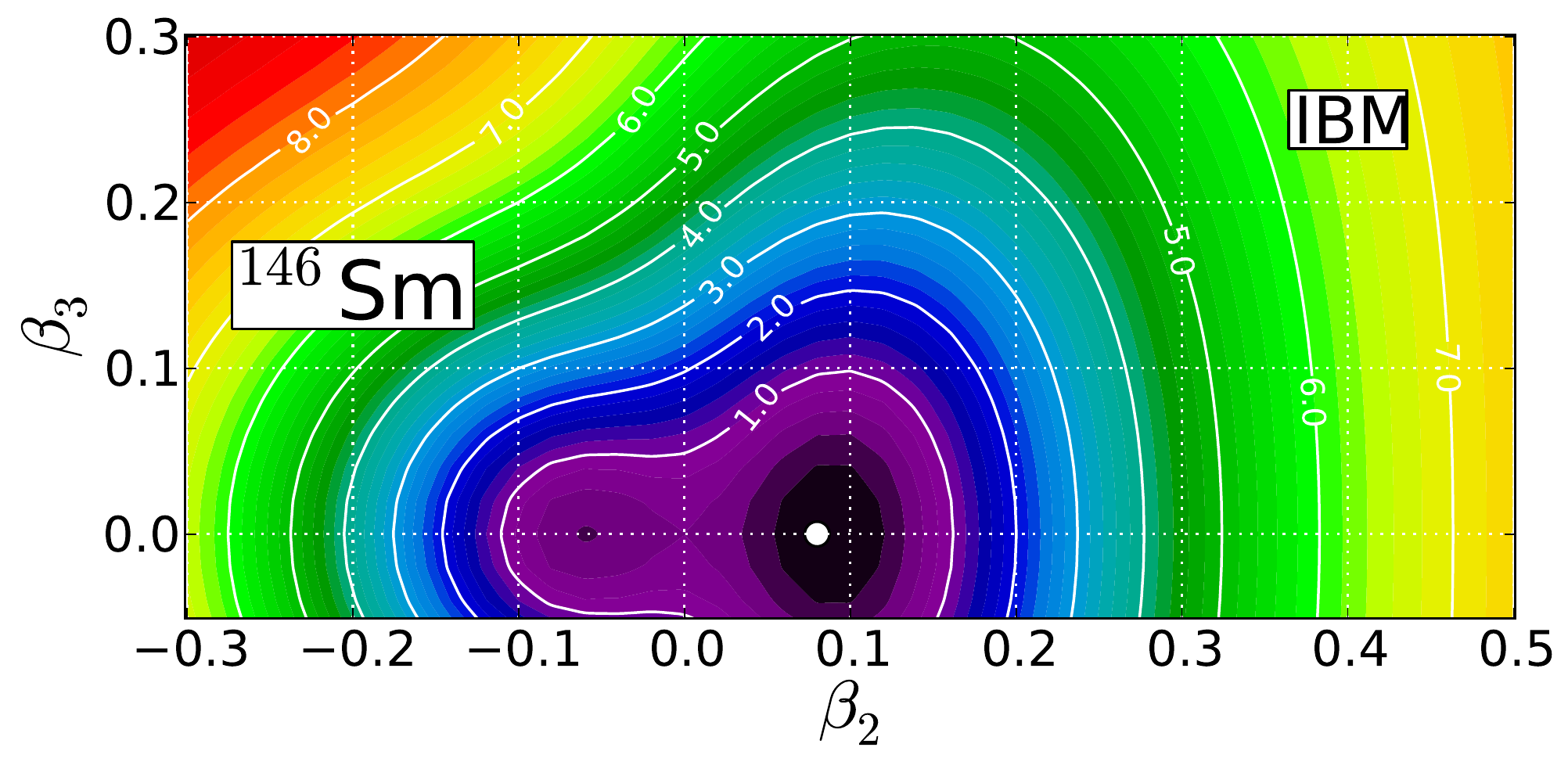} &
\includegraphics[width=5.7cm]{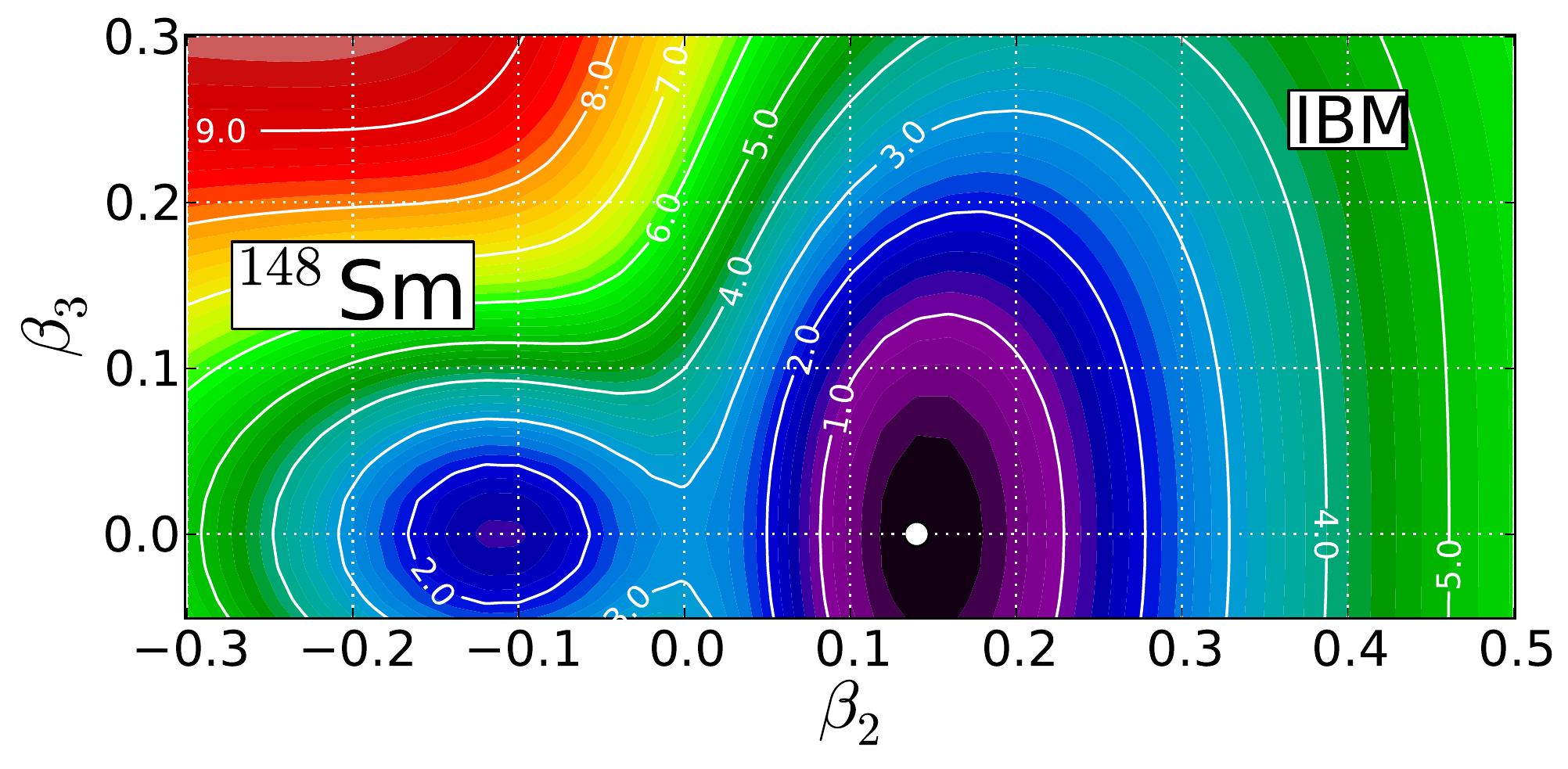} &
\includegraphics[width=5.7cm]{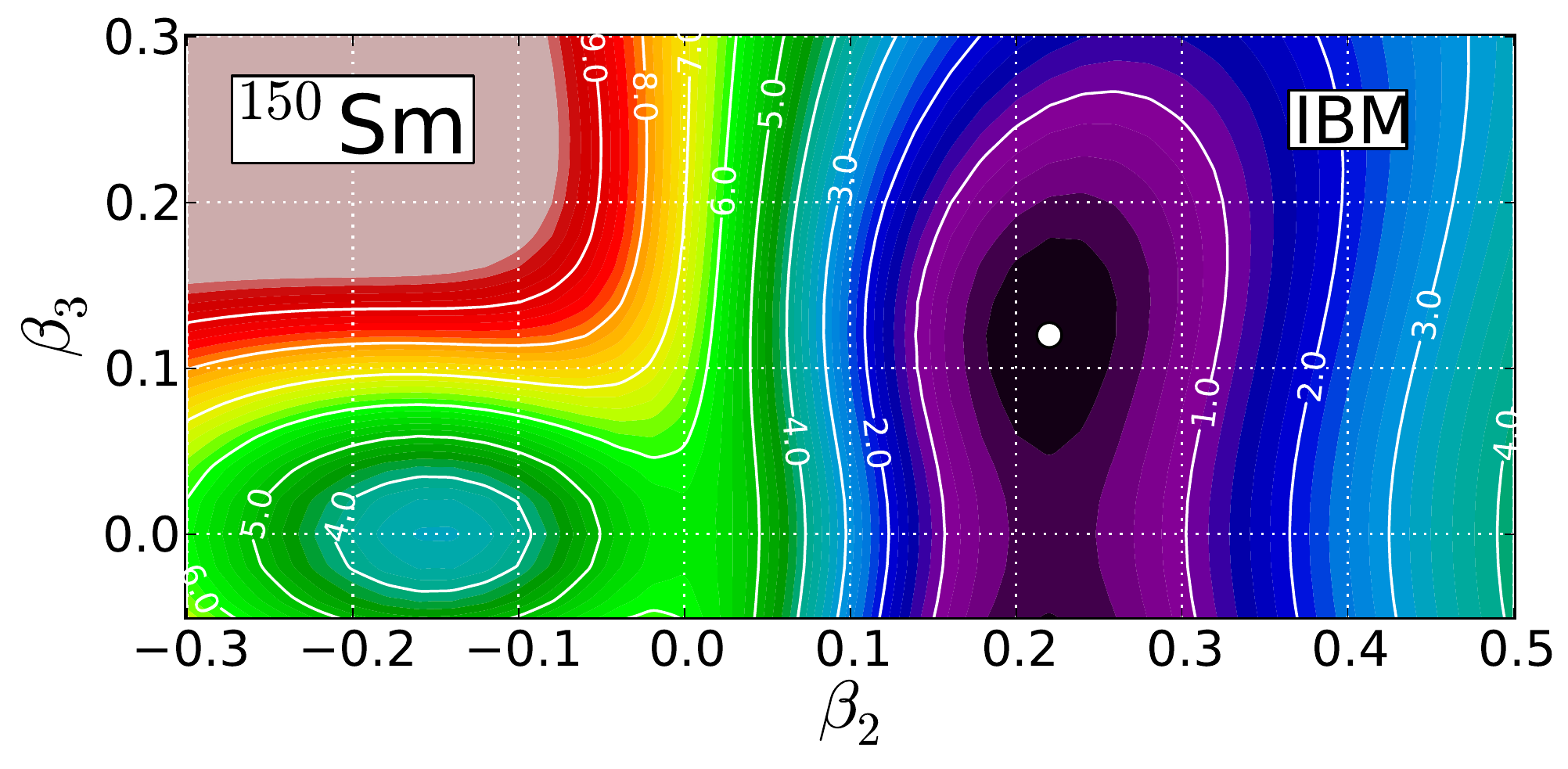} \\
\includegraphics[width=5.7cm]{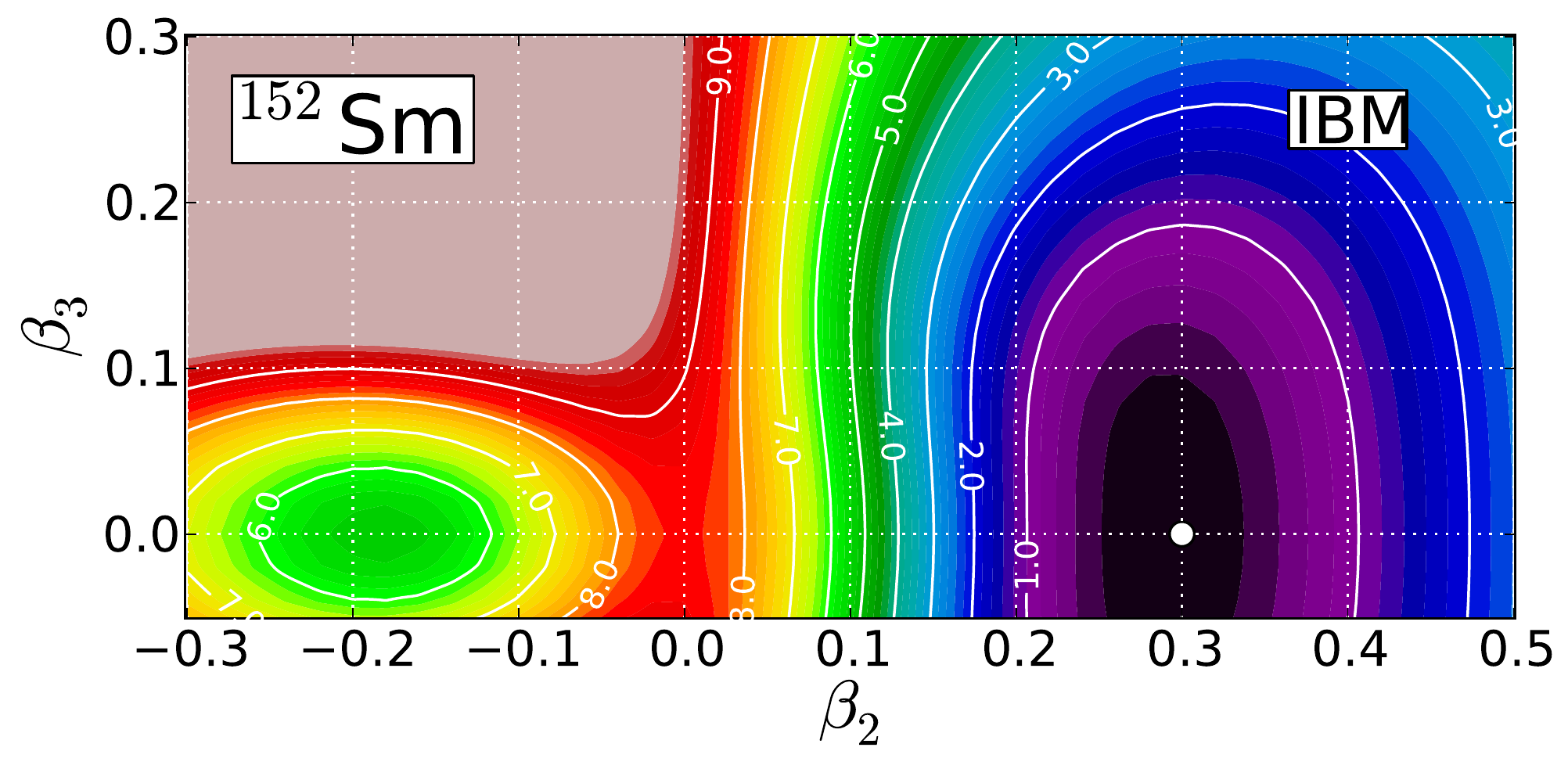} &
\includegraphics[width=5.7cm]{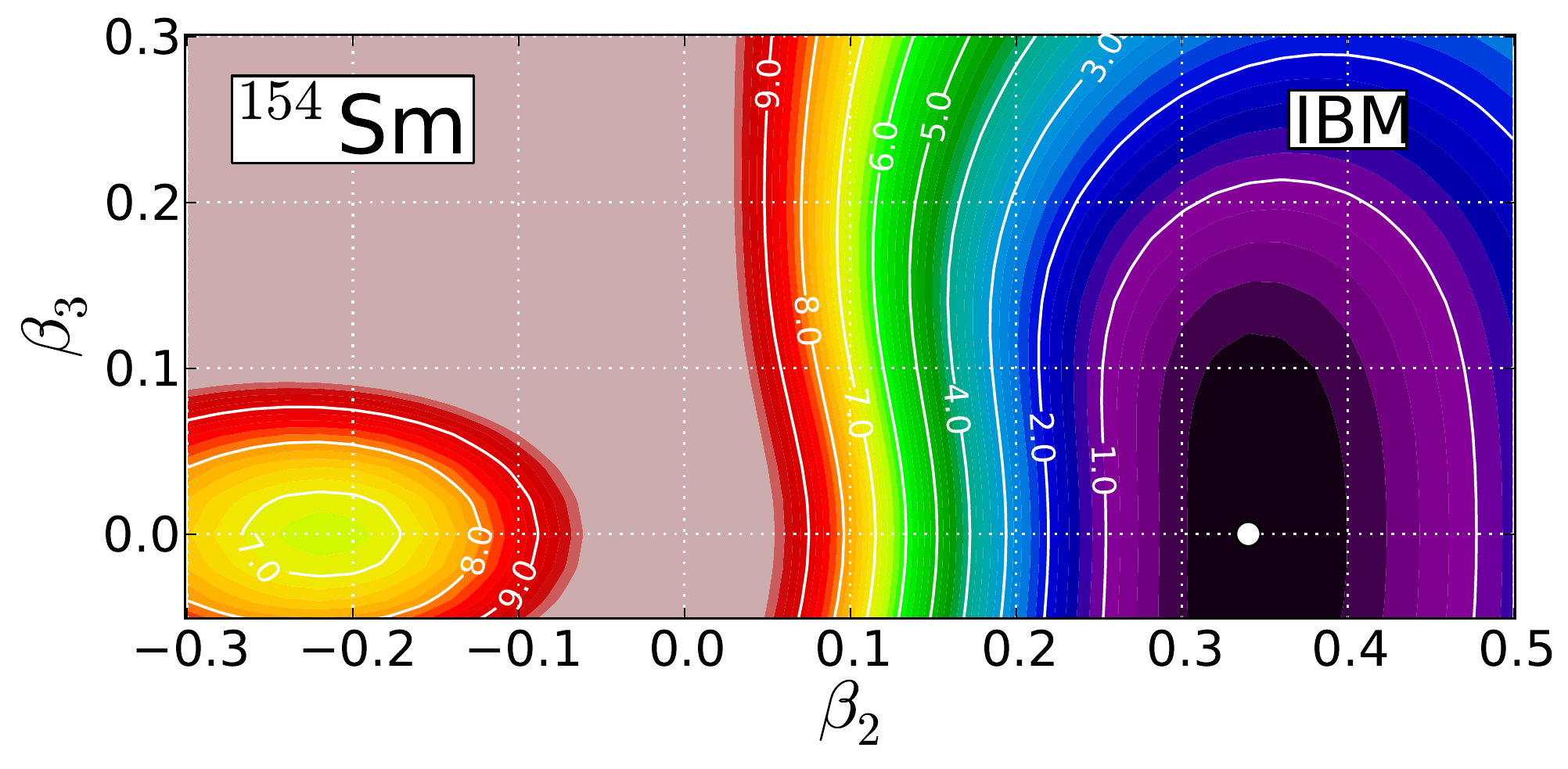} &
\includegraphics[width=5.7cm]{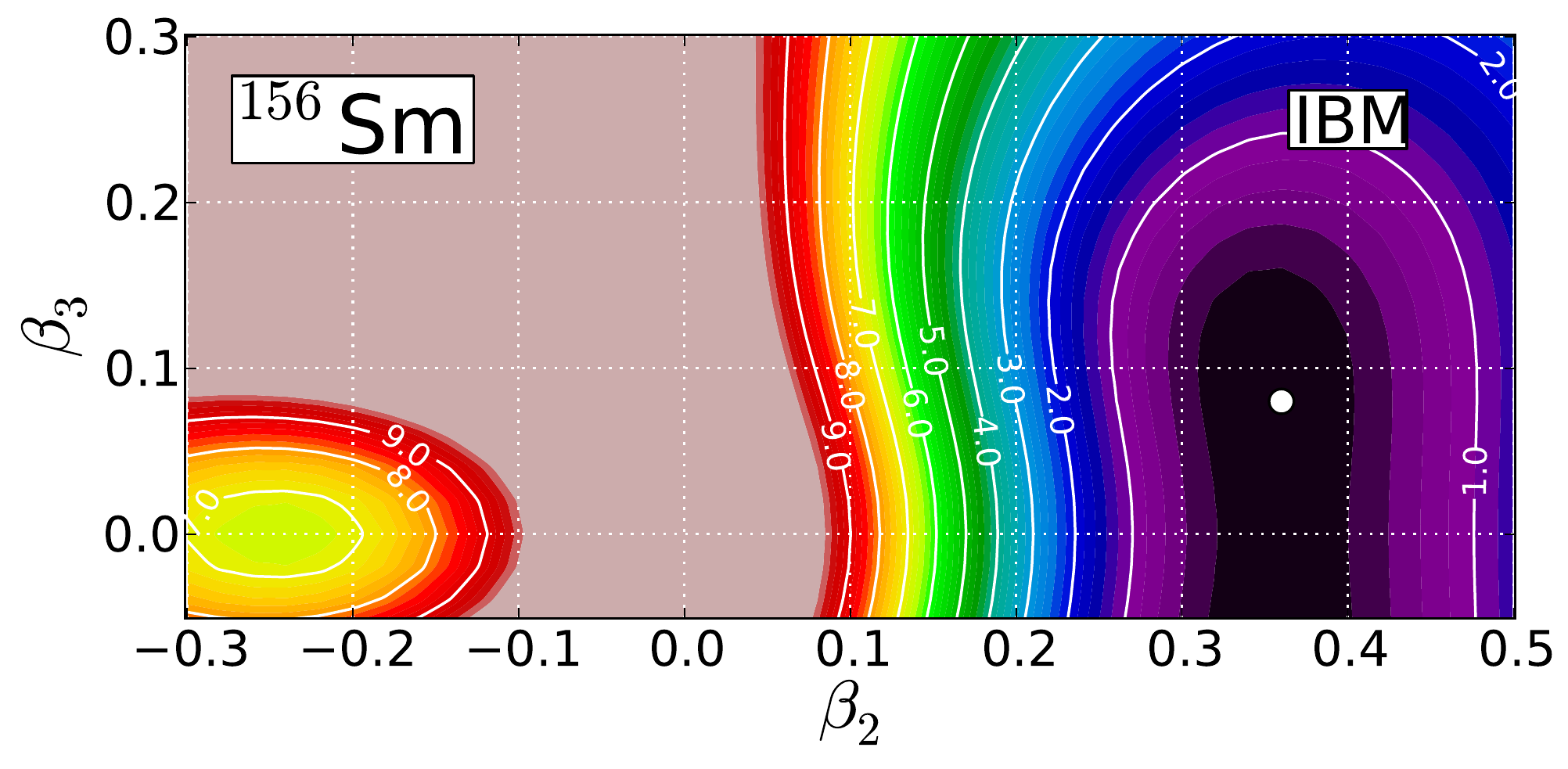} \\
\end{tabular}
\caption{(Color online) Same as the caption to Fig.~\ref{fig:pes_th},
 but for the mapped IBM energy surfaces of $^{146-156}$Sm. }
\label{fig:pes_sm_mapped}
\end{center}
\end{figure*}

\begin{figure*}[ctb!]
\begin{center}
\begin{tabular}{ccc}
\includegraphics[width=5.7cm]{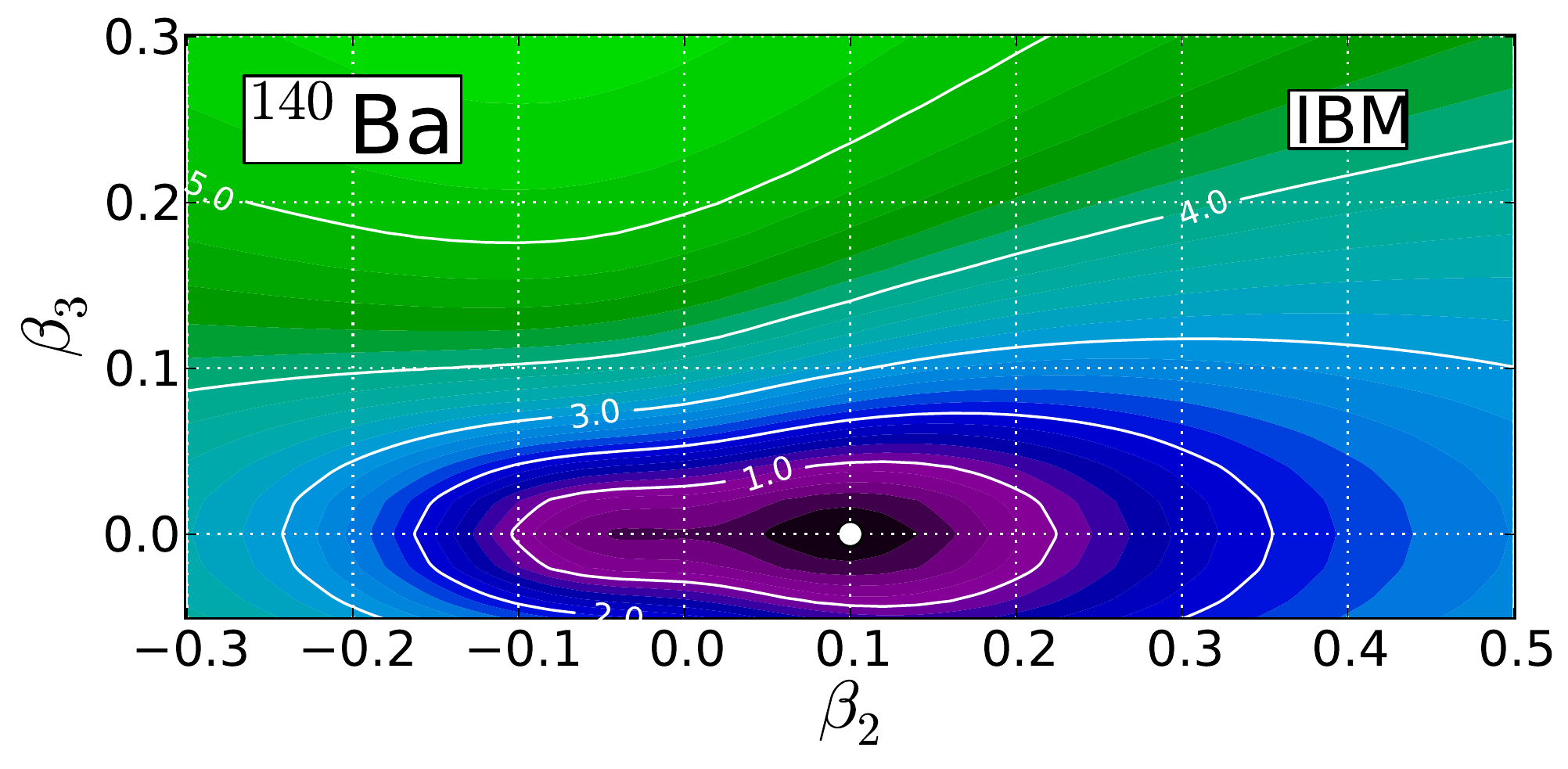} &
\includegraphics[width=5.7cm]{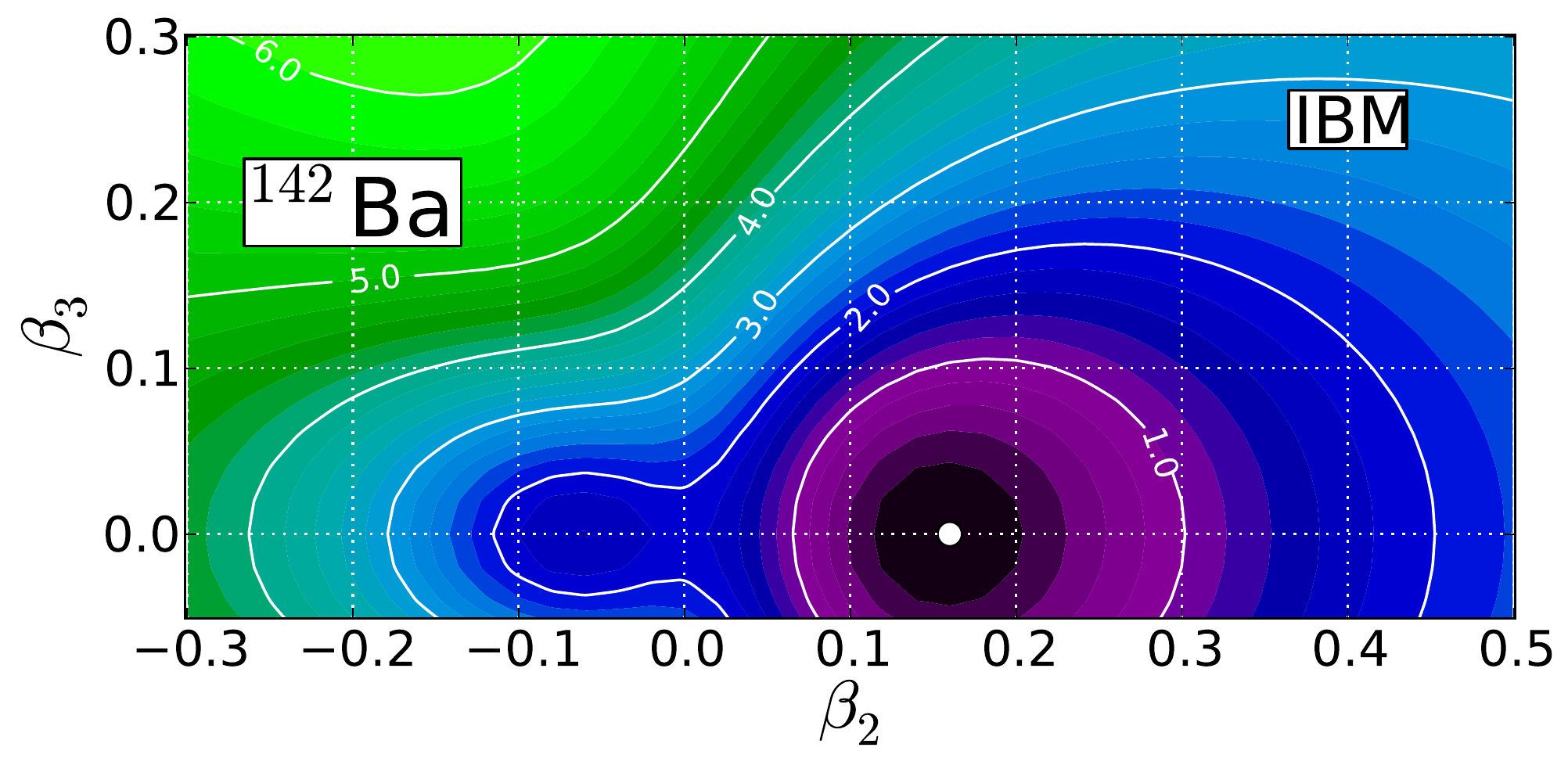} &
\includegraphics[width=5.7cm]{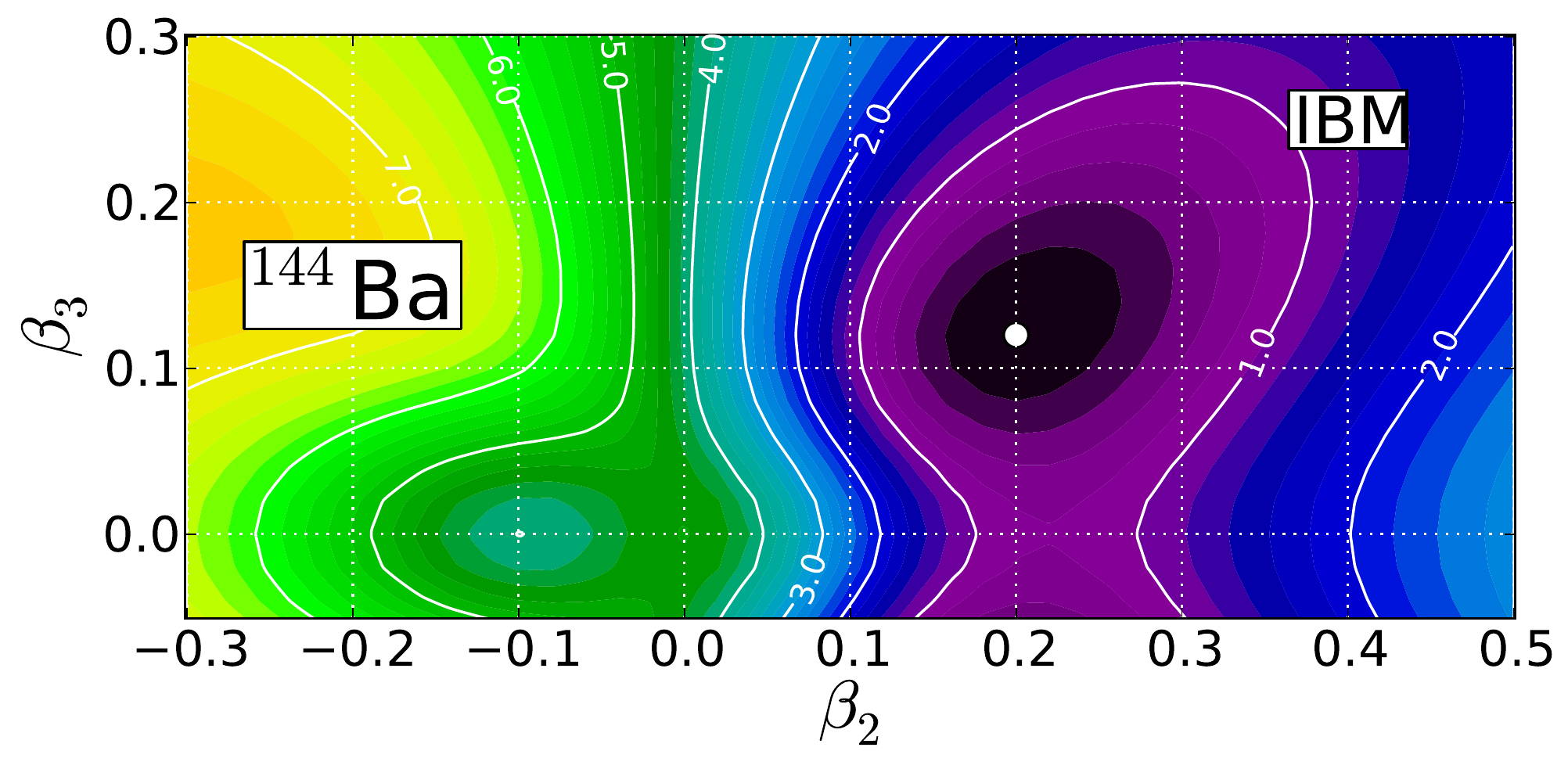} \\
\includegraphics[width=5.7cm]{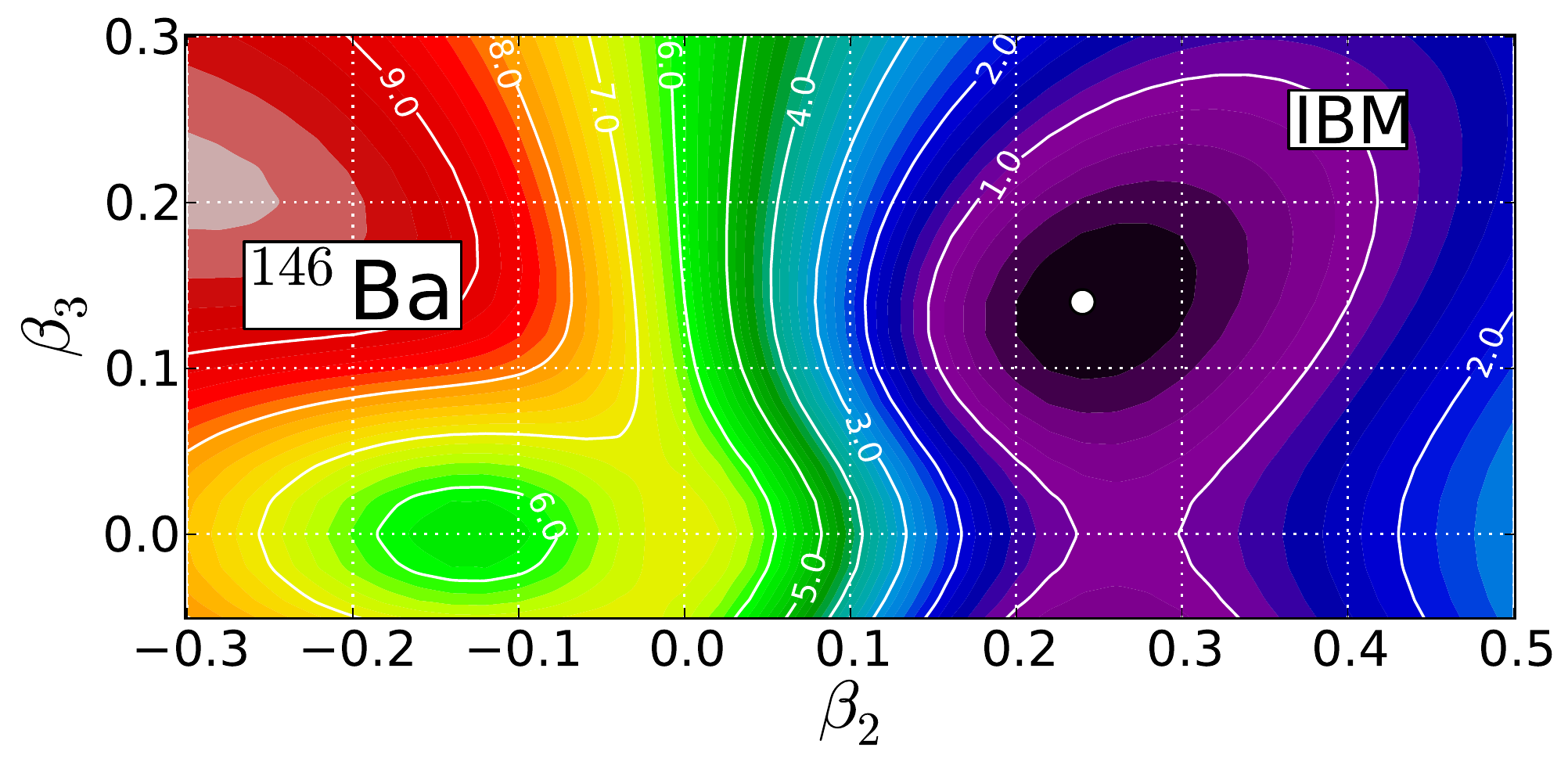} &
\includegraphics[width=5.7cm]{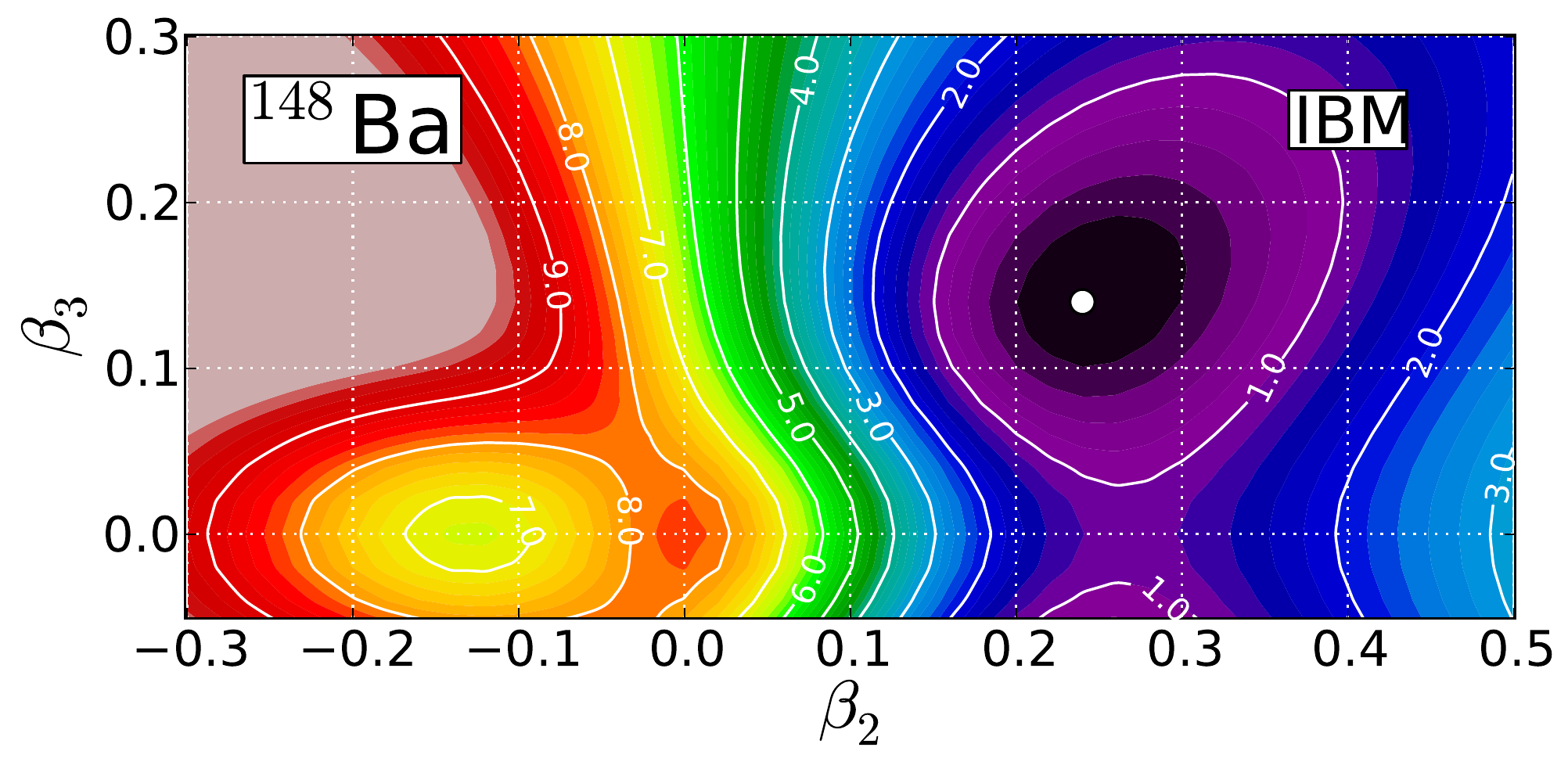} &
\includegraphics[width=5.7cm]{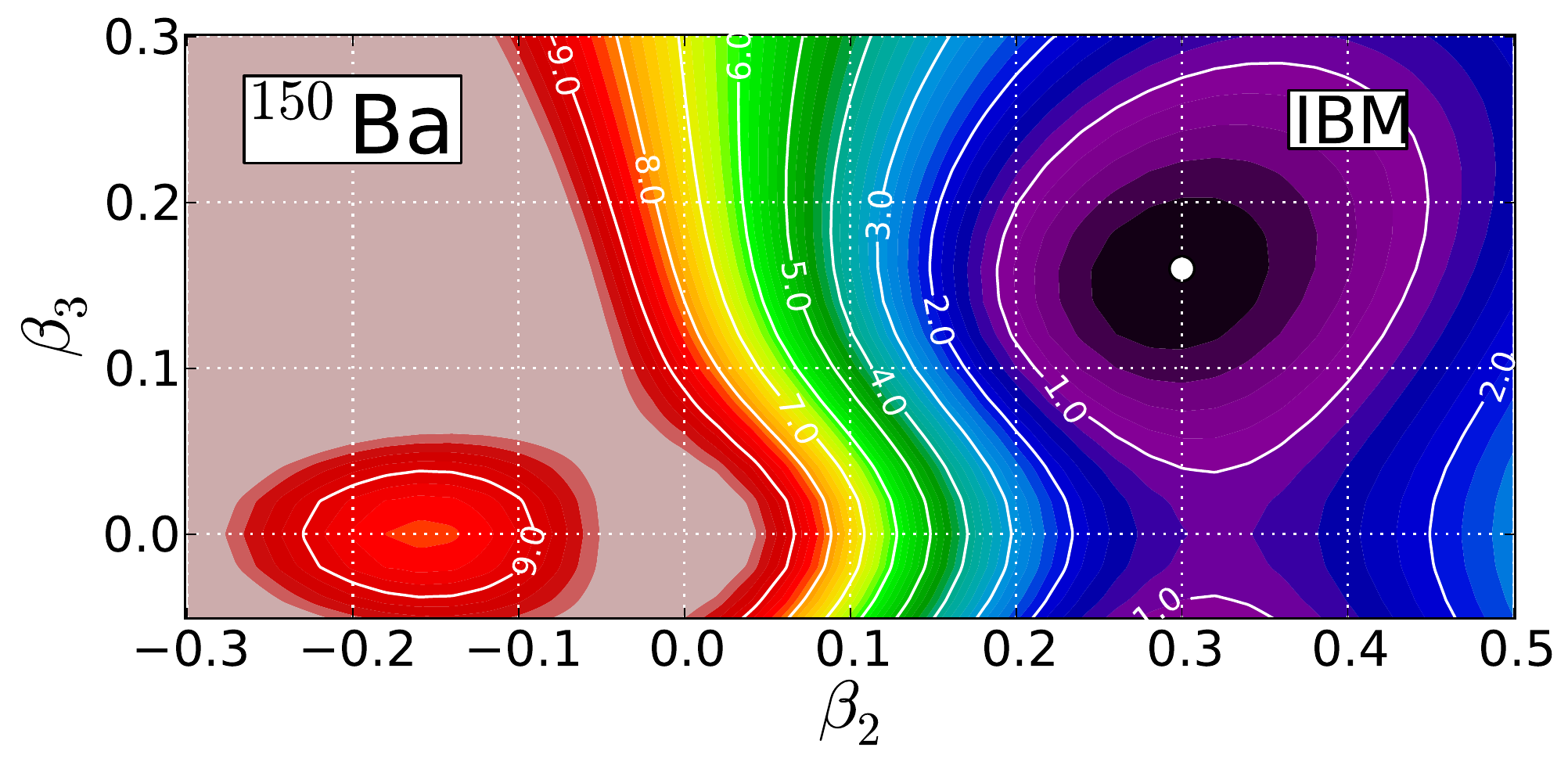} 
\end{tabular}
\caption{(Color online) Same as the caption to Fig.~\ref{fig:pes_th},
 but for the mapped IBM energy surfaces of $^{140-150}$Ba. }
\label{fig:pes_ba_mapped}
\end{center}
\end{figure*}

Figures \ref{fig:pes_th}, \ref{fig:pes_ra}, \ref{fig:pes_sm} and
\ref{fig:pes_ba} display the axially-symmetric deformation energy
surfaces in ($\beta_2,\beta_{3}$) plane, calculated with the constrained
RHB using the microscopic functional DD-PC1, for the isotopes
$^{222-232}$Th, $^{218-228}$Ra, $^{146-156}$Sm and $^{140-150}$Ba,
respectively. 
Each energy surface is plotted up to 10 MeV excitation above its absolute
minimum, and is symmetric with respect to the $\beta_{3}=0$ axis. 
We note that, in Figs.~\ref{fig:pes_th}
and \ref{fig:pes_th_mapped}, the energy surface of $^{220}$Th is not
shown, as it is very similar to the one  
calculated for the adjacent nucleus $^{222}$Th. 

In Fig.~\ref{fig:pes_th} we note that already at the mean-field level
the RHB calculation predicts a very interesting structural evolution in
Th isotopic chain. 
A soft energy surface is calculated for $^{222}$Th, with the minimum in
the region $(\beta_{2},\beta_{3})\approx (0,0)$, and this will typically
lead to low-lying quadrupole vibrational spectra. 
Quadrupole deformation becomes significant in $^{224}$Th, and one also
notices the emergence of octupole deformation. The energy minimum is
found in the $\beta_{3}\neq 0$ region, located at
$(\beta_{2},\beta_{3})\approx (0.15,0.1)$. 
From $^{224}$Th to $^{226,228}$Th the occurrence of a rather strongly
marked octupole minimum is predicted. The deepest octupole minimum is
calculated in $^{226}$Th whereas, starting from $^{228}$Th, the minimum
becomes softer in $\beta_{3}$ direction.  
Octupole-soft surfaces are obtained for 
$^{230,232}$Th, the latter being completely flat in $\beta_{3}$. 
In previous calculations of Th isotopes with the Nilsson-Strutinski method 
that used a deformed Woods-Saxon potential \cite{naza84b,naza85}, a 
quadrupole vibrational shape was obtained for
$^{220}$Th, and a stable octupole deformation was predicted to occur in
$^{222-226}$Th. 
In those studies, the most pronounced octupole minimum was calculated in
$^{224}$Th, a soft octupole shape was obtained for $^{228}$Th and, finally,
a shape without equilibrium octupole deformation for $^{230}$Th. 

The DD-PC1 energy surfaces of Ra isotopes, shown in Fig.~\ref{fig:pes_ra},
display a more gradual evolution of octupole deformation as a function
of mass number, 
and the most pronounced octupole minimum is predicted 
in $^{224}$Ra in the region ($\beta_{2}$,$\beta_{3})\approx
(0.15-0.2,0.10-0.15)$: absolute minimum is found at 
($\beta_{2}$,$\beta_{3})\approx (0.2,0.15)$, while the second minimum,
which is very close in energy, locates at
($\beta_{2}$,$\beta_{3})\approx (0.15,0.10)$. 
These values, particularly the latter one, are rather consistent with
the equilibrium deformation 
$\beta_{2}=0.154$ and $\beta_{3}=0.097$, extracted from the 
experimentally determined intrinsic moments $Q_2$ and $Q_3$, respectively,
in Ref.~\cite{gaffney13}.
The minimum becomes softer in $\beta_{3}$ for $^{226}$Ra, and a
completely $\beta_{3}$-soft potential is predicted in $^{228}$Ra,
characteristic for the onset of octupole vibrations. 
We note that previous mean-field calculations based on the
Nilsson-Strutinski method \cite{naza84b}, and the constrained
Hartree-Fock-Bogoliubov method \cite{robledo10} 
with the Gogny D1S \cite{D1S} and the Barcelona-Catania-Paris (BCP)
\cite{baldo08} effective interactions, predicted the occurrence of the
most prominent octupole equilibrium deformation in $^{222}$Ra. 

In the other mass region considered in the present study, the 
RHB results for   
Sm isotopes, shown in Fig.~\ref{fig:pes_sm}, exhibit a 
simultaneous evolution of both quadrupole and octupole 
deformations with increasing neutron number. A  stable
octupole minimum is predicted to occur in $^{150}$Sm, 
and the deformation energy surface becomes soft in  
$\beta_{3}$ for the heavier isotopes.   
A similar systematic trend was also obtained in a recent analysis that
used the relativistic mean-field plus BCS model \cite{zhang10}, 
based on the PK1 parameter set \cite{long04}. 
In Fig.~\ref{fig:pes_ba} we plot the DD-PC1 deformation energy surfaces
for Ba isotopes. 
Pronounced octupole minima are predicted starting from
$^{144,146}$Ba but, in contrast to the evolution of $\beta_{3}$-softness 
in Sm nuclei, the deformation of the energy surface does not become 
much softer in $\beta_{3}$ for heavier isotopes. Shallow octupole 
equilibrium minima are obtained in $^{146}$Ba -- $^{150}$Ba. 
The evolution of quadrupole and octupole deformations in the Ba isotopes 
predicted by the functional DD-PC1, and in particular the octupole
minimum in 
$^{144,146}$Ba, is consistent with previous mean-field
calculations based on the Nilsson-Strutinski method \cite{naza84b} and
the Gogny D1S effective interaction \cite{robledo10}. 
Both studies, however, predicted the disappearance of the octupole minimum 
in $^{150}$Ba, whereas a shallow minimum with a non-zero $\beta_{3}$
value is predicted in the present calculation.

Figures \ref{fig:pes_th_mapped} -- \ref{fig:pes_ba_mapped} display
the corresponding IBM energy surfaces of Th, Ra, Sm and Ba, mapped 
from the self-consistent mean-field results shown in 
Figs.~\ref{fig:pes_th} -- \ref{fig:pes_ba}.  
As an illustrative example we discuss in more detail the Th
isotopes. Fig.~\ref{fig:pes_th_mapped} shows that the mapping 
reproduces the evolution of octupole minima  
as a function of neutron number: the quadrupole
minimum for $^{222}$Th, the onset of octupole 
minimum in $^{224}$Th, the pronounced octupole deformation in $^{226,228}$Th,
and the soft-octupole potential for $^{230,232}$Th, 
originally displayed by the microscopic energy surfaces in Fig.~\ref{fig:pes_th}. 
If one considers in detail each individual nucleus, particularly the
heavier Th isotopes with $A\geq 226$, it can be noticed that the topology of the
mapped IBM energy surface reproduces the original 
microscopic surface in the region close to the minimum, whereas more 
pronounced differences between the microscopic and
IBM energy surface are observed for $\beta_{2}\le 0.0$. 
The reason is that the
IBM Hamiltonian is constructed in such a way that it reproduces only the 
neighborhood of the minimum of the microscopic
energy surface, which is most relevant 
for the low-lying collective states considered in this work. The 
IBM energy surface of Eq.~(\ref{eq:oct.pes}) cannot reproduce all the 
details of a more complex topology of the microscopic
energy surface. 
One also notices that generally in regions very far from the 
minimum the mapped energy surfaces are more smooth in comparison
to the microscopic ones, which is again due to the restricted boson model
space of the IBM compared to the fermion space of the RHB framework. 
This is a general feature of the IBM approach and  
cannot be modified by simply readjusting the model parameters. 
In particular the deviations are more apparent for isotopes 
like $^{222,224}$Th, that are characterized by a smaller
boson number. 
For the other isotopic chains shown from Figs.~\ref{fig:pes_ra_mapped}
to \ref{fig:pes_ba_mapped}, one should find more or less the same extent
of similarity as for the Th chain between the DD-PC1 and the IBM energy
surfaces. 
Nevertheless, for the Ra (Ba) nuclei, the IBM energy surfaces are shallower
than for the Th (Sm) nuclei because the number of bosons is more
limited. 

\begin{figure}[ctb!]
\begin{center}
\includegraphics[width=8.6cm]{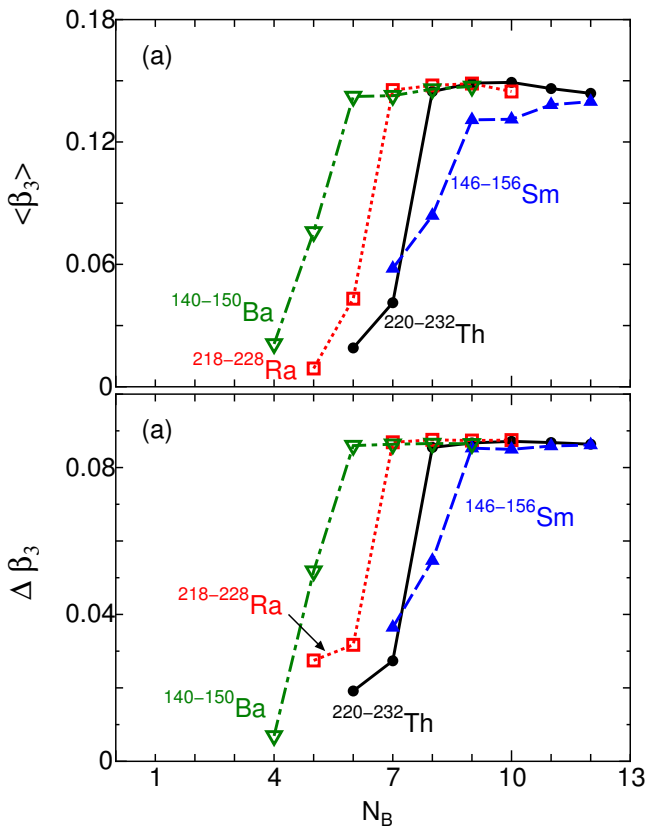}
\caption{(Color online) Mean value of the octupole deformation
parameter $\langle\beta_{3}\rangle$ for $^{220-232}$Th,
 $^{218-228}$Ra, $^{146-156}$Sm and $^{140-150}$Ba, 
computed with the IBM intrinsic state (a), and the corresponding variance 
$\Delta\beta_{3}=\sqrt{\langle\beta_{3}^{2}\rangle-\langle\beta_{3}\rangle^{2}}$ (b), 
as functions of the boson number $N_{B}$. }
\label{fig:bet3}
\end{center}
\end{figure}

Figure~\ref{fig:bet3} displays the mass dependence of the
average value of the octupole 
deformation $\langle\beta_{3}\rangle$ and the variance
$\Delta\beta_{3}=\sqrt{\langle\beta_{3}^{2}\rangle-\langle\beta_{3}\rangle^{2}}$,
for $^{220-232}$Th, $^{218-228}$Ra, $^{146-156}$Sm and $^{140-150}$Ba, 
obtained from the mapped IBM energy surfaces. 
The average octupole deformation and the variance are 
calculated over a region on the 
$(\beta_{2},\beta_{3})$ plane that extends from the absolute 
mean-field minimum up to approximately 2 MeV excitation energy 
above the minimum. 
A prominent feature in Fig.~\ref{fig:bet3} (a) is the sudden increase of  
$\langle\beta_{3}\rangle$ from $^{222}$Th to $^{224}$Th, from $^{220}$Ra
to $^{222}$Ra, from $^{148}$Sm to $^{150}$Sm, and from $^{142}$Ba to
$^{144}$Ba. 
For these nuclei the energy minimum is suddenly displaced from the
$\beta_{3}\approx 0$ axis to the $\beta_{3}\neq 0$ region. For each isotopic chain
the value of $\langle\beta_{3}\rangle$ stabilizes at $\beta_{3}\approx
0.15$ in heavier nuclei.  
The mass dependence of the variance $\Delta\beta_{3}$ in 
Fig.~\ref{fig:bet3}(b) reflects the fluctuation in 
the octupole deformation between the non-octupole deformed shape and
octupole deformation. 

\subsection{Parameters of the quadrupole-octupole IBM Hamiltonian}

\begin{figure*}[ctb!]
\begin{center}
\includegraphics[width=17cm]{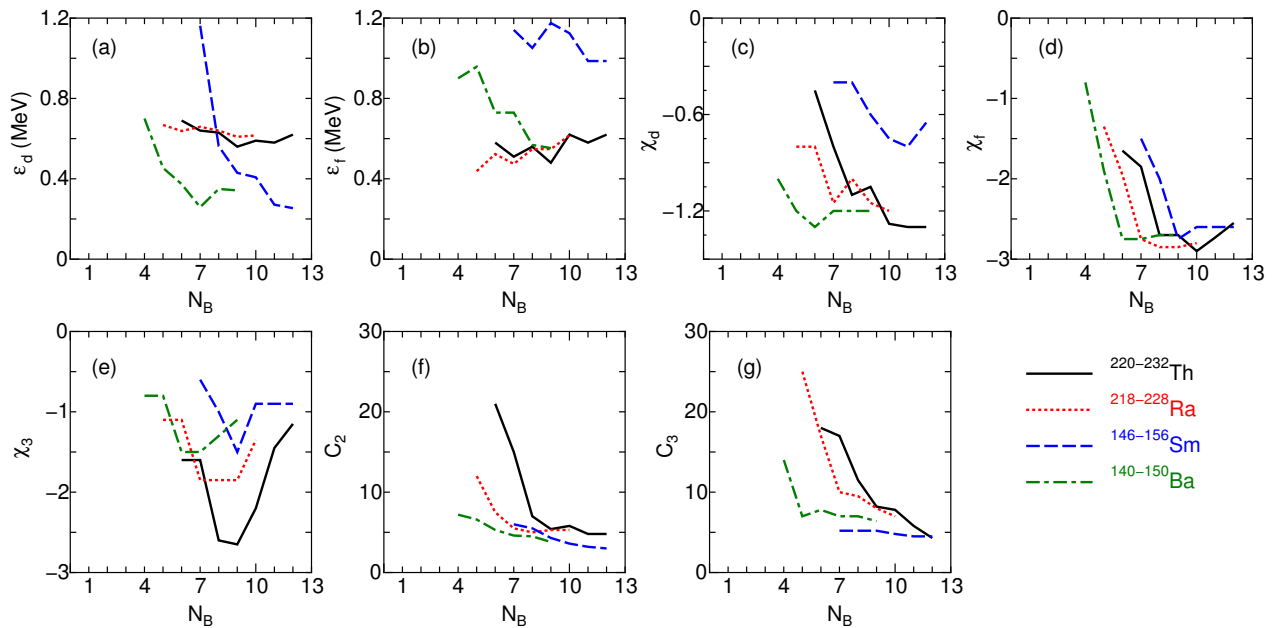}
\caption{(Color online) Total boson number $N_{B}$ dependence of the parameters for the $sdf$ IBM
 Hamiltonian Eq.~(\ref{eq:ham}): 
 $\epsilon_{d}$ (a), $\epsilon_{f}$ (b), $\chi_{d}$ (b), $\chi_{f}$ (c),
 $\chi_{f}$ (d) and 
 $\chi_{3}$ (e), and the coefficient of the shape variables
 $C_{2}$ (f) and $C_{3}$ (g), determined by mapping the 
 microscopic RHB energy surfaces onto the corresponding 
 expectation values of the IBM Hamiltonian. Solid, dotted, dashed, and 
 dot-dashed lines connect the parameters of the Th, Ra, Sm, and 
 Ba isotopic chains, respectively.}
\label{fig:para}
\end{center}
\end{figure*}

In Fig.~\ref{fig:para} we plot the adopted parameters for the $sdf$ IBM 
 Hamiltonian Eq.~(\ref{eq:ham}): 
 $\epsilon_{d}$ (a), $\epsilon_{f}$ (b), $\chi_{d}$ (b), $\chi_{f}$ (c),
 $\chi_{f}$ (d) and 
 $\chi_{3}$ (e), and the coefficient of the shape variables
 $C_{2}$ (f) and $C_{3}$ (g), determined by mapping the 
 microscopic RHB energy surfaces onto the corresponding 
 expectation values of the IBM Hamiltonian in the boson 
 condensate state Eq.~(\ref{eq:coherent}). 
One notices a certain trend in a variation of each parameter 
as a function of boson number, that correlates with the variation of the
intrinsic shape (Figs.\ref{fig:pes_th}-\ref{fig:pes_ba_mapped}). 
The decrease of the $d$ (Fig.~\ref{fig:para}(a)) and $f$
(Fig.~\ref{fig:para}(b)) single-boson energies with $N_{B}$ reflects the
evolution of quadrupole and octupole collectivity, respectively. 
An interesting feature is that, while for Sm and Ba isotopes generally 
$\epsilon_{d}\le \epsilon_{f}$ (consistent with
the study of Ref.~\cite{babilon05} for Sm isotopes), 
the values $\epsilon_{d}$ and $\epsilon_{f}$ are similar 
for the Th and Ra isotopic chains. This indicates that for the 
latter mass region octupole and quadrupole collectivity are comparable
in magnitude. 

Similarly to $\epsilon_{d}$ and $\epsilon_{f}$, the values of
the parameters $\chi_{d}$ (Fig.~\ref{fig:para}(c)) and $\chi_{f}$ 
(Fig.~\ref{fig:para}(d)) decrease with $N_{B}$ and appear to
saturate around $\chi_{d}=-1$ and $\chi_{f}= -2.5$. 
The former value is close to the SU(3) limit for the $sd$ sector, 
$\chi_{d}=- \sqrt{7}/2$ \cite{IBM}. For each chain the
parameter $\chi_{3}$ reaches a maximum 
in magnitude for a particular isotope, that is, for 
$^{226}$Th ($N_{B}=9$), $^{224}$Ra ($N_{B}=8$), $^{150}$Sm ($N_{B}=9$)
and $^{144,146}$Ba ($N_{B}=6,7$ (Fig.~\ref{fig:para}(e)). 
As expected from the analyses in Sec.~\ref{sec:pes},  
octupole deformation is most pronounced in these nuclei. 
By further increasing the number of bosons the magnitude of $\chi_{3}$  
generally decreases, reflecting the evolution of octupole softness
(cf. Figs.~\ref{fig:pes_th}, \ref{fig:pes_ra}, \ref{fig:pes_sm} and
\ref{fig:pes_ba}).  
The scale parameters $C_{2}$ (Fig.~\ref{fig:para}(f)) and $C_{3}$ 
(Fig.~\ref{fig:para}(g)) evolve monotonically with $N_{B}$ according to
the change of the location of the energy minimum on the RHB energy
surfaces.  

The values of the remaining parameters $\kappa_{2}$, $\kappa_{3}$
and $\alpha$ are taken to be almost constant with respect to neutron
number: 
$\kappa_{2}\approx -0.06$ MeV for the Th-Ra isotopes, $\kappa_{2}\approx
-0.088\sim -0.08$ MeV for Sm, and $\kappa_{2}\approx 
-0.11\sim -0.09$ for Ba; $\kappa_{3}\approx 0.06$ MeV for
Th-Ra, $\kappa_{3}\approx 0.02$ MeV for Sm, and $\kappa_{3}\approx 0.04$
MeV for the Ba isotopes; $\alpha\approx -0.02$ MeV for Th-Ra, $^{152-156}$Sm
and $^{146-150}$Ba, $\alpha\approx -0.01$ MeV for $^{146-150}$Sm and
$^{140-144}$Ba.

\begin{figure*}[ctb!]
\begin{center}
\begin{tabular}{cc}
\includegraphics[width=14.0cm]{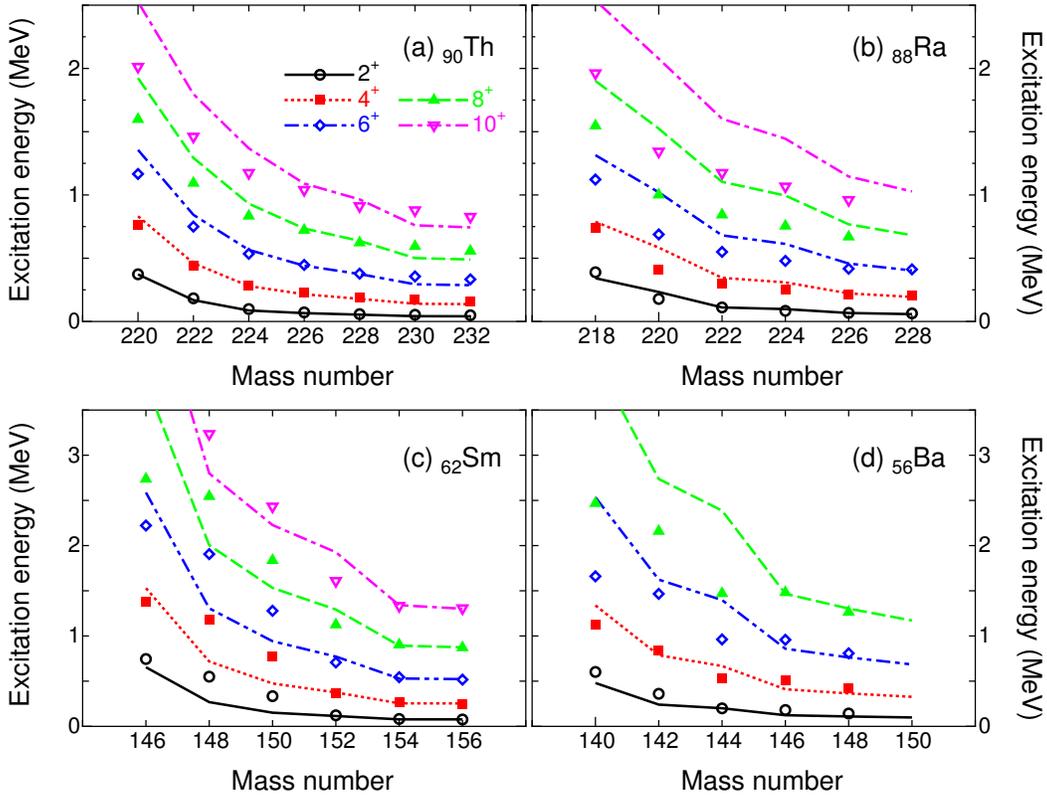}&
\end{tabular}
\caption{(Color online) Excitation energies of low-lying yrast positive-parity
 collective states of $^{220-232}$Th, $^{218-228}$Ra, $^{146-156}$Sm
 and $^{140-150}$Ba, as functions of the 
 mass number. In each panel lines and symbols denote the 
 theoretical and experimental 
 \cite{data} values, respectively. Figure legends are found in panel
 (a). }
\label{fig:pspec}
\end{center}
\end{figure*}

\begin{figure*}[ctb!]
\begin{center}
\begin{tabular}{cc}
\includegraphics[width=14.0cm]{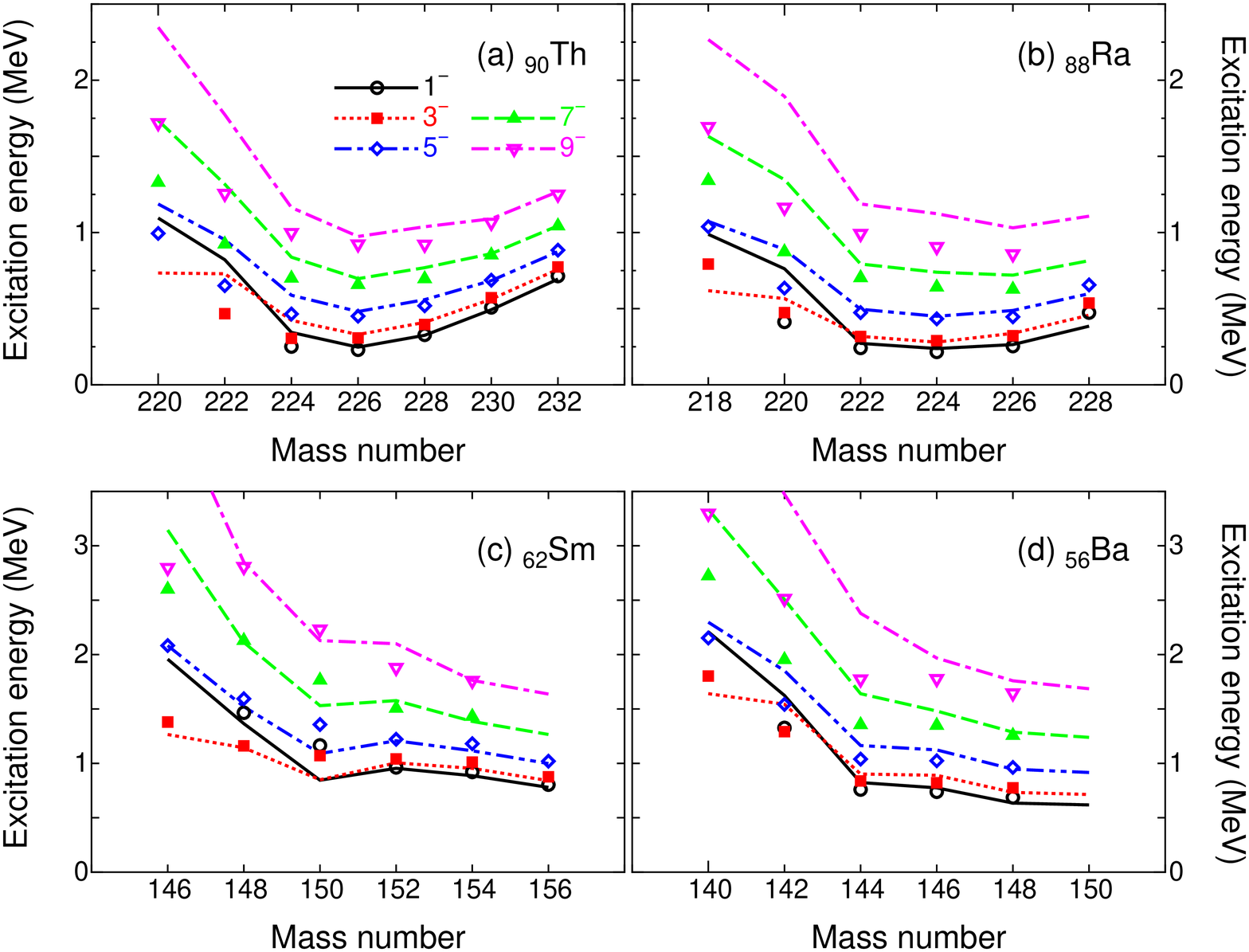}&
\end{tabular}
\caption{(Color online) Same as in the caption to Fig.~\ref{fig:pspec} 
 but for the negative-parity states. }
\label{fig:nspec}
\end{center}
\end{figure*}

\section{Spectroscopic properties \label{sec:results}}

A signature of static reflection-asymmetric shapes is a negative-parity 
$K^\pi =0^-$ band with $J^\pi =$ $1^-, 3^-, 5^-, \ldots$,
located close in energy to the positive-parity 
ground-state band $J^\pi =$ $0^+, 2^+, 4^+, \ldots$. One could
say that the two sequences form a single alternating-parity band, 
with states connected by strong E1 transitions
\cite{butler96}. In most nuclei, however, the two 
bands are displaced from each other and an approximate alternating-parity band is
experimentally observed only for states with higher spin, typically $J>5$. 
In the case of dynamical octupole deformation the negative-parity band lies at 
considerably higher energy and 
the two sequences with $K^\pi =0^+$ and $K^\pi =0^-$ form separate bands. 
An increase of the excitation energy of the negative-parity band 
relative to that of the positive-parity ground-state band indicates 
a transition from octupole deformation to octupole vibrations
\cite{butler96}. 

In this section we analyze the theoretical spectroscopic properties 
that characterize the evolution of octupole deformation in the four 
isotopic chains. Part of the results for $^{220-230}$Th have been 
already reported in our recent study of Ref.~\cite{nom13oct}, but are
included here for completeness and compared to those obtained for Ra
nuclei. 

\subsection{Systematics of excitation energies}

In Fig.~\ref{fig:pspec} we display the systematics of 
the calculated excitation energies of the positive-parity 
ground-state band ($K^{\pi}=0^{+}$), and 
in Fig.~\ref{fig:nspec} the lowest negative-parity 
($K^\pi =0^-$) sequences in   
$^{220-232}$Th,
$^{218-228}$Ra, $^{146-156}$Sm and $^{140-150}$Ba nuclei, in
comparison with available data \cite{data}. 
Firstly we note that, even without any additional adjustment of the 
parameters to data, that is, by using parameters determined exclusively by the 
microscopic calculation of potential energy surfaces, the IBM quantitatively 
reproduces the mass dependence of the excitation energies of levels 
that belong to the lowest bands of positive and negative parity. 

The excitation energies of positive-parity states systematically decrease 
with mass number, reflecting the increase of quadrupole collectivity. For instance,
$^{220,222}$Th exhibit a quadrupole vibrational structure, whereas 
pronounced ground-state rotational bands with
$R_{4/2}=E(4^{+}_{1})/E(2^{+}_{1})\approx 3.33$ are found in
$^{226-232}$Th. 
For the nuclei $^{224,226}$Th, located in the transitional region 
between quadrupole vibrators and axially deformed rotors, the experimental 
$R_{4/2}$ values are 2.90 and 3.14, respectively, the former being
close to the value 2.91 predicted by the X(5) model \cite{X5} for 
reflection-symmetric axially-deformed nuclei. 
In the calculation $R_{4/2}=3.22$ and 3.26 for $^{224,226}$Th,
respectively. 
A similar systematics is found in the other
isotopic chains (Figs.~\ref{fig:pspec}(b-d)), that is, the evolution of quadrupole
collectivity characterized by the lowering of the positive-parity
ground-state band. 
However, the theoretical predictions for the positive-parity states
overestimate the experimental values.
The discrepancies are larger for the Ra and Ba isotopes because the
boson model space is more restricted in comparison to the
neighboring Th and Sm isotopes.

In the present analysis, however, we are more concerned with negative-parity states. 
For the states of the negative-parity band in Th isotopes 
the excitation energies display a parabolic 
structure centered between $^{224}$Th and $^{226}$Th (Fig.~\ref{fig:nspec}(a)). 
The approximate parabola of $1^{-}_{1}$ states has a minimum 
at $^{226}$Th, in which the octupole deformed minimum 
is most pronounced (cf. Figs. \ref{fig:pes_th} and \ref{fig:bet3}). 
Starting from $^{226}$Th the energies of negative-parity 
states systematically increase and the band becomes more 
compressed. A rotational-like collective band based on the octupole vibrational
state, i.e., the $1^{-}_{1}$ band-head, develops. 

For the Ra isotopes shown in Fig.~\ref{fig:nspec}(b) a similar trend,
that is, an approximate parabola of negative-parity yrast
states, is predicted particularly for states with spin
$J^{\pi}=1^{-}$, $3^{-}$ and $5^{-}$. One notices that  
the parabolic dependence is not as pronounced as in the Th case. 
The model predicts that the excitation energy of the $1^{-}_{1}$ state is
lowest in $^{224}$Ra. This result is consistent with the evolution of the experimental
low-spin negative-parity states with neutron number \cite{data}, and also 
with the recent experimental study of stable octupole deformation in $^{224}$Ra \cite{gaffney13}. 
Namely, the experimental states $1^{-}_{1}$ ($3^{-}_{1}$) are observed at
242 (317) keV, 216 (290) keV, and 254 (322) keV, in $^{222,224,226}$Ra, respectively \cite{data}. 
On the other hand, in both the positive and negative parity bands some
high-spin states, particularly for the lighter isotopes, are predicted at much 
higher energies compared to the data \cite{data}. One of the reasons 
is certainly the restricted valence space from which boson states are built.
The recent Hartree-Fock-Bogoliubov calculations
\cite{robledo10,robledo13}, based on the Gogny D1S and D1M \cite{D1M} functionals, 
reproduced the parabolic-like dependence of the $1^{-}_{1}$
state in the Ra isotopes as a function of $N$, similarly to the result obtained in the present study. 

For the Sm (Fig.~\ref{fig:nspec}(c)) and Ba (Fig.~\ref{fig:nspec}(d))
isotopes, the mass dependence of negative-parity yrast states is 
more monotonic. For Sm the calculated excitation energies of both 
positive- and negative-parity states show a very weak variation with mass 
number starting from $^{152}$Sm or $^{154}$Sm. The octupole
vibrational structure reflects the systematics of the
deformation energy surfaces shown in Fig.~\ref{fig:pes_sm}. 
The yrast states of Ba isotopes display no significant
structural change  
starting from $^{144}$Ba or $^{146}$Ba, that is, the excitation energies of 
both positive- and negative-parity yrast states look almost constant
with mass (neutron) number. 
Note, however, that the calculated energy levels for Ba isotopes exhibit
a more abrupt change from $^{144}$Ba to $^{146}$Ba, especially
for higher-spin states.  
Other EDF-based approaches have predicted an isotopic dependence of the
$1^{-}_{1}$ level of Ba isotopes (decrease in energy from $^{140}$Ba to
$^{144}$Ba or $^{146}$Ba) similar to the present result \cite{robledo10}, but also 
a parabolic behavior for the Sm chain with a minimum at $^{150}$Sm \cite{rayner12}, 
rather different from the present trend shown in Fig.~\ref{fig:nspec}(c) and
from the experimental excitation energies \cite{data}. 

\begin{figure*}[ctb!]
\begin{center}
\begin{tabular}{cc}
\includegraphics[width=8.6cm]{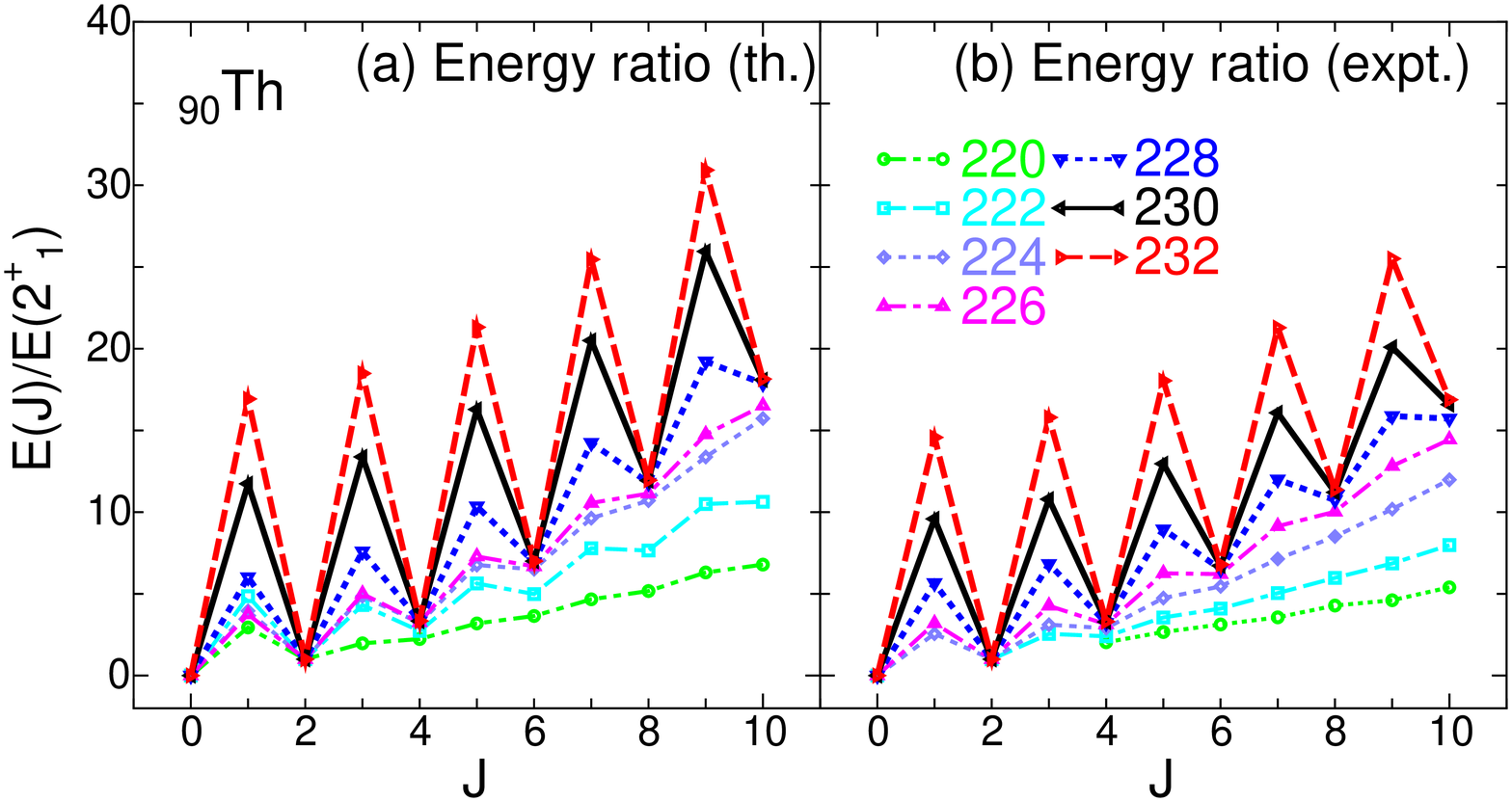} & 
\includegraphics[width=8.6cm]{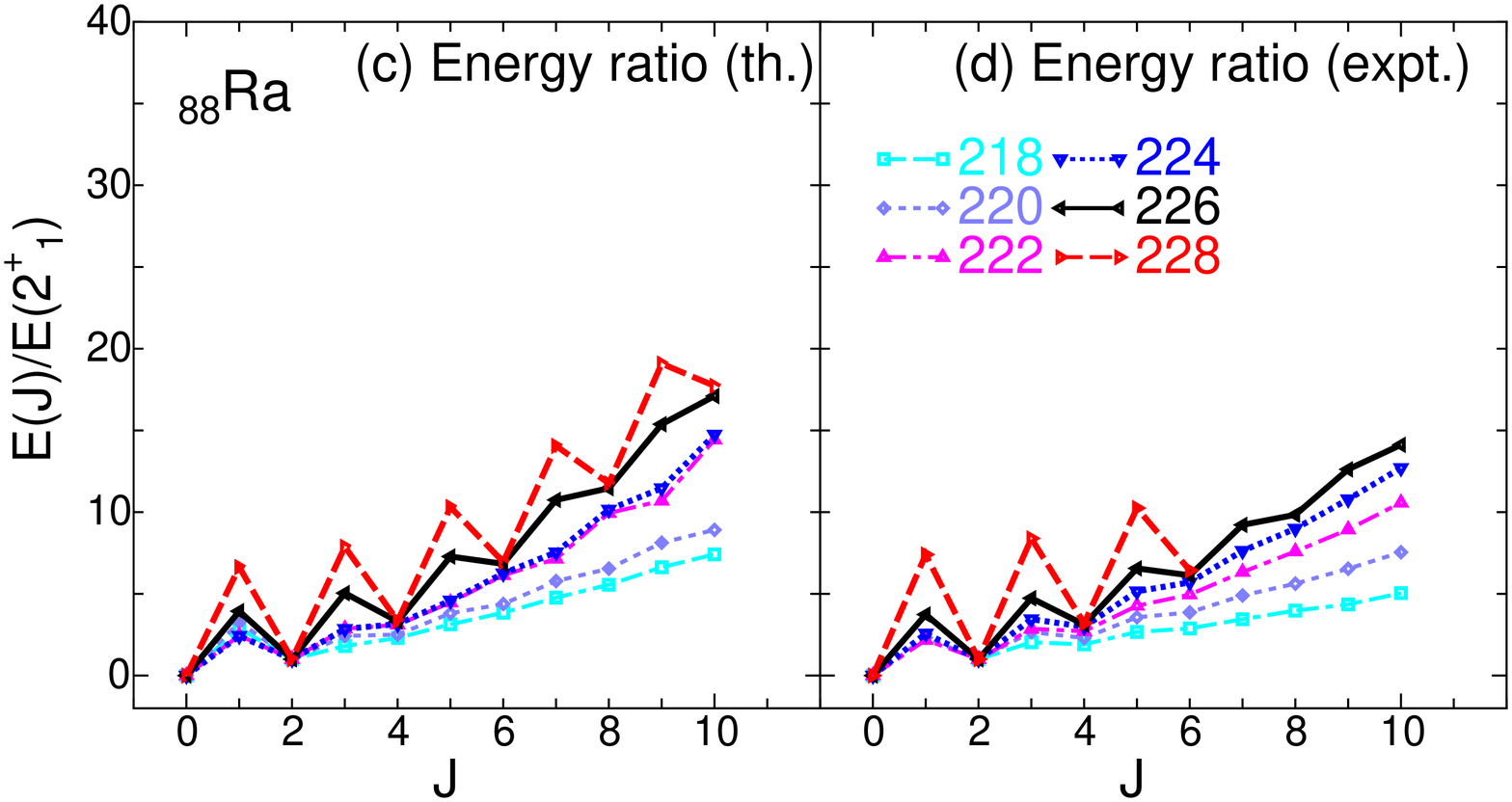} \\
\includegraphics[width=8.6cm]{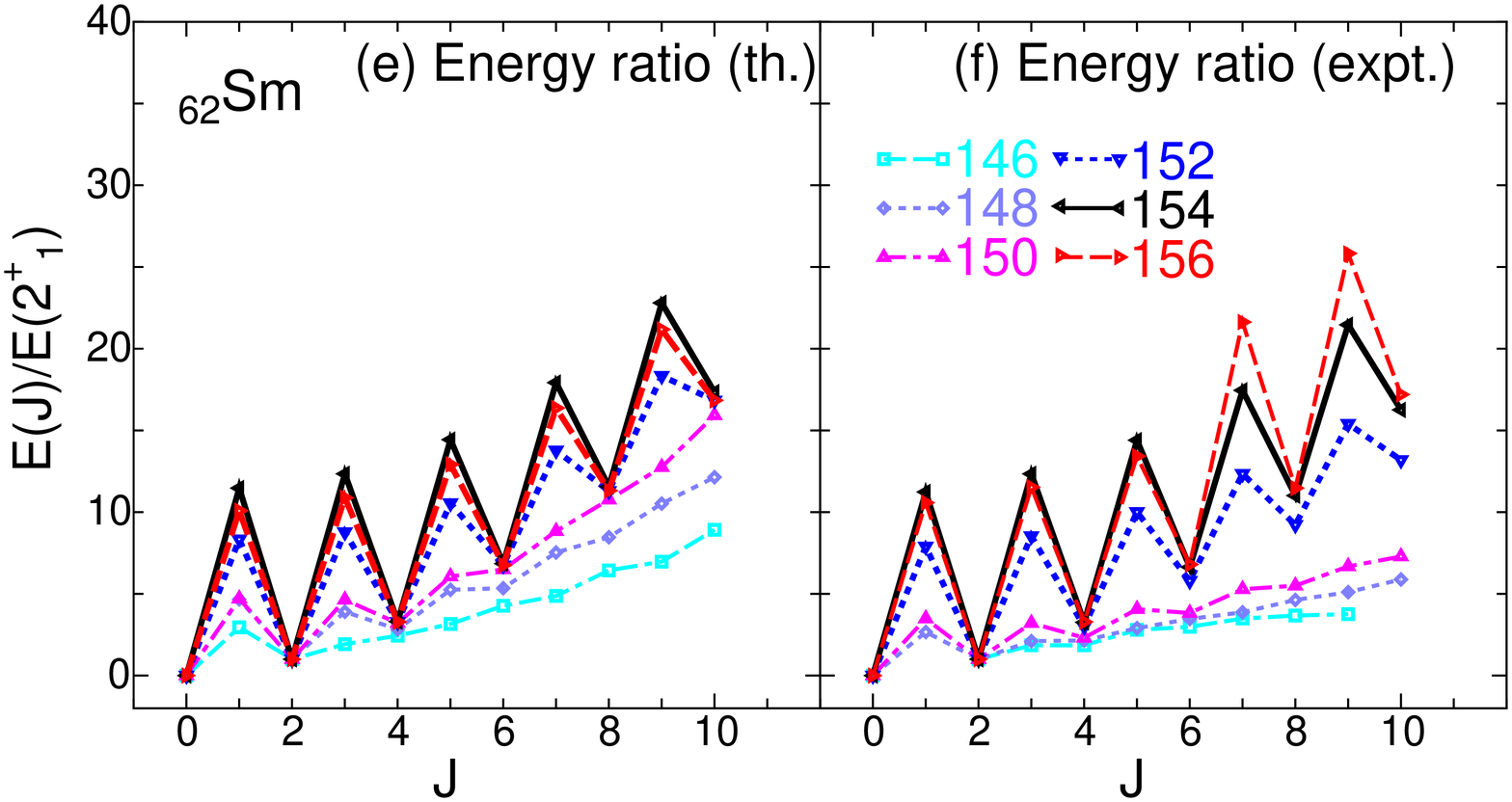} & 
\includegraphics[width=8.6cm]{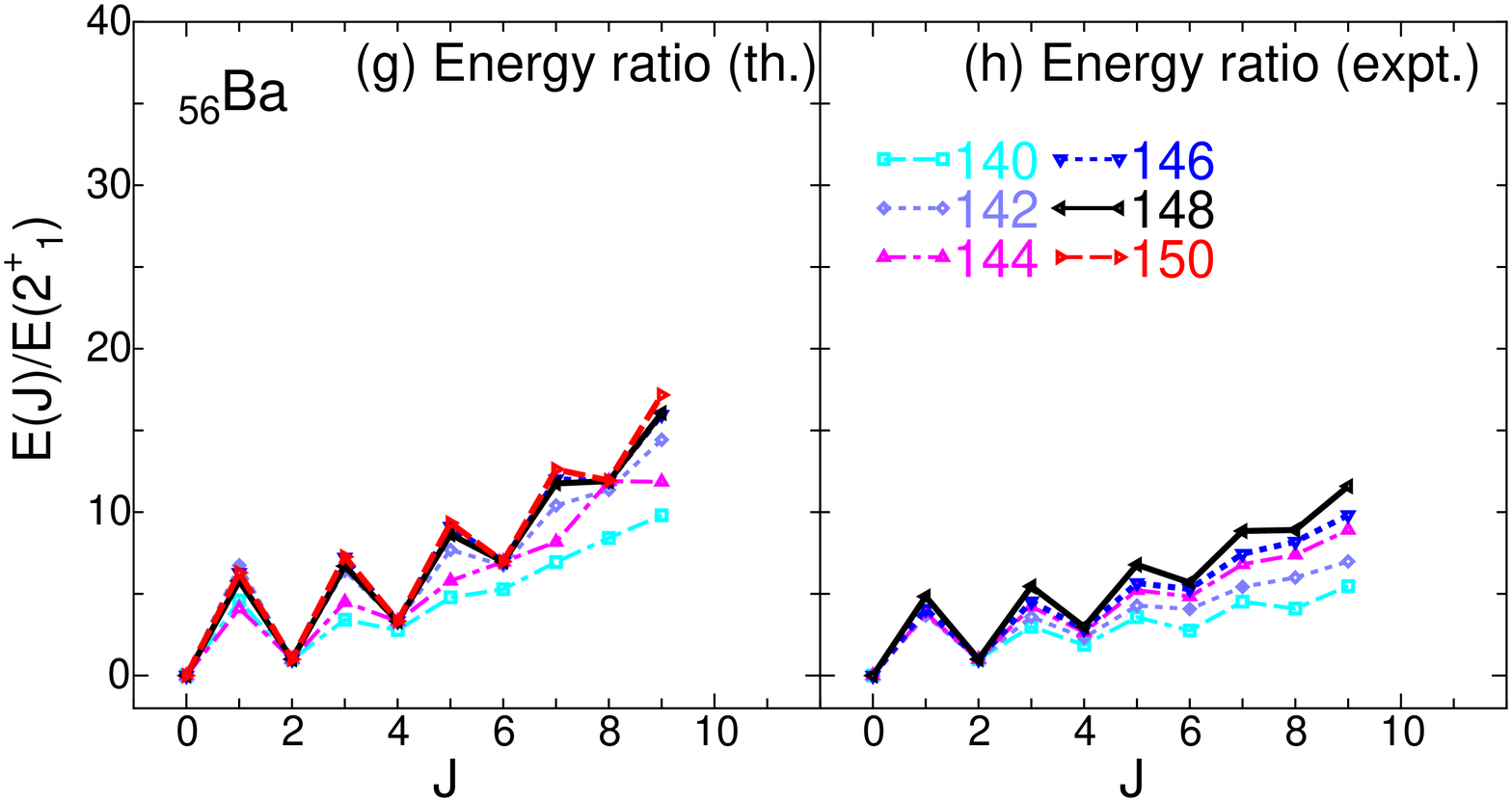}
\end{tabular}
\caption{(Color online) Theoretical and experimental \cite{data} energy ratios $E(J^{\pi})/E(2^{+}_{1})$ for
 states of the positive-parity ground-state band ($J$ even) and the lowest 
 negative-parity band ($J$ odd), as functions of the angular momentum
 $J$, for $^{220-232}$Th (a,b), $^{218-228}$Ra (c,d), 
 $^{146-156}$Sm (e,f), and $^{140-150}$Ba (g,h). }
\label{fig:staggering}
\end{center}
\end{figure*}

\subsection{Transition between static octupole deformation and octupole vibrations}

Another indication of a transition between octupole 
deformation and octupole vibrations for $\beta_3$-soft potentials 
is the odd-even staggering shown in Fig.~\ref{fig:staggering}. 
For both positive and negative parity, we plot the calculated ratios 
$E(J)/E(2^+_1)$ for the yrast states of $^{220-232}$Th, $^{218-228}$Ra,
 $^{146-156}$Sm and $^{140-150}$Ba nuclei as 
functions of the angular momentum $J$, in comparison to data. 
For Th isotopes with $A \leq {226}$ the 
odd-even staggering is negligible, indicating that positive- and 
negative-parity states belong to a single band, that is, the two bands 
are located close in energy. The staggering only 
becomes more pronounced starting from $^{228}$Th, and this 
means that negative-parity states form a separate rotational-like 
collective band built on the octupole vibrational state. 
One notices that the predicted staggering 
of yrast states is in very good agreement with data \cite{data}.
For the Ra isotopes the staggering becomes visible starting with $^{226}$Ra, 
but is much less pronounced compared to the 
Th nuclei with the same number of neutrons. 

Similarly, the odd-even staggering is more pronounced in Sm than in Ba.
For the Sm isotopes the effect becomes significant starting from $^{152}$Sm, but is
negligible for $A \leq {150}$. Note that $^{150}$Sm exhibits the deepest octupole
minimum in the deformation energy surface (Fig.~\ref{fig:pes_sm}). 
The staggering pattern in Sm isotopes, shown in
Figs.~\ref{fig:staggering}(e) and \ref{fig:staggering}(f), appears
slightly different from the one obtained for the Th isotopes
(Figs.~\ref{fig:staggering}(a) and \ref{fig:staggering}(b)). 
For the latter case the amplitude of the staggering keeps increasing 
with mass number, whereas it shows very little variation in 
the Sm chain starting from $^{152}$Sm. This feature reflects the fact 
that the $\beta_{3}$-softness of the potential does not show significant
variations between $^{152}$Sm and $^{156}$Sm. 
A similar trend, but with much less prominent staggering, is observed
for the Ba nuclei in Figs.~\ref{fig:staggering}(g) and
\ref{fig:staggering}(h). 

\begin{figure}[ctb!]
\begin{center}
\includegraphics[width=8.6cm]{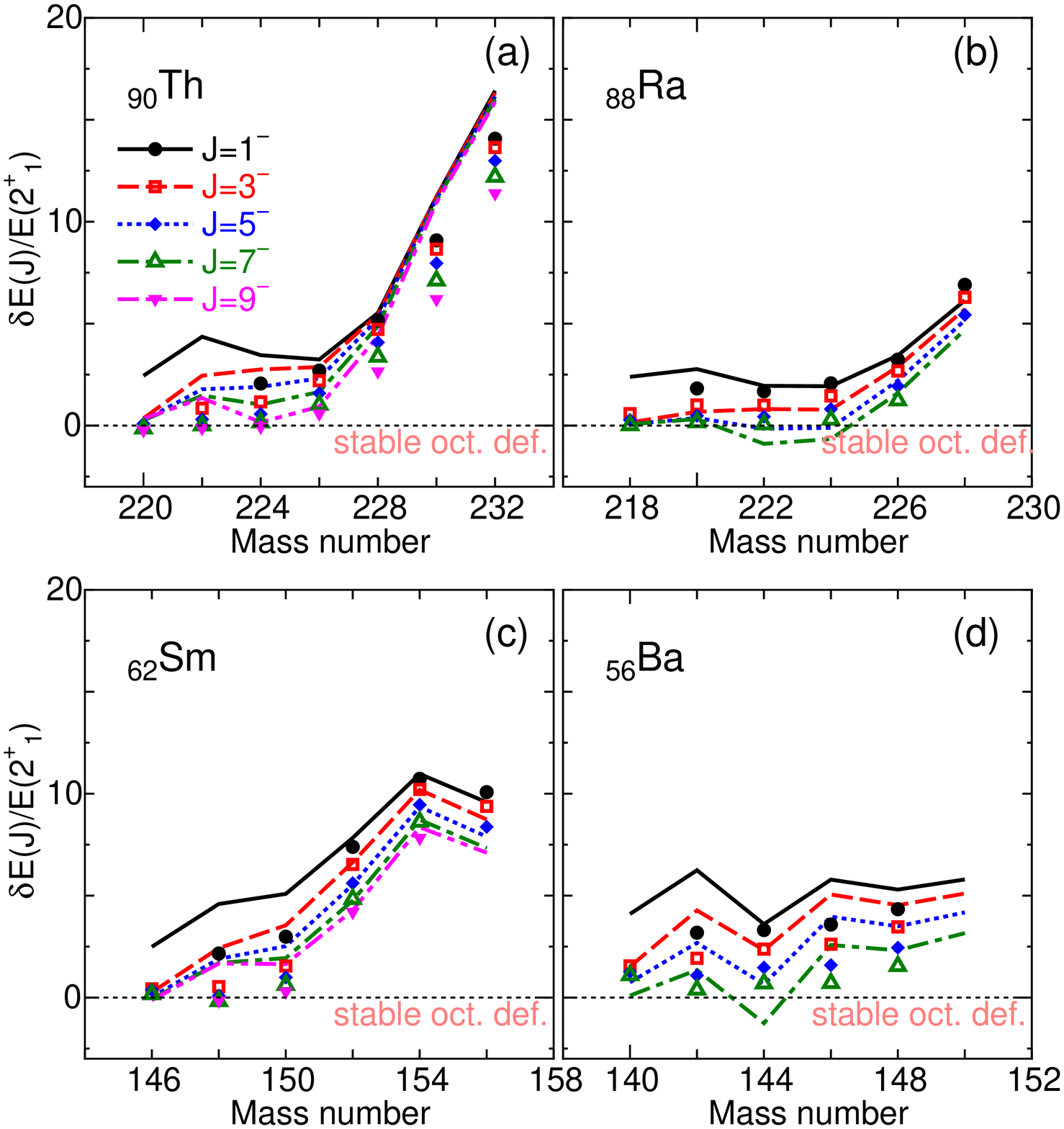}
\caption{(Color online) The energy displacement $\delta E(J)$ (defined
 in Eq.~(\ref{delta})) normalized with respect to the excitation energy
 of $2^{+}_{1}$, as a  function of the mass number. 
The limit of stable octupole deformation $\delta E(J)=0$ is indicated by
 the dotted horizontal line. 
Experimental data, represented symbols in each panel, are taken from
 Ref.~\cite{data}. } 
\label{fig:de}
\end{center}
\end{figure}

The emergence of octupole vibrational states can be further analyzed 
by computing the energy displacement $\delta E(J)$ between 
positive- and negative-parity bands, defined as \cite{naza84b}: 
\begin{eqnarray}
 \delta E(J)=E(J^{-})-\frac{\{E((J+1)^{+})+E((J-1)^{+})\}}{2}
 \label{delta}
\end{eqnarray}
For a stable octupole deformed nucleus, in which the positive- and
negative-parity yrast states form an alternating-parity band, 
$\delta E(J)$ should be approximately zero. 
The deviation from the limit $\delta E=0$ indicates the decoupling of states 
with positive and negative parity, thus pointing to the occurrence of 
octupole vibrational states. 

Figure \ref{fig:de} displays the calculated $\delta E(J)$ values for the
four isotopic chains, normalized with 
respect to the  excitation energy of $2^{+}_{1}$, as functions of the mass
number. The theoretical values are compared to data from Ref.~\cite{data}.
As one would expect, the $\delta E(J)/E(2^{+}_{1})$ values begin to show a 
significant increase starting from a specific isotope in the Th, Ra and Sm
chains, that is, from $^{226}$Th, $^{224}$Ra and $^{150}$Sm, for which a 
stable octupole deformation appears on the corresponding DD-PC1 energy
surface (cf. Figs.~\ref{fig:pes_th}, \ref{fig:pes_ra} and
\ref{fig:pes_sm}), and is also reflected in the plot of excitation
energies (cf. Figs.~\ref{fig:pspec}(a-c) and \ref{fig:nspec}(a-c)). 
For the Ba isotopes, on the other hand, no significant increase is 
observed in the variation of the $\delta E(J)/E(2^{+}_{1})$ value with
mass number. 

\subsection{Systematics of E1 and E3 transitions \label{sec:E1E3}}

\begin{figure*}[ctb!]
\begin{center}
\includegraphics[width=17cm]{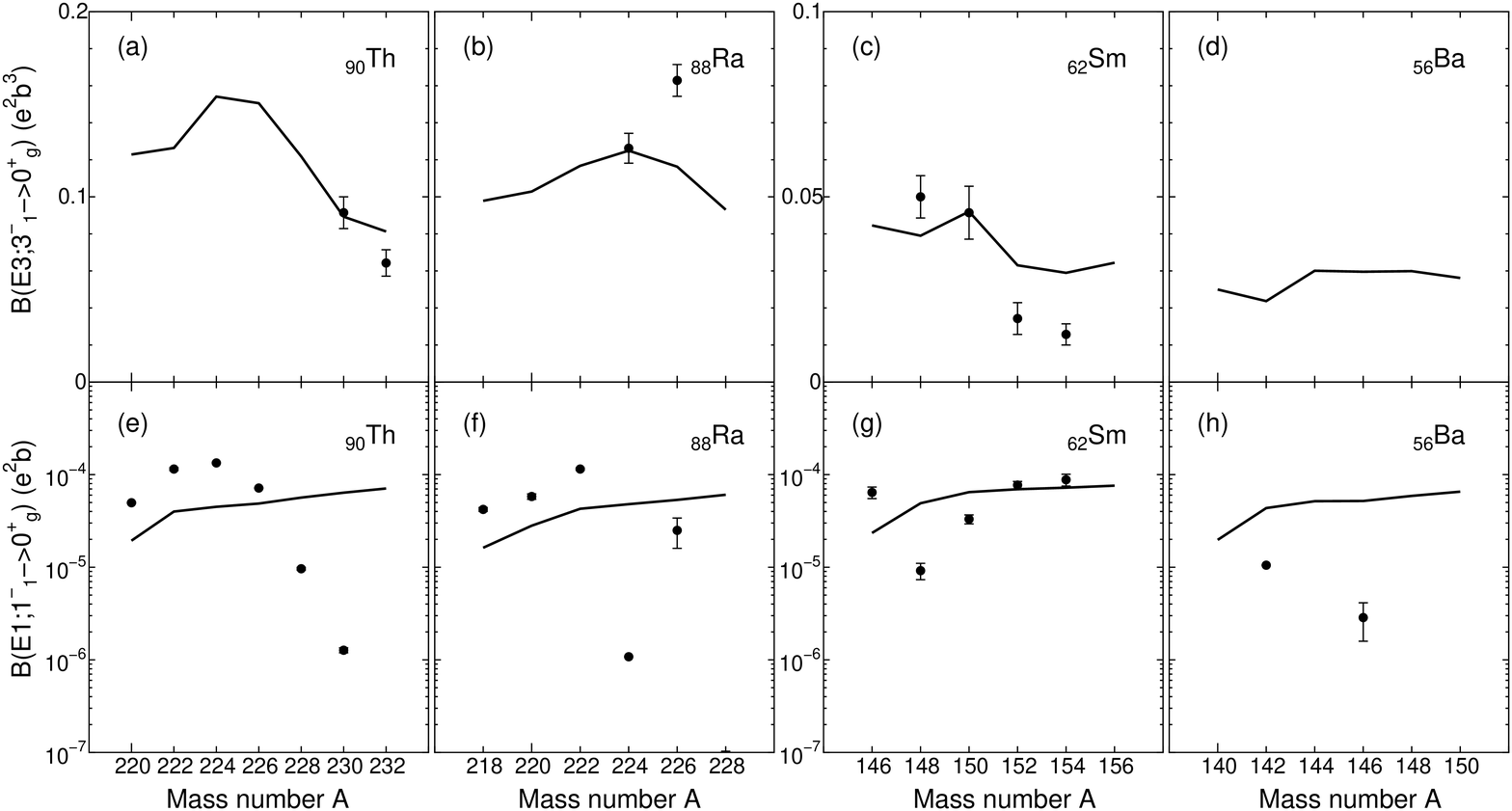}
\caption{(Color online) Isotopic dependence of the $B$(E3;
 $3^{-}_{1}\rightarrow 0^{+}_{1}$) (panels (a) to (d)) and $B$(E1;
 $1^{-}_{1}\rightarrow 
 0^{+}_{1}$ ) (panels (e) to (h)) values for $^{220-232}$Th,
 $^{218-228}$Ra, $^{146-156}$Sm, and $^{140-150}$Ba. 
 The vertical axes for panels (e)-(h) are in
 logarithmic scale. Solid lines connect calculated values, symbols
 denote data taken from Refs.~\cite{butler91,butler96,metzger76,data,gaffney13,kibedi02,wollersheim93}}
\label{fig:E1E3}
\end{center}
\end{figure*}

In addition to the comparison between the calculated and the
experimental energy spectra, in this section we examine the E3 and
E1 properties, which also provide signatures for the onset of
octupole deformation. 
In Fig.~\ref{fig:E1E3} we display the isotopic dependence of the
$B$(E3; $3^{-}_{1}\rightarrow 0^{+}_{1}$) (panels (a) to (d)) and
$B$(E1; $1^{-}_{1}\rightarrow 0^{+}_{1}$) (panels (e) to (h)) values
for $^{220-232}$Th, $^{218-228}$Ra, $^{146-156}$Sm, and $^{140-150}$Ba,
in comparison to available data
\cite{butler91,butler96,metzger76,data,gaffney13,kibedi02,wollersheim93}. 

The E3 transition strength, in particular, can be regarded as a good
measure of octupole collectivity \cite{butler96}. 
The upper part of Fig.~\ref{fig:E1E3} (panels from (a) to (d)) shows that 
the present calculation yields 
$B$(E3; $3^{-}_{1}\rightarrow 0^{+}_{1}$) values that are consistent
with the experimental trend. 
In the Th, Ra and Sm isotopic chains the theoretical  
$B$(E3; $3^{-}_{1}\rightarrow 0^{+}_{1}$) values reach a maximum for
$^{226}$Th, $^{224}$Ra and $^{150}$Sm, respectively, that is, for isotopes 
in which the octupole deformation minimum is most prominent (Figs.~\ref{fig:pes_th},
\ref{fig:pes_ra} and \ref{fig:pes_sm}). 
In the Ba chain the theoretical $B$(E3) values remain almost
constant with increasing neutron number, reflecting the fact that no 
shape transition is predicted by the calculation. 

For all the considered isotopic chains, in contrast to the systematics of
the energy surfaces, the excitation spectra, and the
$B$(E3;$3^{-}_{1}\rightarrow 0^{+}_{1}$) values, the 
calculated $B$(E1; $1^{-}_{1}\rightarrow 0^{+}_{1}$) values exhibit only
a monotonic increase as functions of the mass number. 
The corresponding experimental values seem to suggest more significant
structural changes in the isotopic sequences. 
In fact, only for the $^{148-154}$Sm nuclei the present calculation
qualitatively reproduces the experimental trend in heavier isotopes (Fig.~\ref{fig:E1E3}(g)). 
On the other hand, much larger $B$(E1; $1^{-}_{1}\rightarrow 0^{+}_{1}$) 
values, by $\approx 10^{2-4}$ compared to experimental ones, are
predicted for the heavier Th and Ra isotopes, as well as for the Ba
nuclei.  

There are several possible reasons for the systematic discrepancy 
between theoretical and empirical $B$(E1) values. 
Firstly, the choice of the fixed effective charge $e_{1}=0.01$
$e$b$^{1/2}$ \cite{babilon05} might be too restricted and, in principle,
one could allow a variation of the effective charge with mass number. 
For the Ra isotopes (Fig.~\ref{fig:E1E3}(f)), for instance, a sudden
decrease of the experimental $B$(E1; $1^{-}_{1}\rightarrow 0^{+}_{1}$)
value from $^{222}$Ra to $^{224}$Ra could reflect a complex change of
structure. 
Such an effect could easily be absorbed in the variation of the
effective charge. 
Moreover, the present model considers only isoscalar properties, that
is, there is no distinction between proton and neutron bosons, whereas
for E1 transitions isovector components could play an important role. 
Secondly, the form of the E1 operator $\hat T^{\textnormal{(E1)}}$ in
Eq.~(\ref{eq:E1}) may need to be extended. 
In fact, it has been shown that in the $sdf$ IBM two body terms
should be included in the E1 operator \cite{barfield89}. 
Such an extension invokes additional parameters for the E1 operator, 
but these would be difficult to determine uniquely. 
Most likely, however, the model space of $s$, $d$, and $f$ bosons is by
construction not sufficient to describe the E1 systematics. 
This implies that the inclusion of the $p$
($J^{\pi}=1^{-}$) boson might be necessary. 
Actually, as shown by Otsuka \cite{taka86}, intrinsic wave functions of 
quadrupole-octupole deformed nuclei could contain a large fraction of the
dipole nucleon pair. An extension of our model along these lines 
could be an interesting subject for a future study. 
 
In some previous phenomenological $sdf$ (or $spdf$) IBM
studies, the $B$(E1; $1^{-}_{1}\rightarrow 0^{+}_{1}$) values for Sm
isotopes were reproduced slightly better than in the present calculation 
(see, e.g., \cite{scholten78}).  
In Ref.~\cite{scholten78}, however, two-body terms have been added
to the E1 operator. 
The recent two-dimensional (2D) GCM calculation for Sm isotopes \cite{rayner12}, 
based on the Gogny effective interaction, has reproduced 
the experimental $B$(E1; $1^{-}_{1}\rightarrow 0^{+}_{1}$) values with a
more or less similar level of agreement as the present work.  
Other beyond-mean-field studies of Ra and Ba isotopes
\cite{robledo10}, and Ra-Th isotopes \cite{robledo13}, that used 
the collective Hamiltonian approach based on
the Gogny D1S and the BCP functional, and the 2D GCM approach with the
Gogny D1S and D1M interactions, respectively, 
have reported results for  $B$(E1; $1^{-}_{1}\rightarrow 0^{+}_{1}$)
that are more consistent with the experimental systematics.

\subsection{Detailed level schemes}

\begin{figure*}[ctb!]
\begin{center}
\includegraphics[width=12.0cm]{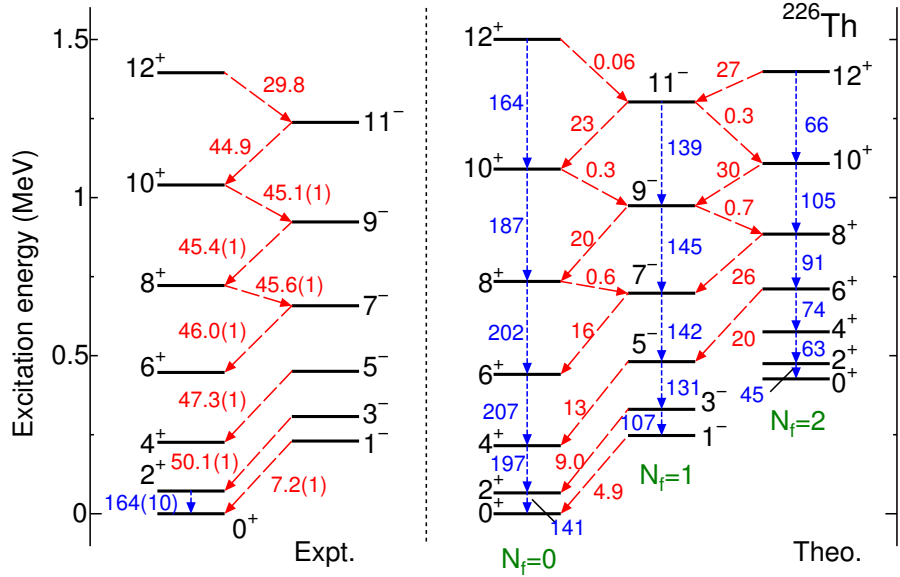}
\caption{(Color online) Partial level scheme of $^{226}$Th. 
The theoretical low-lying spectra, in-band $B$(E2) values (in 
 Weisskopf units, dotted arrows), and
 inter-band $B$(E1) values (in units of $e^{2}$b$\times 10^{-5}$,  
dashed arrows), are compared to
available data \cite{data,butler91,butler96}. } 
\label{fig:226Th}
\end{center}
\end{figure*}

\begin{figure*}[ctb!]
\begin{center}
\includegraphics[width=15.0cm]{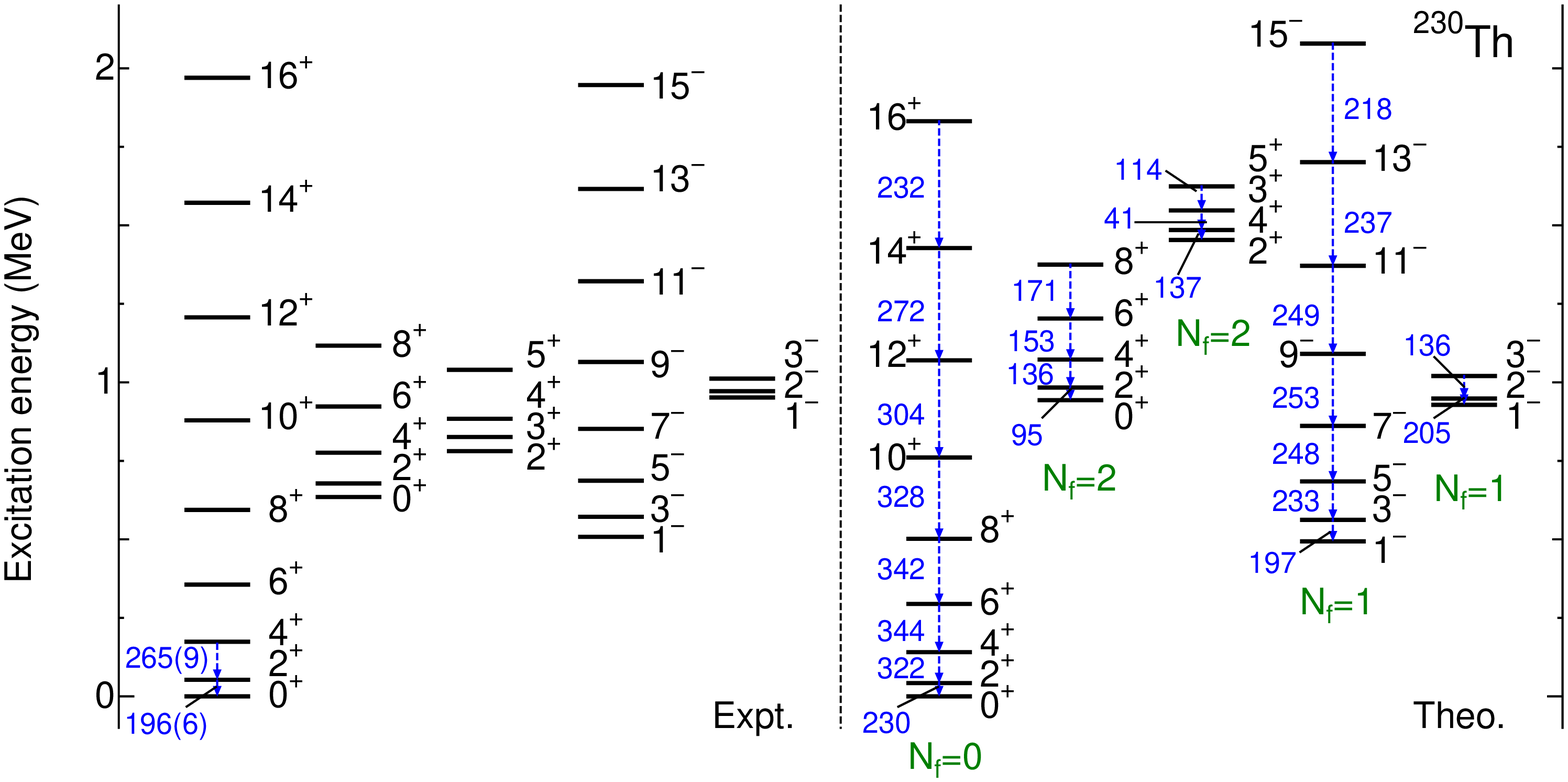}
\caption{(Color online) 
Partial level scheme of $^{230}$Th. 
The theoretical low-lying spectra and in-band $B$(E2) values (in 
 Weisskopf units, dotted arrows) are compared to
available data \cite{data}. 
}
\label{fig:230Th}
\end{center}
\end{figure*}

To illustrate in more detail the level of quantitative agreement between
our microscopic model calculation and data, 
we analyze the low-lying energy spectra of positive- and  
negative-parity states, the $B$(E2) values for in-band transitions, and 
the inter-band $B$(E1) values, for the octupole-deformed nucleus 
$^{226}$Th (Fig.~\ref{fig:226Th}), and the
octupole-soft nucleus $^{230}$Th (Fig.~\ref{fig:230Th}). 

The level scheme of $^{226}$Th (Fig.~\ref{fig:226Th})
shows that the lowest negative-parity band (composed of 
one-$f$-boson states) is located close in energy to the ground-state
positive-parity band (comprised of positive-parity bosons only, that is,
$N_{f}=0$). 
One notices that the lowest positive- and negative-parity bands form a 
single, alternating-parity band, starting with angular momentum $J = 5$. 
Overall, a very good agreement between theory and experiment is obtained for
the excitation spectrum of $^{226}$Th. 
Strong E1 transitions are predicted from odd-$J$ states of the
negative-parity band to the even $(J-1)$ states of the ground-state
positive-parity band, in agreement with data. However, the calculated E1 
transitions in the inverse direction (from even-$J$ to the 
odd $(J-1)$ states) are considerably weaker, whereas the experimental
values are of the same order as for the former transitions. 
The opposite systematics is predicted for the E1 transitions between 
the negative-parity band $N_f =1$ and the second-lowest positive-parity
band $N_f =2$ (built on two-$f$ boson states), that is, the E1
transitions from even-$J$ states to the odd $(J -1)$  states dominate. 

For the nucleus $^{230}$Th we notice in Fig.~\ref{fig:230Th} 
that the present calculation reproduces very well the experimental 
\cite{data} energy levels of the positive-parity
ground-state ($K^{\pi}=0^{+}$) band, 
including the E2 transition strengths, and those of
the two lowest ($K^{\pi}=0^{-}$ and $1^{-}$) negative parity
bands. Compared to $^{226}$Th, the band $K^{\pi}=0^{-}$ 
is found at much higher excitation energy, consistent with the 
picture of octupole vibrations. However, the theoretical 
positive-parity bands built on the
$0^{+}_{2}$ state and the $2^{+}_{3}$ state, are predicted 
high above their experimental counterparts. 
The reason could be the too large value for the quadrupole-quadrupole 
interaction strength $\kappa_{2}$, which is relevant for the bandhead
energies of the side bands. 
In fact, the present $\kappa_{2}$ value is more than three times larger 
than the one used in the previous IBM phenomenological study
\cite{cottle98}. 
This value reflects the pronounced quadrupole deformation minimum 
predicted by the RHB energy surface. 
E2 transitions between the states of
the positive-parity ground-state band and the two 
side bands built on $0^{+}_{2}$ and $2^{+}_{3}$ vanish, and this means
there is no mixing between the corresponding configurations. 
In fact, for $^{230}$Th the IBM model predicts that the states in the
ground-state are composed of $s$ and $d$ bosons only, that is, 
$N_{f}=0$, whereas those of 
the two positive-parity side bands are built on two-$f$ boson states (Fig.~\ref{fig:230Th}). 
The low-lying negative-parity states are, of course, of one-$f$ boson nature.

\subsection{The nucleus $^{224}$Ra}

Finally, we compare the results of the present microscopic calculation 
with very recent data for the octupole deformed nucleus $^{224}$Ra, 
obtained in the Coulomb excitation experiment of Ref.~\cite{gaffney13}. 
Table~\ref{tab:224Ra-T} lists all the experimental $B(\textnormal{E}\lambda)$
values included in Ref.~\cite{gaffney13}, in comparison with our model results.
For the E2 transition rates a very good agreement
is obtained between the experimental and the calculated values, possibly with
the exception of the $5^{-}\rightarrow 3^{-}$ transition which is
underestimated in the calculation. 
We also note the nice agreement of the $B(\textnormal{E}3; J\rightarrow
J-3)$ values, but the calculated $B(\textnormal{E}3; 1^{-}\rightarrow
2^{+})$ is considerably smaller than the experimental value. 
On the other hand, the theoretical $B(\textnormal{E}1)$ values are 
systematically too large, typically by $10^{1} - 10^{2}$, when compared 
with data. Possible reasons for this discrepancy have been addressed in 
Sec.~\ref{sec:E1E3}. 

\begin{figure}[ctb!]
\begin{center}
\includegraphics[width=8.0cm]{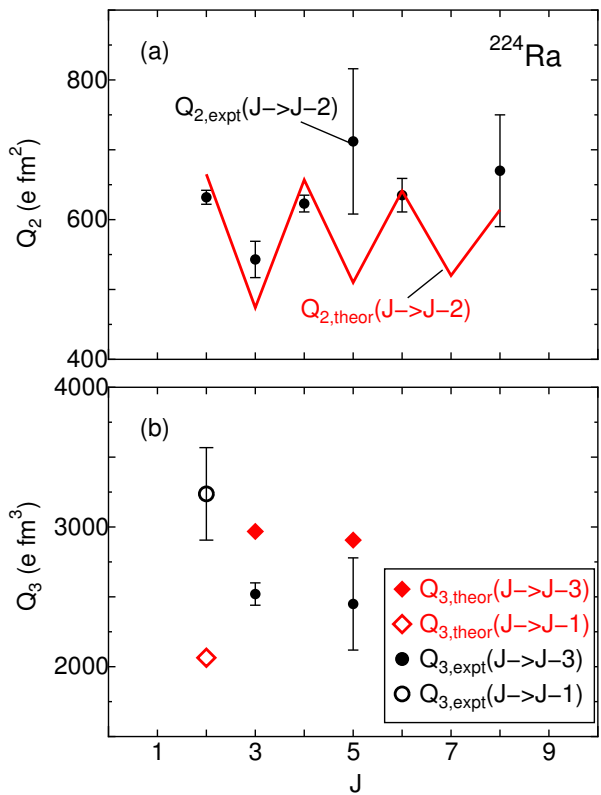}
\caption{(Color online) Comparison between the experimental
 \cite{gaffney13} and calculated quadrupole and octupole intrinsic
 moments of $^{224}$Ra, as functions of the angular momentum $J^{\pi}$ 
 ($\pi=+1$ and $-1$ for $J$
 even and odd, respectively). }
\label{fig:224Ra-Q}
\end{center}
\end{figure}

In Fig.~\ref{fig:224Ra-Q} we plot the E2 and E3 intrinsic moments determined 
from the $B(\textnormal{E}2)$ and the $B(\textnormal{E}3)$ values listed in
Table~\ref{tab:224Ra-T}, using the relation given in Eq.~(\ref{eq:qt}). 
One notes a staggering in the calculated $Q_2(J\rightarrow J-2)$ values, 
and their average value $\approx 600$ $e$fm$^{2}$ is consistent 
with the measured value \cite{gaffney13}. 
The same observation applies to the octupole intrinsic moment $Q_3$. 
The mean value of the three $Q_3$ moments is approximately 2500 $e$fm$^{3}$, 
in agreement with experiment.


\begin{table}[cb!]
\caption{\label{tab:224Ra-T} Comparison between experimental
 \cite{gaffney13} and theoretical $B(\textnormal{E}\lambda)$ values
 for transitions in $^{224}$Ra (in Weisskopf units). All transitions shown in the table,
 except for the E2 transition from the band-head of the band built on
 the $2^{+}_{2}$ state to the
 $0^{+}$ ground state, are between yrast states. }
\begin{center}
\begin{tabular}{ccc}
\hline\hline
\textrm{} &
\textrm{Expt. (W.u.)}&
\textrm{Theor. (W.u.)}\\
\hline
$B({\textnormal{E2}};2^{+}_{}\rightarrow 0^{+}_{})$ & 98$\pm$3 & 109 \\
$B({\textnormal{E2}};3^{-}_{}\rightarrow 1^{-}_{})$ & 93$\pm$9 & 71 \\
$B({\textnormal{E2}};4^{+}_{}\rightarrow 2^{+}_{})$ & 137$\pm$5 & 152 \\
$B({\textnormal{E2}};5^{-}_{}\rightarrow 3^{-}_{})$ & 190$\pm$60 & 97 \\
$B({\textnormal{E2}};6^{+}_{}\rightarrow 4^{+}_{})$ & 156$\pm$12 &
	 159 \\
$B({\textnormal{E2}};8^{+}_{}\rightarrow 6^{+}_{})$ & 180$\pm$60 &
	 153 \\
$B({\textnormal{E2}};2^{+}_{2}\rightarrow 0^{+}_{})$ & 1.3$\pm$0.5
     & 0 \\
$B({\textnormal{E3}};3^{-}_{}\rightarrow 0^{+}_{})$ & 42$\pm$3 & 42 \\
$B({\textnormal{E3}};1^{-}_{}\rightarrow 2^{+}_{})$ & 210$\pm$40 & 85 \\
$B({\textnormal{E3}};3^{-}_{}\rightarrow 2^{+}_{})$ & $<$600 & 46 \\
$B({\textnormal{E3}};5^{-}_{}\rightarrow 2^{+}_{})$ & 61$\pm$17 & 61 \\
$B({\textnormal{E1}};1^{-}_{}\rightarrow 0^{+}_{})$ & $<5\times 10^{-5}$
     & 2.0$\times 10^{-3}$ \\
$B({\textnormal{E1}};1^{-}_{}\rightarrow 2^{+}_{})$ & $<1.3\times 10^{-4}$
     & 1.1$\times 10^{-3}$ \\
$B({\textnormal{E1}};3^{-}_{}\rightarrow 2^{+}_{})$ & $3.9^{+1.7}_{-1.4}\times 10^{-5}$
     & 3.7$\times 10^{-3}$ \\
$B({\textnormal{E1}};5^{-}_{}\rightarrow 4^{+}_{})$ &  $4^{+3}_{-2}\times 10^{-5}$
     & 5.0$\times 10^{-3}$ \\
$B({\textnormal{E1}};7^{-}_{}\rightarrow 6^{+}_{})$ & $<3\times 10^{-4}$
     & 5.8$\times 10^{-3}$ \\
\hline\hline
\end{tabular}
\end{center}
\end{table}

\section{Summary and concluding remarks \label{sec:summary}}

In the present study we have performed a microscopic analysis of octupole
shape transitions in four isotopic chains characteristic for two regions
of octupole deformation and collectivity: Th, Ra, Sm and Ba. 
As a microscopic input we have used the axial quadrupole-octupole 
deformation energy surfaces calculated employing the relativistic 
Hartree-Bogoliubov model based on the universal energy density functional
DD-PC1 which has, of course, not been specifically adjusted to octupole deformed nuclei. 
By mapping the deformation-constrained microscopic energy surfaces onto
the equivalent $sdf$ IBM Hamiltonian, that is, onto the corresponding expectation
values of the IBM Hamiltonian in boson coherent states, the Hamiltonian
parameters have been determined without any specific adjustment to experimental spectra. 
The mapped $sdf$ IBM Hamiltonian has been used to calculate 
low-energy spectra and transition rates for both positive- and
negative-parity states of the four sequences of isotopes. 
The systematics of the axially-symmetric ($\beta_2,\beta_3$) energy
surface (from Figs.~\ref{fig:pes_th} to \ref{fig:pes_ba_mapped}), the
average value of the octupole deformation 
$\langle\beta_{3}\rangle$ (Fig.~\ref{fig:bet3}), the variation of the resulting Hamiltonian parameters
(Fig.~\ref{fig:para}), the calculated excitation spectra
(Figs.~\ref{fig:pspec}, \ref{fig:nspec}, \ref{fig:staggering} and \ref{fig:de}), and the E3 transition
rates (Fig.\ref{fig:E1E3}), show a consistent picture of evolution of octupole correlations 
in the two regions of medium-heavy and heavy nuclei.

The microscopic DD-PC1 energy surfaces suggest a transition from 
non-octupole (quadrupole vibrational) to stable octupole deformed,
and to octupole vibrations characteristic for $\beta_{3}$-soft
potentials, in the Th, Ra and Sm isotopic chains. 
Among all nuclei considered in this work, the Th isotopes 
appear to present the best case 
for the evolution of octupole correlations.
The calculated excitation spectra of all considered isotopes exhibit 
a decrease in energy of states in the positive-parity ground-state band 
with the increase of the neutron number (Fig.~\ref{fig:pspec}), in agreement 
with available data. The evolution of the ground-state bands with neutron number 
clearly shows the transition from spherical vibrators to quadrupole deformed rotors. 
For the Th and the Ra isotopes (Fig.~\ref{fig:nspec}) the states in the lowest 
negative-parity band display a parabolic energy dependence 
on the mass number, with energy minima at 
$^{226}$Th and $^{224}$Ra that correspond to stable octupole
deformations, consistent with the evolution of the microscopic energy surfaces 
(Figs.~\ref{fig:pes_th} and \ref{fig:pes_ra}). 
The approximate parabolas of low-spin negative-parity states can be
identified as signatures of the transition from stable octupole
deformation to octupole vibrations. 
In the case of Sm and Ba isotopes, the evolution of the lowest
negative-parity band is much more moderate, and the excitation energies
of negative-parity states show 
almost no variation for heavier isotopes. This means that the
$\beta_{3}$-soft octupole potentials do not change with neutron number,
and the spectra are those of octupole vibrators.

For most nuclei considered in the present analysis the IBM model based on 
microscopic deformation energy surfaces produces results in a reasonable
agreement with available experimental spectroscopic
properties. Nevertheless, considerable disagreement has been found for 
higher-spin states, particularly for those nuclei for which the space of boson states appear to be too restricted.
In particular, the present model calculation has apparently failed in the 
description of the E1 transition rates (Figs.~\ref{fig:E1E3}(e)-(h) and Fig.~\ref{fig:226Th}). 
We have discussed several possible reasons: (i) the oversimplified 
parametrization of the E1 effective charge, (ii) the restricted 
form of the adopted E1 operator, and (iii) the insufficient $sdf$ model space that 
could be extended by the inclusion of other types of bosons. 
A complete analysis of E1 systematics will be the topic of a future study.

\begin{acknowledgements}
The authors would like to thank R. V. Jolos and J. Zhao for useful
 discussions. 
K. N. acknowledges the support by the Marie Curie
 Actions grant within the Seventh Framework Program of the 
European Commission under Grant 
 No. PIEF-GA-2012-327398. 
Calculations were partly performed 
on the ScGrid of the 
Supercomputing Center, Computer Network Information Center of Chinese Academy of Sciences. 
\end{acknowledgements}

\bibliography{refs}

\end{document}